\documentclass[floats,aps,amsfonts,notitlepage,superscriptaddress,twocolumn,eqsecnum,nofootinbib]{revtex4-1}

\usepackage[retainorgcmds]{IEEEtrantools}

\usepackage{ifpdf}
\usepackage[utf8]{inputenc}
\usepackage{graphicx}
\usepackage[colorlinks]{hyperref}
\usepackage{amsmath, amsthm, amssymb}
\usepackage{multirow}
\usepackage{url}
\usepackage{subfigure}
\usepackage{color}
\usepackage[usenames,dvipsnames,svgnames,table]{xcolor}
\usepackage{tabularx}
\usepackage{mathtools} 

\allowdisplaybreaks[1]

\definecolor{CiteColor}{rgb}{0,0.5,0}
\hypersetup{citecolor=CiteColor}
\definecolor{RefColor}{rgb}{0.55,0,0}
\hypersetup{linkcolor=RefColor}
\definecolor{darkgreen}{rgb}{0.2,0.7,0.2}

\newcommand{\bx}{{\boldsymbol{x}}}

\newcommand{\ab}{{\bar{a}}}
\newcommand{\bb}{{\bar{b}}}
\newcommand{\cb}{{\bar{c}}}
\newcommand{\db}{{\bar{d}}}
\newcommand{\eb}{{\bar{e}}}
\newcommand{\fb}{{\bar{f}}}
\newcommand{\gb}{{\bar{g}}}
\newcommand{\hb}{{\bar{h}}}
\newcommand{\ib}{{\bar{i}}}
\newcommand{\jb}{{\bar{j}}}
\newcommand{\kb}{{\bar{k}}}

\newcommand{\xb}{{\bar{x}}}

\newcommand{\rb}{{\bar{r}}}

\newcommand{\tb}{{\bar{t}}}

\newcommand{\thz}{\bar{\theta}}

\newcommand{\ur}{u^{\bar{r}}}
\newcommand{\uth}{u^{\bar{\theta}}}

\newcommand{\phb}{{\bar{\phi}}}
\newcommand{\thb}{{\bar{\theta}}}

\newcommand{\zrho}{{\rho}}

\newcommand{\adv}{{(adv)}}
\newcommand{\ret}{{(ret)}}
\newcommand{\sing}{{(S)}}
\newcommand{\reg}{{(R)}}

\newcommand{\PhiS}{\Phi^{\rm \sing}}

\newcommand{\lnpow}[1]{[\scalebox{0.85}{-}#1]}

\DeclareMathOperator{\sgn}{{\text{sgn}}}

\begin{document}
\title{Regularization of a scalar charged particle for generic orbits in Kerr spacetime}

\author{Anna Heffernan}
\affiliation{IAC3--IEEC,  Universitat de les Illes Balears, E-07122 Palma, Spain}
\affiliation{${}^*$Department of Physics, University of Guelph, 50 Stone Rd E., Guelph, Ontario, Canada
N1G 2W1}
\affiliation{${}^*$Perimeter Institute of Theoretical Physics, 31 Caroline Street N., Waterloo, Ontario, Canada, N2L 2Y5}
\affiliation{${}^*$Department of Physics, University of Florida, 2001 Museum Road, Gainesville, FL 32611-8440, USA}
\affiliation{${}^*$School of Mathematics and Statistics, University College Dublin, Belfield, Dublin 4, Ireland}


\begin{abstract}
${}^*$Previous affiliations where part of this work was carried out.\\
A scalar charged particle moving in a curved background spacetime will emit a field affecting its own motion; the resolving of this resulting motion is often referred to as the self-force problem. This also serves as a toy model for the astrophysically interesting compact-body binaries, extreme mass ratio inspirals, targets for the future space-based gravitational wave detector, LISA. In the modeling of such systems, a point-particle assumption leads to problematic singularities which need to be safely removed to solve for the motion of the particle regardless of the scenario; scalar, electromagnetic or gravitational. Here, we concentrate on a scalar charged particle and calculate the next order of the Detweiler-Whiting singular field and its resulting regularization parameter when employing the mode-sum method of regularization. This enables sufficiently faster self-force calculations giving the same level of accuracy with significantly less $\ell$ modes. Due to the similarity of the governing equations, this also lays the groundwork for similar calculations for an electromagnetic or mass charged particle in Kerr spacetime and has applications in other regularization schemes like the effective source and matched expansion.
\end{abstract}

\maketitle


\section{Introduction} \label{sec:intro}
The scalar model of the two-body problem, a legitimate problem in itself, also serves as a toy model towards solving the motion of two massive particles. The motion of two masses in a vacuum has garnered a new generation of attention due to the rather recent field of gravitational wave astronomy. Current ground-based detectors, Advanced LIGO \cite{LIGOScientific:2014pky}, Advanced VIRGO \cite{VIRGO:2014yos}, GEO600 and KAGRA \cite{KAGRA:2018plz}, are live, and to date have produced a catalogue of 90 detections \cite{LIGOScientific:2020ibl, LIGOScientific:2021djp}. Indeed the future ESA-led space-based detector, LISA \cite{2017arXiv170200786A} is due to launch in 2034 opening the window to a new frequency band of gravitational waves.

A key source for LISA, with exceptional science reward, are Extreme Mass Ratio Inspirals (EMRIs) \cite{Berry:2019wgg} -- when a `small' stellar mass compact body falls into the grasp of a massive black hole $(10^7-10^9 M_{\odot})$. LISA is expected to see anywhere from several to thousands of these during its mission \cite{Gair:2017ynp, Babak:2017tow} through multiple formation channels \cite{Pan:2021oob}. To enable detection and disentangle the cacophony of signals expected from LISA, one must have knowledge of the possible waveforms. In modeling EMRI's, current numerical relativity has not quite got to the mass ratios required ($\sim 1/100$) \cite{Sperhake:2011ik} while post-Newtonian approximations will break down as the particles approach \cite{LeTiec:2014oez}, leaving self-force as the current state of the art.

In the self-force regime, one perturbs the Einstein field equations in the mass ratio; at zero order, the particle follows a geodesic of the background spacetime, usually taken as Schwarzschild (nonspinning) or Kerr (spinning). At first order, the particle's effects on its local curvature result in the particle moving off this geodesic, hence the so-called self-force. When modeling the self-force it has become standard to use a point-particle description; although point particles do not exist in nature, one considers a small enough distribution of mass or charge so that, to the desired order, the point particle suffices \cite{Quinn:1996am, Mino:Sasaki:Tanaka:1996}.

One issue that immediately arises, from the point-particle assumption, is the singular structure of the potential - singularities are obviously not ideal for numerics, as well as being unnatural. For efficient computation of the nonsingular or regular potential and resulting equations of motion, one must safely remove this singularity. The first successful regular-singular split implementation was produced by Barack and Ori \cite{Barack:1999wf} via a mode-sum decomposition; however, their regular and singular fields were not independent solutions to the homogeneous and sourced wave equations respectively. This more physically intuitive concept was later introduced by Detweiler and Whiting \cite{Detweiler:2002mi}, and has been the practical gold standard since (and what we use here). The full physical picture was later completed by Harte \cite{Harte:2008xq, Harte:2009uy, Harte:2014wya}, who illustrated that regularization by the removal of the Detweiler-Whiting singular field, although arose as a band-aid to the point-particle assumption, quite beautifully is the point-particle limit to a more complete set of laws of motion that govern the more physically relevant system of an extended body, or nonsingular distribution of charge, moving in curved spacetime.

Although mode-sum was the first, and to date the most successful, regularization scheme, the effective source \cite{Vega:2007mc, Barack:Golbourn:2007} and matched expansions \cite{Casals:2013mpa} have also been successfully implemented.  In developing self-force techniques, it has become standard to initiate the calculations in the toy model of a scalar charged particle, and increase complexity in several directions; upgrading to the gravitational case of two masses, moving from Schwarzschild to Kerr spacetime and tackling more challenging orbits like eccentric or inclined. To this extent, mode-sum and the effective source have been implemented for a massive particle on an eccentric orbit and circular orbit in Schwarzschild spacetime respectively \cite{Barack:Sago:2010, Dolan:2012jg}. In Kerr spacetime, mode-sum was used for generic orbits of a massive particle \cite{vandeMeent:2017bcc} while the effective source is still restricted to circular orbits in this scenario \cite{Isoyama:2014mja}. 

In the case of a scalar charged particle in Schwarzschild spacetime, mode-sum has grown from circular \cite{Detweiler:2002gi} to eccentric \cite{Haas:2007}, while in Kerr, calculations have evolved through circular equatorial \cite{Warburton:2010eq}, eccentric equatorial \cite{Warburton:2011hp}, inclined spherical \cite{Warburton:2014bya} to fully generic \cite{Nasipak:2019hxh}. The effective source has gone through a similar development for a scalar field, with Schwarzschild circular \cite{Vega:2007mc} leading to eccentric \cite{Vega:2013wxa} and Kerr equatorial circular \cite{Wardell:2011gb} growing to equatorial eccentric \cite{Thornburg:2016msc}.The matched-expansions method, although the least successful thus far, as a semianalytic method, is the most powerful. To date, it has been used to calculate eccentric orbits for a scalar charged particle in Schwarzschild spacetime \cite{Casals:2013mpa}.  

It should be noted for waveform generation, it is also necessary to evolve the orbit, while parameter estimation will require second order self-force contributions. Orbit evolution for a scalar charge with higher orders was first successfully implemented borrowing from effective field theory \cite{Galley:2010xn, Galley:2011te}. This is closely related to evolving the Green function methods (analogous to matched expansions) \cite{Wardell:2014kea} that have been successfully implemented for eccentric orbits in Schwarzschild and which have received a recent boost in numerical efficiency \cite{OToole:2020ejc}. Evolving orbits for the gravitational case of massive particles using osculating geodesics with first order mode-sum self-force calculations have been accomplished for equatorial eccentric orbits in both Schwarzschild \cite{Warburton:2011fk} and Kerr \cite{Lynch:2021ogr}. More recently the first waveforms for a massive particle with second order contributions have been successfully generated for a circular orbit in Schwarzschild spacetime via a two-timescale expansion \cite{Wardell:2021fyy}, with work on a spinning secondary well under way \cite{Mathews:2021rod}.

One of the nice benefits of the mode-sum scheme was observed by Detweiler et al. when they noted through circular orbits of a scalar particle in Schwarzschild \cite{Detweiler:2002gi}, higher accuracy of the singular field leads to faster convergence when summing over the $\ell$ modes via high-order {\sl regularization parameters}. Whereas Barack and Ori, in their pioneering work, provided the first two orders for generic orbits of scalar, electromagnetism and mass charged particles in both Schwarzschild and Kerr spacetime \cite{Barack:2002mha, Barack:2002bt, Barack:2002mh}, the higher terms had been somewhat neglected. This led to further techniques in calculating high-order expressions of the Detweiler-Whiting singular field, with Haas and Poisson initially extending their work to eccentric orbits for a scalar charge in Schwarzschild \cite{Haas:2006ne}. Heffernan et al. exploited various techniques to produce higher parameters for eccentric orbits of a scalar, electromagnetic and mass charged particle in Schwarzschild \cite{Heffernan:2012su}, eccentric equatorial orbits in Kerr \cite{Heffernan:2012vj} and nongeodesic motion of a scalar charged particle in Schwarzschild \cite{Heffernan:2017cad}. We build on this previous work by calculating the first higher-order expressions for a generic orbits in Kerr for a scalar charged particle. In the case of mode-sum, we make available the fourth order regularization parameter, which reduces computation time of the scalar self-force by an order of magnitude, via a Mathematica package on Zenodo \cite{heffernan_anna_2022_6282572}. This has already proven important particularly when resonances occur \cite{Nasipak:2021qfu, Nasipak:2022xjh} with large uncertainties propagating without their inclusion. 

With the governing equations of the scalar self-force so closely relating to those of electromagnetic or mass charged particles to first order in the mass ratio, we thus are laying the necessary ground work for the more physically relevant massive particles. Indeed with the ultimate goal of producing a wave bank of computationally expensive gravitational waveforms, increasing computer efficiency is of the utmost importance. Add in the more recent application of self-force waveforms to binary sources of less extreme mass ratios \cite{Rifat:2019ltp}, i.e., those used by the ground-based detectors, and the applications of this work are extensive. 

This article is organised by the following: Sec.~\ref{sec:scalar} gives the background of the scalar self-force, from deriving the wave equation and equations of motion to the first regularization procedure by Quinn \cite{Quinn:2000wa}. Sec.~\ref{sec:dw} recaps the emergence of the Detweiler-Whiting singular field and its superseding of Quinn's regularization; here we also outline the technique in producing high-order coordinate invariant expressions for both the scalar field and the scalar self-force. Sections~\ref{sec:dw:calc} and \ref{sec:ms} are the main work of this article where in \ref{sec:ms} we illustrate how to calculate higher-order mode-sum regularization parameters for generic orbits in Kerr spacetime. The results are illustrated in Sec.~\ref{sec:result} and discussed in Sec.~\ref{sec:discuss}.

\begin{figure} 
\includegraphics[width=6cm]{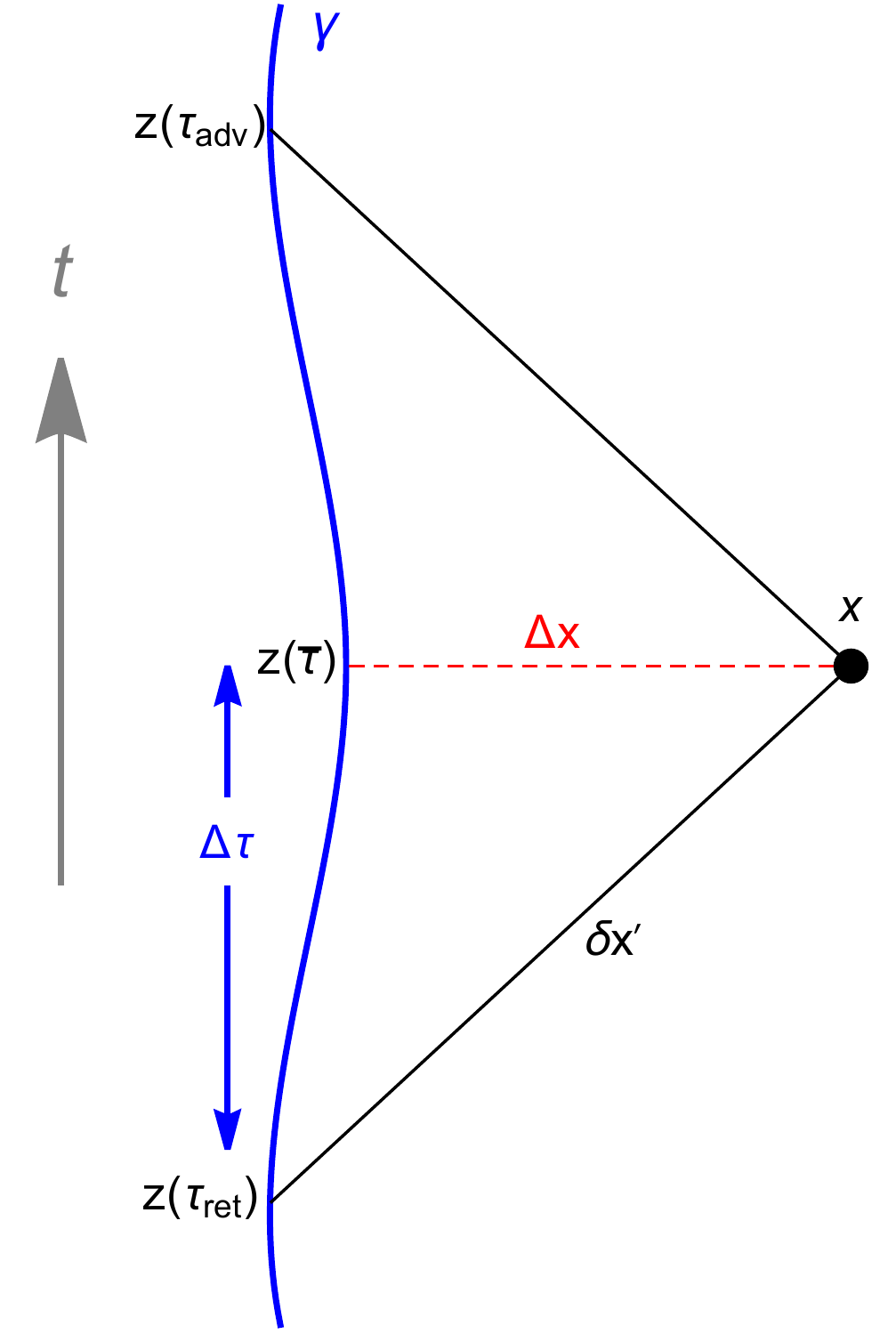} 
\caption{The worldline $\gamma$ of the point-particle source $z(\tau)$ where $x$ is the field point and $\xb \equiv z(\bar{\tau})$ is that point on the worldline that lives on the same constant time hypersurface as the field point $x$ with $\Delta x=x-\xb$ as their spatial separation. $\delta x'=x-x'$ is the separation of the field point $x$ and a point $x'$ on the worldline (often taken to be the advanced $z(\tau_{\text{adv}})$ or retarded point $z(\tau_{\text{ret}})$ - those points on the worldline connected by a null geodesic to the field point $x$).}
\label{fig:wl2D}
\end{figure}

Throughout this paper, we use units in which $G=c=1$ and adopt the sign conventions of
\cite{Misner:Thorne:Wheeler:1974}. We denote symmetrization of indices using parenthesis, $(ab)$, antisymmetrization using square brackets, $[ab]$, and exclude indices
from (anti)symmetrization by surrounding them by vertical
bars, $(a | b | c)$ and $[a | b | c]$. For spatial and four-velocity vectors we use the notation,
$x^a x^b \dots \equiv x^{a b c \dots}, u^a u^b \dots \equiv u^{ a b \dots}$ or $\dot{z}^{a b c \dots} \equiv \dot{z}^a \dot{z}^b \dot{z}^c \dots$, while biscalars with indices imply covariant differentiation, e.g., $\nabla^a \nabla^b \sigma(x,x') \equiv \sigma^{a b }$. We also make reference to several points; to clarify we direct the reader to Fig.~\ref{fig:wl2D} where 
\begin{itemize}
\item[-] $z(\tau)$ refers to a point on the worldline $\gamma$ of the source, parametrized by proper time $\tau$, with shorthand notation $x'\equiv z(\tau')$.
\item[-] $x$ refers to a field point off the worldline.
\item[-] $z(\bar{\tau})$ refers to a fixed point on the worldline living on the same constant time spacelike hypersurface as the field point $x$, also shortened to $\xb \equiv z(\bar{\tau})$.
\item[-] $z(\tau_{\text{ret}})$ and $z(\tau_{\text{adv}})$ are the retarded and advanced points respectively - those points on the worldline connected by a null geodesic to the field point $x$.
\item[-] $\Delta x=x-\xb$ refers to the distance between the field point $x$ and $\xb$ on the worldline.
\item[-] $\delta x'=x-x'$ refers to the distance between the field point $x$ and $x'$ on the worldline.
\item[-] $\Delta \tau=\tau'-\bar{\tau}$ refers to the difference in proper time between t$\xb$ and another point on the worldline $x'$, often chosen to be the retarded or advanced points.
\item[-] $\epsilon \sim \Delta x \sim \delta x' \sim \Delta \tau$ is a dimensionless parameter introduced to keep track of the order of each expansion.
\end{itemize}


\section{A scalar charged particle moving in curved spacetime} \label{sec:scalar}


\subsection{Equations of motion for a point charge} \label{sec:eom}
The motion of a scalar charge $q$ on a worldline $\gamma$, with affine parameter $\lambda$, in curved spacetime is described by its position vector $z^a(\lambda)$. The action, $S$, of such a system allows a Lagrangian density description of the (assumed point) particle, its generated scalar potential $\phi$ and the interaction between the two, that is
\begin{equation}
S=\int \left(\hat{\mathcal{L}}_{particle}+\hat{\mathcal{L}}_{field}+\hat{\mathcal{L}}_{interaction} \right) \sqrt{-g} d^4 x,
\end{equation}
where
\begin{eqnarray}
\hat{\mathcal{L}}_{particle}&=&-m_0 \int_\gamma \delta_4 (x,z) d \tau, \\
\hat{\mathcal{L}}_{field}&=&-\frac{1}{8 \pi}  \left(g^{ab} \phi_a \phi_b+\xi R \phi^2\right),\\
\hat{\mathcal{L}}_{interaction}&=&q \int_\gamma \phi(x) \delta_4 (x,z) d \tau,
\end{eqnarray}
$g$ and $R$ are the determinant and Ricci scalar of the metric respectively, $m_0$ is the bare mass of the particle, $d \tau=\sqrt{-g_{ab} \dot{z}^a \dot{z}^b} d \lambda$ is the differential of proper time, the overdot refers to differentiation with respect to $\lambda$, the scalar field $\phi_a \equiv \nabla_a \phi$ is the usual covariant derivative of the scalar potential, $\xi$ is a dimensionless coupling factor linking the scalar potential $\phi$ to the curved spacetime, and $\delta_4(x,z)$ is an invariant Dirac functional in curved spacetime as defined in Eq.~(13.2) of \cite{Poisson:2011nh}; this is related to the standard coordinate Dirac distribution functional $\delta_4(x-x')$ by
\begin{equation}
\delta_4(x, x')=\frac{\delta_4(x-x')}{\sqrt{-g}} = \frac{\delta_4(x-x')}{\sqrt{-g'}}.  
\end{equation}

Varying this action with respect to the potential $\phi$ and finding the stationary point leads to the usual associated Euler-Lagrange equation,
\begin{equation}
\frac{\partial \hat{\mathcal{L}}}{\partial \phi}-\nabla_a\left( \frac{\partial \hat{\mathcal{L}}}{\partial \dot{\phi}}\right)=0.
\end{equation}
Inserting the Lagrangian density of the field and interaction terms results in the scalar wave equation,
\begin{eqnarray} \label{eqn:field}
\left(\Box -\xi R\right) \phi&=&-4 \pi q \int_\gamma \delta_4(x,z(\tau)) d\tau, \\
	&\equiv& -4\pi \mu(x), \nonumber
\end{eqnarray}
where we have defined the source,
\begin{equation} \label{eqn:mu}
\mu(x) = q \int_\gamma \delta_4(x,z(\tau)) d\tau.
\end{equation}

For varying the action with respect to the position $z^a(\lambda)$, one need only consider the particle and interaction terms, resulting in a simplified system,
\begin{equation}
S=\int_\gamma \left(L_{particle}+L_{interaction} \right)d\lambda,
\end{equation}
where
\begin{eqnarray}
L_{particle}&=&-m_0 \left[-g_{ab}(z) \dot{z}^a \dot{z}^b\right]^{1/2}, \\
L_{interaction}&=&q  \phi(z) \left[-g_{ab}(z) \dot{z}^a \dot{z}^b\right]^{1/2}.
\end{eqnarray}
The associated Euler-Lagrange equations,
\begin{equation}
\frac{\partial L}{\partial z^a}-\frac{d}{d \lambda}\left( \frac{\partial L}{\partial \dot{z}^a}\right)=0,
\end{equation}
on taking $\lambda$ as proper time, lead to the equations of motion,
\begin{flalign} \label{eqn:eom}
&\frac12 \left(g_{de}\dot{z}^{de}\right)^{-1/2}\left[ m(\tau) \dot{z}^{bc} g_{bc,a} +2 q \phi_a (z) g_{bc} \dot{z}^{bc}\right] &\nonumber \\ 
&\qquad \quad -\frac{d}{d \lambda} \left[m(\tau) g_{ab} \dot{z}^b \left(-g_{cd} \dot{z}^{cd}\right)^{-1/2}\right]=0,  &\nonumber\\
&\Rightarrow
m(\tau) \left(\frac{D \dot{z}^a}{d \tau} \right)=q \phi_b (z) (\dot{z}^{ab}+g^{ab}),& \nonumber \\
&\Rightarrow
F^a\equiv \frac{D}{d \tau} \left[ m(\tau)  \dot{z}^a \right]=q \phi^a (z),&
\end{flalign}
where one takes $m(\tau)\equiv m_0-q \phi(z)$ to be a time-dependent mass with
\begin{equation}
\frac{d m}{d \tau}=-q \phi_a (z) \dot{z}^a.
\end{equation}

We now have a set of coupled equations; Eq.~\eqref{eqn:field} describes the field produced by the movement of the scalar charged particle, while Eq.~\eqref{eqn:eom} depicts the motion of that scalar charged particle, which in turn is influenced by the field through the presence of $\phi_a$. In this manner, the field generated by the particle's motion is seen to produce a force that affects the motion of that particle, hence the so-called self-force. Fortunately, these equations bear a very close resemblance to those generated by the gravitational perturbation in the mass ratio to first order for a point mass in curved spacetime; therefore the techniques developed here can be extended to the astrophysically interesting gravitational case.

Before one attempts to solve this coupled system, it is important to note that the scalar wave Eq.~\eqref{eqn:field} describes a field that diverges on the worldline - a consequence of our point-particle assumption. In reality, in particular when considering the electromagnetic or gravitational counterparts of electric or `mass' charged point particles, one can imagine that the charged particle does have an extended body (or distribution of charge), and in turn an extended body would not result in this singularity on the worldline. Extended body calculations are hard; fortunately it has been shown the resulting equations of motion for an extended charge distribution are identical to that of the (regularized) point-particle assumption for a sufficiently small charge distribution \cite{Quinn:2000wa, Harte:2008xq, Harte:2014wya}; hence the point-particle assumption has been validated. 

In place of a full extended-body calculation, the self-force modeling program resolves to treat the field as the summation of two parts, a regular and singular field. In this scenario for a scalar potential $\phi$, we write
\begin{equation}
\phi = \phi^\reg+\phi^\sing.
\end{equation}
By design, as will be described in the next section, the regular potential, $\phi^\reg$, is a solution to the homogeneous version of the wave Eq.~\eqref{eqn:field}, and hence smooth on the worldline; it captures all effects on the particle's motion. The singular potential, $\phi^\sing$ is a solution to the sourced wave Eq.~\eqref{eqn:field}, capturing the singularity of the field but with no impact on the motion of the particle. That is
\begin{eqnarray}
\left(\Box -\xi R\right) \phi&=&\left(\Box -\xi R\right) (\phi^\reg+\phi^\sing),\nonumber \\
&=&0-4 \pi q \int_\gamma \delta_4(x,z) d\tau,
\end{eqnarray}
and
\begin{equation}
F_a\equiv F_a^{\reg} =q \phi_a^\reg (z).
\end{equation}
Such a decomposition allows a simpler calculation of the resulting motion as one can safely remove problematic singularities. It also allows the more physical interpretation of a smooth or regular field as would be expected from an extended particle, although this is not quite what is calculated here. 

Harte has illustrated that this regularization, a requirement due to the point-particle approximation, is in fact the point-particle limit of more general laws of motion that affect a nonsingular, hence more realistic, distribution of charge (or extended body) \cite{Harte:2008xq, Harte:2014wya}. Alternatively, if one ignores the required regularization and rethinks the problem as separating the field into two; one that affects the motion and one that does not, Harte has illustrated this (Detweiler-Whiting) procedure extends to the scenario of a distribution of charge and is nonsingular. The point-particle limit of this separation results in the Detweiler-Whiting singular-regular field split.


\subsection{Regularization in flat spacetime}

\begin{figure} 
\includegraphics[width=6.5cm]{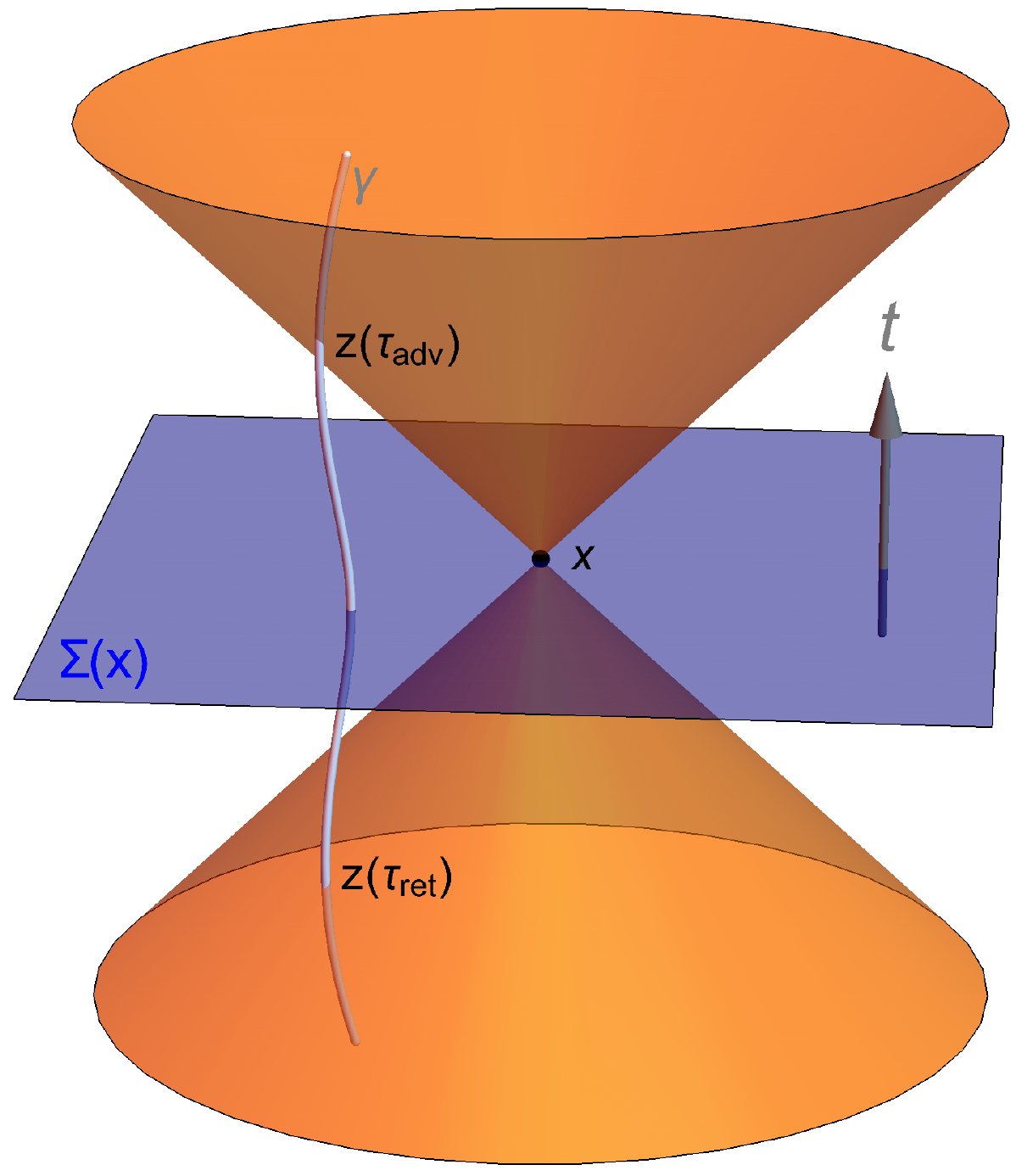} 
\caption{The worldline $\gamma$ of the point-particle source $z(\tau)$ in flat spacetime. The retarded potential at $x$ depends only on expressions evaluated at $\tau_{ret}$, where its past light cone intersects the worldline of the source. Similarly, the advanced potential only has support at $\tau_{adv}$. The singular and regular potentials depend on both.}
\label{fig:Gflat}
\end{figure}

To understand the complications that arise for regularization procedures in curved spacetime, it is beneficial to recall the regularization procedures adopted for the Klein Gordon equation in flat spacetime, 
\begin{equation} \label{eqn:field_flat}
\Box \phi(x)=-4 \pi \mu (x),
\end{equation}
where $\mu$ is prescribed source as described by Eq.~\eqref{eqn:mu}. Using a Green function we have
\begin{eqnarray} \label{eqn:field_flat_G}
\phi_{\pm}(x)&=& \int G_{\pm} (x, x') \mu(x') d^4x', \\
&=&  q \int_\gamma G_{\pm} (x, z(\tau)) d\tau,
\end{eqnarray}
where
\begin{equation} 
\Box G_\pm(x,x')=-4 \pi \delta(x-x'),
\end{equation}
and $z(\tau)$ describes the worldline $\gamma$ of the source point parametrized by proper time $\tau$ as illustrated in Fig.~\ref{fig:Gflat}. 

The explicit Green functions that formulate the solution of Eq.~\eqref{eqn:field_flat_G}, which can be derived via a Fourier transform, are the well-known retarded $G_{+}$ and advanced $G_{-}$ flat spacetime Green functions,
\begin{eqnarray} \label{eqn:Gflat}
G_\pm(x,x')&=&\frac{\delta(c t - c t' \mp |\bx-\bx'|)}{|\bx-\bx'|},\\
&=&2\Theta[\pm(c t - c t')] \delta\left[ (c t-c t')^2-|\bx-\bx'|^2\right], \nonumber \\
&=&\delta_\pm (\sigma),
\end{eqnarray}
where $\Theta$ is the usual Heaviside step function; $\sigma=\frac12 \eta_{ab} \Delta x^{ab}$ is the flat spacetime Synge world function, equivalent to half the distance squared in flat spacetime (hence zero on the null cone); and $\Delta x$ is the spacelike distance (same coordinate time) between our field point and our source point. 
We have made use of the scaling and factorizing properties of the Dirac delta function,
\begin{eqnarray}
\delta(a x)&=&\frac1{|a|}\delta (x), \\
\delta(x^2-z^2)&=&\frac1{2|z|}\left[\delta(x-z)+\delta(x+z)\right].
\end{eqnarray}

We also introduced 
\begin{equation}
\delta_\pm(\sigma(x,x'))=\Theta_\pm(\Sigma(x),x') \delta(\sigma(x,x')),
\end{equation} 
where $\Theta_\pm(\Sigma(x),x')$ is a generalised step function; $\Theta_+(\Sigma(x),x')$ is equal to 1 when $x'$ is in the past of the spacelike hypersurface $\Sigma(x)$ that connects the field point $x$ to the worldline of $x'$ as shown in Fig.~\ref{fig:Gflat};  $\Theta_-(\Sigma(x),x')=1-\Theta_+(\Sigma(x),x')$ is equal to 1 when $x'$ is in the future of $\Sigma(x)$. In this manner, $\delta_\pm(\sigma)$ is only nonzero when the two points are connected by a null geodesic with 
the support of $\delta_+$ restricted to 
where the worldline of $x'$ coincides with the past-directed null cone of $x$, known as the retarded time, $\tau_{ret}$. Similarly $\delta_-=0$ everywhere except where the worldline of $x'$ coincides with the future-directed null cone of $x$, that is the advanced time $\tau_{adv}$. Therefore, when one carries out the integration over $\gamma$ of Eq.~\eqref{eqn:field_flat_G}, $\phi(x)$ 
only depends on the source at 
$\tau_{ret}$ or $\tau_{adv}$, depending on whether we take the retarded or advanced solution accordingly (Fig.~\ref{fig:Gflat}).

However, the solution Eq.~\eqref{eqn:field_flat_G} becomes problematic when one considers the equations of motion for the source particle. These will contain the generated field of $\phi$ via its gradient, which due to our point-particle assumption, will contain singularities. As discussed previously, singularities are often considered nonphysical, and in this scenario, a direct consequence of the point-particle assumption. They therefore need to be removed in a careful regularization procedure that has no impact on the motion of the particle.

Fortunately, regularization in flat spacetime is straight forward. One designs the singular potential as the averaged sum of the retarded and advanced potentials,
\begin{equation} \label{eqn:phiSingFlat}
\phi_\sing=\frac12 \left( \phi_\ret + \phi_\adv \right).
\end{equation}
As the retarded potential, $\phi_\ret \equiv \phi_+$ is associated with outgoing radiation from the source particle, and the advanced potential $\phi_\adv \equiv \phi_-$ is associated with incoming radiation, the singular potential can be interpreted as a standing wave, evident by the reciprocity relation of the Green functions,
\begin{equation}
G_+ (x,x')=G_- (x',x).
\end{equation}
The singular potential therefore leads to no net gain or loss of energy to the system, and is a solution of the wave equation, Eq.~\eqref{eqn:field_flat}. Taking the retarded solution as the causal satisfying physical solution, we can then define the regular potential as
\begin{equation}
\phi_\reg=\phi_\ret-\phi_\sing=\frac12 \left( \phi_\ret - \phi_\adv \right) .
\end{equation}
As the singular potential results in no net loss or gain of energy, it can be safely removed without affecting the motion of the particle. The resulting regular potential, $\phi_\reg$, is a solution to the homogeneous wave equation; by design, it is smooth on the worldline of the charged particle and captures all effects on the particle's motion.


\subsection{Problems regularizing in curved spacetime} \label{sec:reg_curve}

In curved spacetime the solution of the scalar potential wave Eq.~\eqref{eqn:field}, like flat spacetime is constructed via Green functions, that is
\begin{eqnarray} \label{eqn:field_curve_G}
\phi_{\pm}(x)&=& \int G_{\pm} (x, x') \mu(x') \sqrt{-g} d^4x', \nonumber \\
&=&  q \int_\gamma G_{\pm} (x, z(\tau)) d\tau, 
\end{eqnarray}
where
\begin{equation} \label{eqn:G}
 \left( \square'  -\xi R \right) G = -4 \pi \delta_4(x,x').
\end{equation}
However the Green functions here account for curved spacetime, and are given, at least locally, by the Hadamard construction \cite{Hadamard,DeWitt:1960},
\begin{equation} \label{eqn:green}
G_\pm (x,x')=U(x,x') \delta_\pm (\sigma) - V (x,x') \Theta_\pm (-\sigma)
\end{equation}
where $x'$ is constrained to the normal convex neighbourhood of $x$, $\mathcal{N}(x)$; $U$ and $V$ are smooth biscalars. The Synge world function $\sigma$ in curved spacetime is
\begin{eqnarray}
\sigma (x'',x')&=&\frac12 \left( \lambda''-\lambda' \right) \int_{\lambda'}^{\lambda''} g_{ab}(z)  \frac{dz^a}{d\lambda}\frac{dz^b}{d\lambda} d\lambda,  \\
&=& \begin{cases}
		-\frac12 \Delta \tau^2 & \text{timelike } \beta,\\
		0  & \text{lightlike } \beta,\\
		\frac12 \Delta s^2 &\text{spacelike } \beta,
	\end{cases} 
\end{eqnarray}
where going from the first line to the second we assumed $x''$ and $x'$ are connected by a geodesic $\beta$ (parametrized by $\lambda$). By placing our retarded Green function composition of Eq.~\eqref{eqn:green} into its governing equation, Eq.~\eqref{eqn:G}, as shown in \cite{Poisson:2011nh, Heffernan:2012xlf}, one retrieves
\begin{eqnarray} 
U(x,x') &=& \Delta^{1/2}(x,x'),  \label{eqn:UVV}\\
\left( \square'  -\xi R \right)  V(x,x')&=&0, \label{eqn:boxV} \\
2 \sigma^{a} V_{,a} + \left( \sigma^{a}_a - 2\right) V &=& -\left( \square'  -\xi R \right) U \big|_{\sigma=0}, \label{eqn:Vrecur} 
\end{eqnarray}
where $\Delta^{1/2}(x,x')$ is the Van Vleck determinant. The second of these illustrates that our $V$ potential is a solution to the homogeneous wave equation, while the third leads to a recursion relationship to obtain an expression for $V$ \cite{Decanini:Folacci:2005a}. It should be noted here that our $V$ differs slightly from that of \cite{Poisson:2011nh} by a minus sign as we used the convention of \cite{Decanini:Folacci:2005a, Detweiler:2002mi}.
\begin{figure} 
\includegraphics[width=7cm]{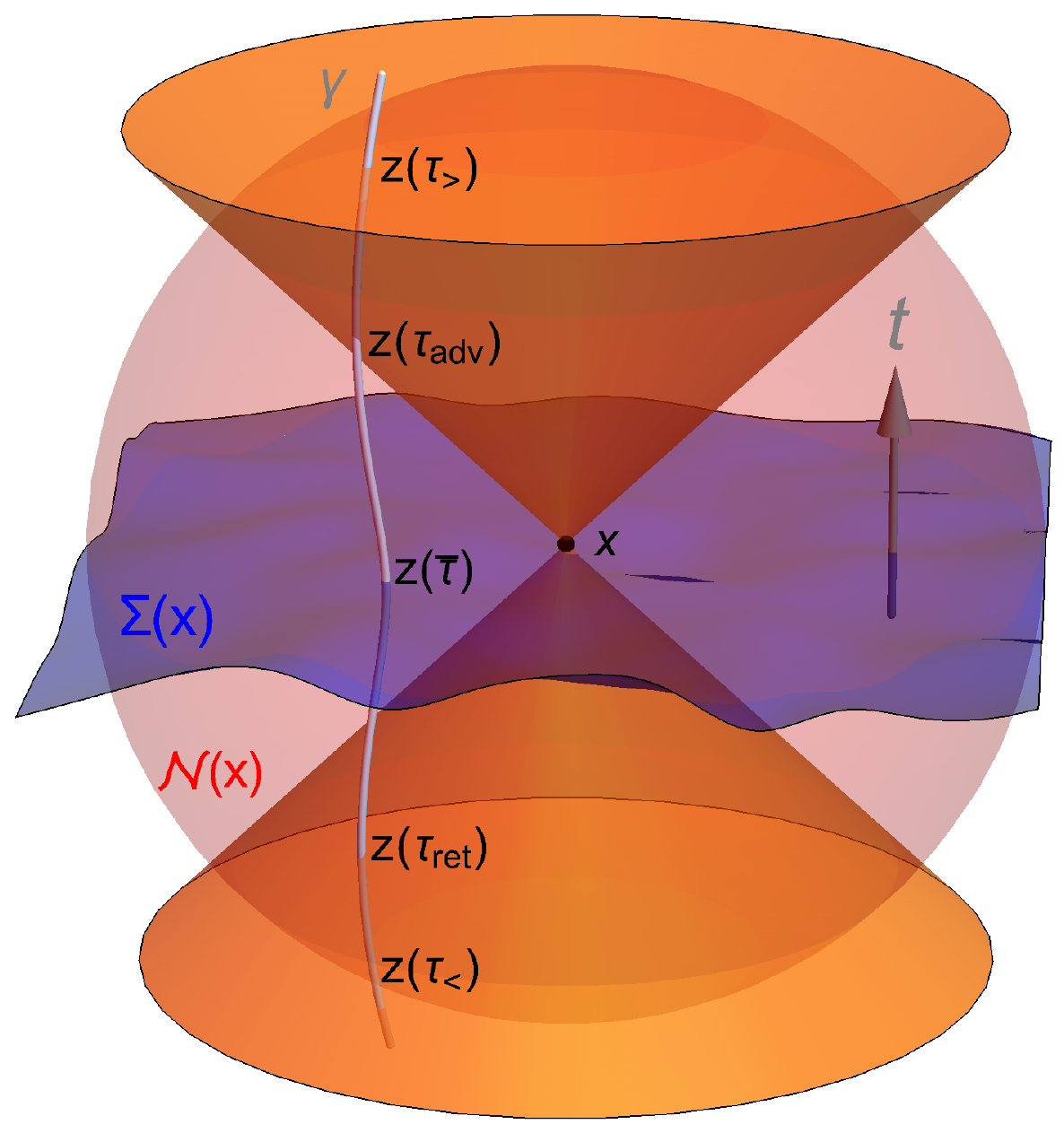} 
\caption{The worldline $\gamma$ of the source point particle $z(\tau)$ in curved spacetime. The point $\xb = z(\bar{\tau})$ lives on the same hypersurface as the field point $x$ with $\Delta x=x-\xb$ representing their separation. The retarded potential at $x$ depends on expressions evaluated at $\tau_{ret}$, where its past light cone intersects the worldline of the source as well as the worldline history previous to $\tau_{ret}$. Similarly, the advanced potential only has support at $\tau_{adv}$ and its future afterwards. The singular average of both depends on the entire future and history of the charged particle.}
\label{fig:Gcurve}
\end{figure}

As in flat spacetime, the Green functions of Eq.~\eqref{eqn:green} have a `direct' part
, $U$, that is only supported on the light cone. In addition there is now a `tail' potential $V$, resulting from the radiation scattering off curvature and thus arriving later; in this manner the advanced and retarded Green functions $G_\pm$ depend on the entire future and past of the source respectively. Explicitly, Eq.~\eqref{eqn:field_curve_G} with Eq.~\eqref{eqn:green} becomes
\begin{eqnarray} \label{eqn:phipm}
\phi_{\pm}(x)&=& \pm q \int_{\tau_{</>}}^{\tau_{\pm}} U(x,z(\tau)) \delta_\pm (\sigma) - V (x,z(\tau)) \Theta_\pm (-\sigma) d\tau \nonumber \\
&&\pm q\int_{\mp \infty}^{\tau_{</>}} G_\pm (x, z(\tau)) d\tau, \nonumber \\ 
&=& \pm q \frac{U(x,z(\tau_{\pm}))}{(\sigma_a u^a)|_{\tau_{\pm}}} \mp q \int_{\tau_{</>}}^{\tau_{\pm}} V (x,z(\tau))  d\tau \nonumber \\
&&\pm q\int_{\mp \infty}^{\tau_{</>}} G_\pm (x, z(\tau)) d\tau, \\
&=& \pm q \frac{U(x,z(\tau_{\pm}))}{(\sigma_a u^a)|_{\tau_{\pm}}} \pm \lim_{\tilde{\epsilon} \rightarrow 0} q\int_{\mp \infty}^{\tau_{\pm} \mp \tilde{\epsilon}} G_\pm (x, z(\tau)) d\tau, \nonumber
\end{eqnarray}
where $\sigma_a \equiv \nabla_a \sigma$, $u^a$ is the four-velocity of the charged particle, $\tau_<$ and $\tau_>$ mark the intersection of the worldline of the charged particle with the boundary of the normal neighbourhood of $x$, $\mathcal{N}(x)$ as illustrated in Fig.~\ref{fig:Gcurve}; '$+$' and `$-$' refer to retarded and advanced respectively. We used
\begin{equation} \label{eqn:dtau}
d \tau = \left(\frac{d \tau}{d \sigma} \right) d \sigma = \frac{d \sigma}{\dot{\sigma}} =  \frac{d \sigma}{\sigma_a u^a},
\end{equation}
where Eq.~\eqref{eqn:dtau} holds because of the direction of the $d \tau$ integral in Eq.~\eqref{eqn:phipm}; in the retarded and advanced case $\tau$ is moving in the direction from where $\sigma$ is timelike to null-like to spacelike, i.e., $\sigma$ is increasing with $\tau$ even though $\tau$ is increasing in the retarded case and decreasing in the advanced case.  In the second integral, $\tilde{\epsilon}$ is introduced as a small parameter to move off the light cone of $x$, that is
\begin{equation} \label{eqn:VasG}
\int_{\tau_{</>}}^{\tau_{\pm}} V (x,z(\tau))  d\tau = \lim_{\tilde{\epsilon} \rightarrow 0} \int_{\tau_{</>}}^{\tau_{\pm} \mp \tilde{\epsilon}}  G_\pm (x, z(\tau)) d\tau.
\end{equation}
If one now forms a symmetric singular potential as in flat spacetime, by averaging the advanced and retarded Green functions, one gets a singular, and resulting regular field, that depend on the entire past and future of the source particle; clearly this is not a realistic solution.

The resulting diverging forces are then given by
\begin{eqnarray} \label{eqn:fpm}
\partial_a \phi_{\pm}(x)&=& \pm q \frac{\partial_a U(x,z(\tau))}{\sigma_b u^b}\bigg|_{\tau_{\pm}} 
\mp  q\frac{U(x,z(\tau)) \partial_a (\sigma_b u^b)}{(\sigma_b u^b)^2} \bigg|_{\tau_{\pm}} \nonumber \\
 \pm &q& \bigg[ 
	\frac{\dot{U}(x,z(\tau))}{\sigma_b u^b}  - \frac{U(x,z(\tau)) \partial_\tau (\sigma_b u^b)}{(\sigma_b u^b)^2} 
\nonumber \\  && 
 	- V(x,z(\tau))
\bigg]_{\tau_{\pm}} \partial_a (\tau_{\pm}) 
\mp q \int_{\tau_{</>}}^{\tau_{\pm}} \partial_a V (x,z(\tau))  d\tau \nonumber \\
\pm &q&\int_{\mp \infty}^{\tau_{</>}} \partial_a G_\pm (x, z(\tau)) d\tau,
\end{eqnarray}
where one has 
\begin{eqnarray}
0=\delta_a [\sigma (x, x(\tau_\pm))]&=&\sigma_a(x,x(\tau_\pm))+ \dot{\sigma}(x,x(\tau_\pm)) \partial (\tau_\pm), \nonumber \\
\Rightarrow \partial_a (\tau_\pm)&=&\pm \frac{\sigma_a(x,x(\tau_\pm))}{(\sigma_b u^b)|_{\tau_\pm} }.
\end{eqnarray}


\subsection{The Quinn-Wald regularization procedure} \label{sec:quinn}
The first correct expression for a regularised self-force in the case of a scalar charge was produced by Quinn \cite{Quinn:2000wa}, which used the same regularization procedure developed by Quinn and Wald \cite{Quinn:1996am}, often referred to as the axiomatic approach. Here one considers two worldlines, that of the particle and that of a field point; these can exist in two different spacetimes. Initially the field point is considered to be the worldline of another charged particle, producing its own field. If these two particles have the same singular structure, Quinn asks and answers; can you subtract the forces generated by one from that of the other so the divergences cancel exactly, leaving a finite result?

Following \cite{Quinn:1996am}, they suggest the finite subtraction be carried out via
\begin{equation} \label{eqn:fdiff}
f^a_1 - \tilde{f}^a_2 = \lim_{r \rightarrow 0} q \left \langle \nabla^a \phi_1 - \tilde{\nabla}^a \phi_2  \right \rangle,
\end{equation}
where $\tilde{}$ indicates a different spacetime, $r=(\sigma_b u^b)_{\tau_+}$ and the angular brackets are introduced to average out any directional dependence. Although technically a time delay, $r$ is an affine parameter on the null geodesic connecting the field point $x$ to the retarded point $z(\tau_{\text{ret}})$ on the worldline $\gamma$ (Fig.~\ref{fig:Gcurve}) and equates to a distance when $c=1$; thus it is often referred to as the {\it retarded distance} between $x$ and $\gamma$ \cite{Poisson:2011nh} .
When considering forces of the type in Eq.~\eqref{eqn:fpm}, apart from the integral extending to infinity, we can examine the nature of the more local terms by means of an expansion in $r$. 
We can safely assume $r\sim \epsilon \sim \Delta x$ as it is represents the separation of $x$ and $x'$. Such an expansion as carried out in \cite{DeWitt:1960} (or \cite{Poisson:2011nh} for a more recent version), gives
\begin{eqnarray} \label{eqn:UVexp}
U(x,x(\tau_\pm)) &=& 1 + \mathcal{O}(\epsilon)^2, \nonumber \\
V(x,x(\tau_\pm))&=& -\frac1{12}R(x(\tau_\pm)) +\mathcal{O}(\epsilon), \nonumber \\
\sigma_a (x,x(\tau_\pm))&=& \mathcal{O}(\epsilon). \nonumber
\end{eqnarray}
From Eq.~\eqref{eqn:phipm}, we see the leading, and divergent term, is
\[
\phi_{\pm}(x)=\frac{q}r \epsilon^{-1} + \mathcal{O}(\epsilon)^0,
\]
which leads to the resulting force
\begin{eqnarray} \label{eqn:phiar}
\nabla^a \phi_{\pm}(x)&=&- \frac{q}{r^2}\hat{r}^a \epsilon^{-2}+ \mathcal{O}(\epsilon)^{-1},  \\
&\equiv&- \frac{q}{r^2}\hat{r}^a \epsilon^{-2} + \phi^a_{\pm [-1]} r^{-1} \epsilon^{-1} + \phi^a_{\pm [0]} r^{0} + \mathcal{O}(\epsilon)^{1}, \nonumber
\end{eqnarray}
where $\hat{}$ indicates the usual outward pointing normal vector. A suitable parallel transport $g_{a b'}$ to allow the subtraction in Eq.~\eqref{eqn:fdiff} would be by means of Riemann normal coordinates; the two leading divergent terms are only dependent on the scalar charge $q$, the four-velocity $u^a$ and in the case of nongeodesic motion the four-acceleration $a^a$ \cite{Quinn:2000wa, Heffernan:2017cad}. By choosing a particle with equal scalar, four-velocity and four-acceleration, then in Riemann normal coordinates,
\[
g_{a \tilde{b}}=g_{a b} + \frac16 r^2 R_{a c b d} \hat{r}^{c d} \epsilon^2 +\mathcal{O}(\epsilon)^3,
\]
acts on Eq.~\eqref{eqn:phiar}, ensuring the finite subtraction of Eq.~\eqref{eqn:fdiff}. 
That is we get
\begin{eqnarray}
g_{a \tilde{b}}\nabla^{\tilde{b}} \phi_{\pm}(x)&=&- \frac{q}{r^2}\hat{r}_a \epsilon^{-2}+ \phi_{a \pm [-1]}(x) r^{-1} \epsilon^{-1}
 \\ &&
	+ \phi_{a \pm [0]}(x) r^{0} 
	+ \frac{q}{6} R_{a c b d} \hat{r}^{b c d} 
	+ \mathcal{O}(\epsilon), \nonumber
\end{eqnarray}
where the Riemann tensor term will go to zero due to symmetries in $(b,d)$. We can therefore safely carry out the subtraction in Eq.~\eqref{eqn:fdiff} with our divergent terms exactly canceling and no new terms introduced due to the parallel transport (terms $\mathcal{O}(r) \rightarrow 0$ in the limit).

To complete the design of our secondary field, in addition to having the same singular structure, we need it to have a resultant force of zero, that is $\tilde{f}^a_2 =0$ in Eq.~\eqref{eqn:fdiff}. For this, we simply borrow from flat spacetime; the averaged sum of the retarded and advanced fields in flat spacetime results in no force due to its standing wave nature. That is, recalling our flat spacetime Green function, Eq.~\eqref{eqn:Gflat}, we have
\begin{equation}\label{eqn:phiFlat}
\phi_{2}(x) = \frac{q}2 \frac{1}{(\sigma_a u^a)}\bigg|^{\tau_+}_{\tau_-},
\end{equation}
where again we must be careful of the direction of our integration, flipping the limits for $\tau_-$ ensures $\tau$ is decreasing as it crosses the advanced time. Similarly, one can use the curved spacetime counterpart,
 \begin{equation} \label{eqn:phiSelf}
\phi_{2}(x) = \frac{q}2 \frac{U(x,x(\tau))}{(\sigma_a u^a)}\bigg|^{\tau_+}_{\tau_-},
\end{equation}
with the two options of $\phi_{2}(x)$ differing by $\mathcal{O}(r)$ from our local expansion of $U$ above. When substituted into Eq.~\eqref{eqn:fdiff}, we get a finite and nondiverging $f_1^a = \mathcal{O}(r^0)$ where any directional dependence on how one takes the limit is averaged over the 2-sphere. That is, taking our retarded field as the causal sensible field, $\phi_1 = \phi_+$, the self-force is the gradient of 
\begin{eqnarray}
\phi^{(\text{self})} &=& \phi_+ - \frac{q}2 \frac{U(x,x(\tau))}{(\sigma_a u^a)}\bigg|^{\tau_+}_{\tau_-}, \nonumber \\
&=& \frac{q}2 
	\frac{U(x,x(\tau))}{(\sigma_a u^a)}\bigg|^{\tau_-}_{\tau_+}
	+q \lim_{\epsilon \rightarrow 0} \int_{- \infty}^{\tau_+ - \epsilon} G_+ (x, z(\tau)) d\tau, \nonumber
\end{eqnarray}
averaged over the 2-sphere and in the limit $x \rightarrow x'$.


\section{The Detweiler-Whiting Singular Field} \label{sec:dw}


\subsection{The direct and tail fields} \label{sec:dw:fields}

Although the Quinn-Wald regularization was inspired by its flat spacetime counterpart, it doesn't quite capture the same essence; in flat spacetime the singular and regular fields are solutions to the wave equation. Inspired by this, Detweiler and Whiting observed that the $\mathcal{O}(\epsilon)^0$ term that arises from 
\[
- q V(x,z(\tau_+)) \partial_a (\tau_{+}) = -\frac{q}{12} R(x) \hat{r}_a + \mathcal{O}(\epsilon),
\]
in Eq.~\eqref{eqn:fpm}, with the local expansion Eq.~\eqref{eqn:UVexp}, integrates to zero over the 2-sphere. This term arises from the gradient hitting the limit $\int_{-\infty}^{\tau_+}$ on the integral over $V$. However, consider the integral,
\begin{eqnarray}
\lim_{x\rightarrow x'} \nabla_a \frac12 \int^{\tau_-}_{\tau_+} V(x,z(\tau)) d \tau &=& \frac12 V(x,x)\nabla_a  (\tau_- - \tau_+ ) , \nonumber \\
	&=& V(x,x)\nabla_a r + \mathcal{O}(\epsilon), \nonumber \\
	&=& -\frac1{12} R(x) \hat{r}_a + \mathcal{O}(\epsilon), \nonumber
\end{eqnarray}
where $ (\tau_- - \tau_+ ) =r+ \mathcal{O}(r^2)$ and we note this cancels exactly the previous term. This led Detweiler and Whiting to amend the singular field, Eq.~\eqref{eqn:phiSelf}, to
\begin{equation} \label{eqn:DW}
\phi^\sing=\frac{q}2 \frac{U(x,x(\tau))}{(\sigma_a u^a)}\bigg|^{\tau_+}_{\tau_-} + \frac{q}2 \int^{\tau_-}_{\tau_+} V(x,z(\tau)) d \tau.
\end{equation}

Indeed a separate derivation found in \cite{Poisson:2011nh}, notes that by adding an unknown biscalar $H(x,x')$ to the averaged sum of the singular and regular, one can make sensible demands that result in the same Detweiler-Whiting singular field,
\begin{eqnarray}
G^\sing (x,x')&=& \frac12 \left[G^\ret(x,x')+G^\adv (x,x')-H(x,x')\right], \nonumber \\
G^\reg(x,x') &=& G^\ret(x,x') -G^\sing (x,x'), \nonumber \\
&=& \frac12 \left[G^\ret(x,x')-G^\adv (x,x')+H(x,x')\right].\nonumber
\end{eqnarray}
One can determine a suitable $H(x,x')$ by demanding when $x'$ is in the past of $x$, $G^\sing(x,x')$  does not depend on the history of the worldline $z(\tau)$. Similarly when $x'$ is in the future of $x$, $G^\sing(x,x')$  does not depend on the future of $z(\tau)$. From Eq.~\eqref{eqn:green}, we have
\begin{eqnarray}
G^\sing (x,x')&=& \frac12 U(x,x')\delta (\sigma) -\frac12 H(x,x')\nonumber \\
&&+ \frac12 V(x,x')[\Theta_+(-\sigma)+\Theta_-(-\sigma)], \nonumber
\end{eqnarray}
where we used $\delta = \delta_+ + \delta_-$. When $x'$ is in the past of $x$, 
\[
G^\sing (x,x') = \frac12 [U(x,x')\delta(\sigma)+  V(x,x')\Theta_+(-\sigma)-H(x,x')],
\]
similarly, when $x'$ is in the future of $x$, 
\[
G^\sing (x,x') = \frac12 [U(x,x')\delta(\sigma)+ V(x,x')\Theta_-(-\sigma)-H(x,x')].
\]
Both of these can be solved by $H(x,x')=V(x,x')$ (this also preserves reciprocity of the Green functions and is a solution to the homogeneous wave equation). The singular field then becomes,
\begin{eqnarray}
G^\sing (x,x') &=& \frac12 U(x,x')\delta(\sigma)+ \frac12 V(x,x')(\Theta_+(-\sigma)-1), \nonumber \\
&=&\frac12 U(x,x')\delta(\sigma)+ \frac12 V(x,x')\Theta(\sigma),
\end{eqnarray}
which is the Green function that would generate the Detweiler-Whiting singular field above, Eq.~\eqref{eqn:DW}.
 
We can now see the linear combination of $G^\ret (x,x')$, $G^\adv (x,x')$ and $V(x,x)$ that makes up $G^\sing (x,x')$ is a solution to the full scalar wave equation while the combination forming $G^\reg (x,x') $ is a solution to the homogeneous wave equation as desired. This also removes the need for averaging over the 2-sphere, indeed as is shown later, one must simply ensure the limit $x\rightarrow x'$ is taken in the same direction for both the retarded and singular field to ensure a safe subtraction.


\subsection{Explicit calculation of the Detweiler-Whiting singular field} \label{sec:dw:calc}

Our starting description of the singular field Eq.~\eqref{eqn:DW} is
\begin{equation} \label{eqn:PhiS}
\PhiS(x) = \frac{q}{2} \Bigg[ \frac{U(x,x')}{\sigma_{c'} u^{c'}} \Bigg]_{x'=x_{\rm \ret}}^{x'=x_{\rm \adv}}
   + \frac{q}{2} \int_{\tau_{\rm \ret}}^{\tau_{\rm \adv}} V(x,z(\tau)) d\tau,
\end{equation}
where we remind the reader that the term $r=(\sigma_{c'} u^{c'})_{x'=x_{\rm \ret}}$ is the same retarded distance of Sec.~\ref{sec:quinn}, with $r_{adv}=-(\sigma_{c'} u^{c'})_{x'=x_{\rm \adv}}$ often defined as the {\it advanced distance} between the field point $x$ and the worldline $\gamma$ \cite{Poisson:2011nh}. To obtain explicit expressions, we use
\begin{eqnarray} \label{eqn:deltax}
\Delta x^{\ab} &=& x^a - x^\ab,\\
 \delta x^{a'} &=& x^{a} - x^{a'},
 \end{eqnarray}
where $\xb\equiv x^\ab$ is where the worldline intersects the hypersurface of $x$ as shown in Fig.~\ref{fig:Gcurve} and we assume both $\Delta x^a$ and $\delta x^{a'}$ are of similar magnitude $\sim \epsilon$. As previously described in \cite{Heffernan:2012su, Heffernan:2012vj}, one can expand all biscalars of $(x,x')$ in $\delta x^{a'}$ and use governing equations to determine unknown coefficients.

The direct potential $U(x,x')$, which is equivalent to the Van Vleck (Eq.~\eqref{eqn:UVV}), was previously determined to be 
\begin{equation}
U(x,x')=\Delta^{1/2}(x,x')=1+\mathcal{O}(\epsilon)^4,
\end{equation} 
in Eq.~(B5) of \cite{Heffernan:2012su}. Similarly the tail potential $V(x,x')$ described by Eq.~\eqref{eqn:Vrecur} was illustrated to be 
\begin{equation}
V(x,x')=V_0(x,x')+\mathcal{O}(\epsilon)^{2}=\mathcal{O}(\epsilon)^{2},
\end{equation}
in Eq.~(B9) of \cite{Heffernan:2012su}. With the Synge world function $\sigma(x,x')$ having leading order $\mathcal{O}(\epsilon)^2$, we can immediately see from Eq.~\eqref{eqn:PhiS} that the singular field will be of leading order $\mathcal{O}(\epsilon)^{-1}$ through the direct potential. As we are targeting the leading four orders (the next two unknown orders), and the tail potential $V(x,x')$ will pick up an order through integration, we now need only calculate
\begin{equation}
\PhiS(x) = \frac{q}{2} \left[ \frac1{\sigma_{c'} u^{c'}} \right]_{x'=x_{\rm \ret}}^{x'=x_{\rm \adv}} +\mathcal{O}(\epsilon)^3.
\end{equation}
Interestingly this means for the desired order, one does not require / observe the effects of the tail potential, illustrating an even closer link to the flat spacetime singular field of Eqs.~\eqref{eqn:phiSingFlat} and \eqref{eqn:phiFlat}.

To produce expressions for the Synge world function, we have
\begin{eqnarray} \label{eqn:sigma}
\sigma(x,x') &=& \tfrac12 g_{a b}(x) \delta x^{a'} \delta x^{b'} + A_{abc}(x) \delta x^{a'} \delta x^{b'} \delta x^{c'}
\nonumber \\
&&+ B_{abcd}(x) \delta x^{a'} \delta x^{b'} \delta x^{c'} \delta x^{d'} 
 \\
&&+ C_{abcde}(x) \delta x^{a'} \delta x^{b'} \delta x^{c'} \delta x^{d'} \delta x^{e'} 
+ \mathcal{O}(\epsilon)^6. \nonumber
\end{eqnarray}
Taylor expanding functions of $x$ as $(\xb+\Delta x)$ and using $2 \sigma = \sigma_{a'} \sigma^{a'}$, one arrives at the coefficients,
\begin{eqnarray*}
A_{a b c}	&=& -\tfrac12 \Gamma_{(a b ,c)}, \\
 B_{a b c d}&=& \tfrac16 \Gamma_{(a b c, d)}-\tfrac{1}{24} \Gamma^e_{(ab} \Gamma_{|e|cd)} ,\\
C_{a b c d e} &=&-\tfrac1{24} \Gamma_{(a b c, d e)}-\tfrac1{24}\Gamma^f_{(ab} \Gamma_{c d e), f}+\tfrac1{24} \Gamma^f_{(ab} \Gamma_{|f| c d, e)} \\
&&+ \tfrac1{24} \Gamma^f_{(ab} \Gamma_{c d |f|, e)} - \tfrac1{24} \Gamma^f_{(ab} \Gamma^{|g|}_{d e} \Gamma_{|f| c) g}.
\end{eqnarray*}
One can then obtain $\sigma_{a'}$ from a simple partial derivative of Eq.~\eqref{eqn:sigma}.

For the advanced and retarded points on the worldline, we Taylor expand in $\Delta \tau = \tau' - \bar{\tau}$,
\begin{eqnarray} \label{eqn:xp}
x^{{a'}}(\tau') &=&x^{{a'}}(\bar{\tau}+\Delta \tau) \\
&=& x^{\ab} + u^{\ab} \Delta \tau + \tfrac{1}{2!} \dot{u}^{\ab} \Delta \tau^2 + \tfrac{1}{3!} \ddot{u}^{\ab} \Delta \tau^3 +  \cdots,  \nonumber \\
\Rightarrow \delta x^{a'}&=&\Delta x^\ab-u^{\ab} \Delta \tau - \tfrac{1}{2!} \dot{u}^{\ab} \Delta \tau^2 - \tfrac{1}{3!} \ddot{u}^{\ab} \Delta \tau^3 +  \cdots, \nonumber
\end{eqnarray}
where $u^\ab$ is the contravariant four-velocity along the worldline of the particle evaluated at $\xb$ and an overdot denotes differentiation with respect to $\tau$.  Practically implementing Eq.~\eqref{eqn:xp}, we find the resulting expressions greatly reduce in complexity if we rewrite the higher derivatives of the four-velocity in terms of Christoffels,
\begin{eqnarray} 
\label{eqn:4v1}
\dot{u}^\ab&=&-\Gamma^\ab{}_{\bb \cb} u^{\bb \cb},
\\
\ddot{u}^\ab&=&\left( 2 \Gamma^\ab{}_{\bb \eb} \Gamma^\eb{}_{\cb \db} - \Gamma^\ab{}_{\bb \cb,\db} \right) u^{\bb \cb \db},
\\ \label{eqn:4v2}
\dddot{u}^\ab&=& - \big[
\Gamma^\ab{}_{\bb \cb,\db \eb}
- \Gamma^\fb{}_{\bb \cb} \big(
\Gamma^\ab{}_{\db \eb,\fb} + 4 \Gamma^\ab{}_{\db \fb,\eb} - 2 \Gamma^\ab{}_{\fb \gb} \Gamma^\gb{}_{\db \eb}
\big)
\nonumber \\ 
&& \quad
- \Gamma^\ab{}_{\bb \fb} \left(
2 \Gamma^\fb{}_{\cb \db,\eb}
-4 \Gamma^\fb{}_{\cb \gb} \Gamma^\gb{}_{\db \eb}
\right) \big] u^{\bb \cb \db \eb}.  \label{eqn:4v3}
\end{eqnarray}
An expression for $\Delta \tau$ can be found from combining our expression for $\delta x'$ of Eq.~\eqref{eqn:xp} with the Synge world function of Eq.~\eqref{eqn:sigma}. As we define $x'$ to be those points on the worldline connected by a null geodesic to $x$, we use $\sigma(x,x')=0$ to solve for
\begin{equation}
\Delta \tau = \epsilon \tau_1 + \epsilon^2 \tau_2 + \epsilon^3 \tau_3 + \epsilon^4 \tau_4 + \mathcal{O}(\epsilon)^5, 
\end{equation}
to obtain
\begin{eqnarray}
\tau_1&=&-u_\ab \Delta x^\ab \pm \rho, \label{eqn:tau1}\\
\tau_2&=&\frac1{2 \rho}\left( \Delta x^c - u^\cb \tau_1 \right) \Gamma_{\ab \bb \cb} \Delta x^{a b}, \label{eqn:tau2}
\end{eqnarray}
\begin{widetext}
\begin{eqnarray}
\tau_3&=&-\frac1{24 \rho}\Big\{ 
	12\tau_2^2 - \left[
		4 \tau_1^2  u^{\ab \bb} - \left(4 \tau_1 u^\ab -  \Delta x^a \right)  \Delta x^b
	\right] \Gamma_{\eb \cb \db} \Gamma^\eb{}_{\ab \bb}  \Delta x^{c d} \\
&& \quad
	+ 4  \left[ 
		\tau_1^2 \Gamma_{\eb \bb \db} \Gamma^\eb{}_{\ab \cb} u^{\ab \bb} \Delta x^d 
		+ 3 \tau_2 \Gamma_{\ab \bb \cb} u^\ab \Delta x^b + \left( \tau_1 u^\ab - \Delta x^a\right) \Gamma_{\ab \bb \cb, \db} \Delta x^{b d} 
		+ 2 \tau_1^2 \Gamma_{\cb \ab [\bb, \db]} 
		u^{\ab \bb} \Delta x^{d}
	\right] \Delta x^c
\Big\}, 	
\nonumber \\
\tau_4&=& \frac1{24\rho} \Big\{
	\left[ 
		\Gamma_{\ab \bb \cb, \db \eb}+ \Gamma^\fb{}_{\ab \bb} \left( 
			\Gamma_{\fb \cb \ib} \Gamma^\ib{}_{\db \eb} - 2\Gamma_{\cb \db [\fb, \eb]} - \Gamma_{\fb \cb \db, \eb} 
		\right)
	\right] \Delta x^{a b c d e}
	+ 16 \tau_2 \tau_1 \left( 
		 \Gamma_{\cb \ab [\db, \bb]} 
		+ \Gamma^\eb{}_{\cb [\db} \Gamma_{|\eb| \ab] \bb} 
	\right) u^{\ab \bb} \Delta x^{c d}
\\ \label{eqn:tau1} && \quad
	- \tau_1 \left[ 
		4 \Gamma^\fb{}_{\bb \ab} \Gamma^\hb{}_{\cb \db} 
			\Gamma_{(\hb \fb) \eb} 
		+ \Gamma_{\ab \bb \cb, \db \eb} - \Gamma^\fb{}_{\bb \cb} \left( 
			\Gamma_{\fb \ab \hb} \Gamma^\hb{}_{\db \eb} + 2\Gamma_{\ab \db [\fb, \eb]} + \Gamma_{\fb \db \ab, \eb} 
		\right)
		- 2 \Gamma^\fb{}_{\bb \ab} \left( 
			2\Gamma_{\cb \db [\fb, \eb]} + \Gamma_{\fb \cb \db, \eb} 
		\right)
	\right] u^\ab \Delta x^{b c d e}
\nonumber \\ && \quad 
	+\tau_1^2 \Big[
		2\Gamma_{\cb \ab [\db, \bb] \eb} 
		+ \Gamma^\fb{}_{\ab \bb} \left( 
			2 \Gamma_{\cb \db [\fb, \eb]} + \Gamma_{\fb \cb \db, \eb} 
		\right) 
		+ 2 \Gamma^\fb{}_{\cb \ab} \left( 
			2 \Gamma_{\fb \db}{}^\hb \Gamma_{\hb \eb \bb} 
			- 2 \Gamma^\hb{}_{\db \eb} \Gamma_{(\fb \hb) \bb}
			- \Gamma_{\fb \db \bb, \eb} + 2 \Gamma_{\bb \db [\eb, \fb]} 
		\right)
\nonumber \\ && \qquad
		+ \Gamma^\fb{}_{\cb \db} \left(  
			2 \Gamma_{\eb \ab [\fb, \bb]} +2 \Gamma^{[\hb}{}_{\eb] \fb} \Gamma_{\hb \ab \bb} 
			+ \Gamma_{\fb \eb \ab, \bb} 
		\right)
	\Big] \Delta u^{\ab \bb} \Delta x^{c d e} 
	+ 4 \tau_2 \left(
		\Gamma_{\eb \bb \ab} \Gamma^\eb{}_{\cb \db} - \Gamma_{\ab \bb \cb, \db} 
	\right) u^\ab  \Delta x^{b c d}
	-24 \tau_3 \tau_2 
\nonumber \\ && \quad
	+ \tau_1^3 \Big[
		\Gamma^\fb{}_{\db \eb} \left( 
			2 \Gamma_{\ab \bb [\fb, \cb]} + \Gamma_{\fb \ab \bb, \cb} 
			- 3 \Gamma_{\fb \ab}{}^\hb \Gamma_{\hb \bb \cb} - \Gamma_{\hb \cb \fb} \Gamma^\hb{}_{\ab \bb}  
		\right)
		+ 2 \Gamma^\fb{}_{\db \ab} \left( 
			2 \Gamma_{\fb \cb \hb} \Gamma^\hb{}_{\eb \bb} 
			+ \Gamma_{\fb \eb}{}^\hb \Gamma_{\hb \bb\cb} 
			- \Gamma_{\fb \eb \bb, \cb} + 2 \Gamma_{\eb \bb [\cb, \fb]} 
		\right)
\nonumber \\ && \qquad
		+ \Gamma^\hb{}_{\ab \bb} \left( 
			4 \Gamma_{\db \cb [\hb, \eb]} 
			+ \Gamma_{\hb \db \cb, \eb} + 2 \Gamma_{\cb \db [\eb, \hb]} 
			- 2 \Gamma_{\hb \db}{}^\fb \Gamma_{\fb \eb \cb} 
		\right)
		+ 2\Gamma_{\db \ab [\eb, \bb] \cb} 
		+ 2\Gamma_{\ab \bb [\cb, \db] \eb} 
	\Big] u^{\ab \bb \cb} \Delta x^{d e}
	- 12 \tau_3 \Gamma_{\ab \bb \cb} u^\ab \Delta x^{b c} 
\Big\}, \nonumber 
\end{eqnarray}
\end{widetext}
where
\begin{equation} \label{eqn:rhoS}
\rho^2=(g_{\ab \bb}+u_{\ab \bb}) \Delta x^{ab}.
\end{equation}
One can think of $\rho$ simply as the square root for the quadratic solution of
\begin{equation}
\sigma=\frac12 g_{\ab \bb} \left(\Delta x^{a b}-u^{\ab \bb} \tau_1\right)\epsilon^2+\mathcal{O}(\epsilon)^3=0,
\end{equation}
as in Eq.~\eqref{eqn:tau1}. For the retarded solution, we have $\Delta \tau_\ret= \tau_\ret - \bar{\tau}$ with $\tau_1$ corresponding to the negative $\rho$. Similarly, the advanced solution $\Delta \tau_\adv= \tau_\adv- \bar{\tau}$ equates to $\tau_1$ taking the positive root. A more physical picture of $\rho$ is its equivalence to the retarded distance $r(x,x')$, more specifically its leading order  when converted to a coordinate expansion. This type of Riemann normal expansion has benefits over $r(x,x')$ as it allows one to avoid tetrad decompositions - in previous work this has enabled much higher expansions \cite{Heffernan:2012su, Heffernan:2012vj}.

Combining the above for the unknown expressions in Eq.~\eqref{eqn:PhiS}, we arrive at the singular field,
\begin{eqnarray}
\PhiS_{[-1]} (x)&=&\frac1{\rho}, \\
\PhiS_{[0]} (x) &=&-\frac{1}{2 \rho^3} \left( \Gamma_{\ab \bb \cb}+\Gamma^\db{}_{\ab \bb} u_{\cb \db}\right) \Delta x^{a b c},
\end{eqnarray}
\begin{widetext}
\begin{eqnarray}
\PhiS_{[1]} (x)&=&\frac{u^{\ab \bb} \Delta x^{c d} }{3 \rho} \left(
	\Gamma_{\cb \ab [\db, \bb]} + \Gamma^\eb{}_{\cb [\db} \Gamma_{|\eb| \ab] \bb}  
\right) 
+\frac{3\Delta x^{c d e f}}{8 \rho^5}  \Gamma_{\ab \eb \fb} \Gamma_{\bb \cb \db} \left[ 
	\Delta x^{a b} + 2 u^\ab \Delta x^b \left( u \cdot \Delta x \right)  + u^{\ab \bb}\left( u \cdot \Delta x \right)^2
\right]  
\\ &&
+\frac{ \Delta x^{c d}}{24 \rho^3} \bigg\{ 
	\Delta x^{a b} \left( \Gamma_{\eb \cb \db} \Gamma^\eb{}_{\ab \bb} - 4 \Gamma_{\ab \bb \cb, \db} \right)
	+ 4 u^{\ab} \Delta x^b  \left( u \cdot \Delta x \right) \left( \Gamma_{\fb \cb \ab} \Gamma^\fb{}_{\db \bb}  - \Gamma_{\ab \cb \db, \bb} \right)
\nonumber \\ && \quad
 	- u^{\eb \fb} \left[ 3 \Delta x^{a b} \Gamma_{\eb \cb \db} \Gamma_{\fb \ab \bb} 
	+ 8  \left( u \cdot \Delta x \right)^2 \left( 
		\Gamma_{\cb \eb [\fb, \db]} + \Gamma^\ib{}_{\cb [\eb} \Gamma_{|\ib| \db] \fb} 
	\right) 
	\right]
\bigg\},
  \nonumber \\
\PhiS_{[2]} (x)&=& \frac{u^{\ab \bb} \Delta x^{d e} }{24 \rho} 
\bigg\{
	\Delta x_\ib \left( g^{\cb \ib} - 3 u^{\cb \ib} \right) \left[
		\Gamma^{\fb}{}_{\ab \bb}  \Gamma_{\fb \cb \db, \eb}  + \Gamma^{\fb}{}_{\db \eb}  \Gamma_{\fb \cb \ab, \bb} 
		+ \Gamma^{\fb}{}_{\ab \db}  \left(
			\Gamma_{\fb \eb}{}^{\hb} \Gamma_{\hb \bb \cb} - 2 \Gamma_{\fb \eb \bb, \cb} 
		\right)
	\right]
\\ && \quad +
	\Delta x_\ib \left( g^{\cb \ib} + 3 u^{\cb \ib} \right) \bigg[
		\Gamma^{\fb}{}_{\cb \db} \left( 
				2 \Gamma_{\eb \ab [\fb, \bb]} - \Gamma_{\hb \ab \bb} \Gamma_{\fb \eb}{}^{\hb} 
		\right) + 2 \Gamma^{\fb}{}_{\ab \bb} \Gamma_{\cb \db [\fb, \eb]} 
		+ 2 \Gamma_{\cb \ab [\db , \bb] \eb} 
		+ \Gamma^{\fb}{}_{\db \eb} \Gamma^{\hb}{}_{\ab \bb} \Gamma_{\hb \cb \fb}
\nonumber \\ && \qquad	
		+ 2 \Gamma^{\fb}{}_{\ab \cb} \Gamma_{\fb \db}{}^{\hb} \Gamma_{\hb \eb \bb} 
	\bigg]
	+ \Delta x^{c} \bigg[
		\Gamma^{\fb}{}_{\cb \ab} \left(
			\Gamma_{\fb \db}{}^{\hb} \Gamma_{\hb \eb \bb} 
			+ 4 \Gamma_{\bb \db [\eb, \fb]}
		\right) 
		-4 \Gamma^{\fb}{}_{\cb \db} 
			\Gamma^{\hb}{}_{\eb \ab} \Gamma_{(\fb \hb) \bb} 
	\bigg]
\nonumber \\ && \quad
	- 3 \left(u  \cdot \Delta x \right) u^{\cb} \bigg[
		\Gamma^{\fb}{}_{\cb \db} \left(
			4 \Gamma^{\hb}{}_{\eb \bb} \Gamma_{\fb \ab \hb} 
		\right)
		+ 2\Gamma_{\db \ab [\eb, \bb] \cb}
		+\Gamma^{\fb}{}_{\db \eb} \left(
			2 \Gamma_{\ab \bb [\fb, \cb]} 
			- 3 \Gamma_{\fb \ab}{}^{\hb} \Gamma_{\hb \bb \cb} 
		\right)
		+ 4 \Gamma^{\fb}{}_{\ab \bb} 
			\Gamma_{\db \cb [\fb, \eb]} 
	\bigg]
\bigg\}
\nonumber \\ &&
+ \frac{\Delta x^{d e}}{24 \rho^3} 
\bigg\{
	\Delta x_\ib \left( g^{\ab \ib} + u^{\ab \ib} \right) \left[
		\Delta x^c \Delta x_j \Gamma^{\fb}{}_{\bb \cb}  \left(g^{\bb \jb} + 2 u^{\bb \jb} \right)\left(
			2\Gamma_{\ab \db [\fb, \eb]} + \Gamma_{\fb \ab \db, \eb}
		\right)
		- \Delta x^{b c} \Gamma_{\ab \bb \cb, \db \eb}
	\right]
\nonumber \\ && \quad
	- \Delta x^{b c}\Delta x_\ib \left( g^{\ab \ib} - u^{\ab \ib} \right)
	\Gamma^{\fb}{}_{\bb \cb} \Gamma^{\hb}{}_{\db \eb} \Gamma_{\fb \ab \hb}
	-u^{\bb}\Delta x^{c}\Delta x_\ib \left( g^{\ab \ib} - u^{\ab \ib} \right) \left( u \cdot \Delta x \right) \left[
		4 \Gamma^{\hb}{}_{\cb \bb} \Gamma^{\fb}{}_{\eb \db} \Gamma_{(\fb \hb) \ab} 
	\right]
\nonumber \\ && \quad
	+ u^{\bb \cb} \Delta x_\ib \left( g^{\ab \ib} + u^{\ab \ib} \right) \left( u \cdot \Delta x \right)^2 \left(
		2 \Gamma_{\ab \cb [\bb, \db] \eb} + 2 \Gamma^{\fb}{}_{\cb \bb} \Gamma_{\ab \db [\eb \fb]} 
		- 2 \Gamma^{\fb}{}_{\ab \cb} \Gamma_{\fb \db}{}^{\hb} \Gamma_{\hb \eb \bb}
		- \Gamma^{\fb}{}_{\db \eb} \Gamma^{\hb}{}_{\ab \fb} \Gamma_{\hb \cb \bb}
	\right)
\nonumber \\ && \quad
	- u^{\bb \cb} \Delta x_\ib \left( g^{\ab \ib} - u^{\ab \ib} \right) \left( u \cdot \Delta x \right)^2 \left[
		\Gamma^{\fb}{}_{\cb \bb} \Gamma_{\fb \ab \db, \eb}
		+ \Gamma^{\fb}{}_{\eb \db} \left(
			2 \Gamma_{\ab \cb [\fb, \bb] }+ \Gamma_{\fb \ab \cb, \bb} - \Gamma_{\fb \ab}{}^{\hb} \Gamma_{\hb \cb \bb}
		\right)
		+ 2 \Gamma^{\fb}{}_{\db \cb} \Gamma_{\fb \eb}{}^{\hb} \Gamma_{\hb \ab \bb}
	\right]
\nonumber \\ && \quad
	+ 2 u^{\ab \bb} \Delta x^{f h} \left[
		\Delta x^c \Gamma _{\ab \cb \db} \left(
			\Gamma^{\ib}{}_{\fb \hb} \Gamma_{\ib \eb \bb} - \Gamma_{\bb \eb \fb, \hb}
		\right)
		+ 2 \Delta x_i \left(g^{\cb \ib} + 3 u^{\cb \ib} \right)\Gamma_{\cb \db \eb} \left(
			\Gamma^{\ib}{}_{\fb [\ab} \Gamma_{|\ib| \hb] \bb} 
			+\Gamma_{\fb \ab [\bb, \hb]}
		\right)
	\right]
\nonumber \\ && \quad
	+ 2 u^{\ab \bb \cb}  \left( u \cdot \Delta x \right)^3 \bigg[
		\Gamma_{\db \ab [\eb, \bb] \cb} 
		- \Gamma^{\fb}{}_{\db \eb} 
			\Gamma_{\fb \ab}{}^{\hb} \Gamma_{\hb \bb \cb} 
		+ 2\Gamma^{\fb}{}_{\ab \bb} 
			\Gamma_{\db \cb [\fb, \eb]} 
		+ \Gamma^{\fb}{}_{\db \ab} \left(
			2 \Gamma^{\hb}{}_{\eb \bb} \Gamma_{\fb \cb \hb} + 2 \Gamma_{\eb \bb [\cb, \fb]}
			- \Gamma_{\fb \eb \bb, \cb}
		\right)
	\bigg]
\bigg\}
\nonumber \\ &&
+\frac{\Delta x^{d e f h} }{16 \rho^5} \Gamma_{\ab \db \eb} \bigg\{
	\Delta x_j \left(g^{\ab \jb} + u^{\ab \jb} \right)\bigg[
		\Delta x^{b c } \left(
			4 \Gamma_{\bb \cb \fb, \hb} - \Gamma^{\ib}{}_{\bb \cb} \Gamma_{\ib \fb \hb}
		\right)
		+ 4 u^{\bb} \Delta x^{c} \left( u \cdot \Delta x \right)\left(
			\Gamma_{\bb \cb \fb, \hb} - \Gamma^{\ib}{}_{\fb \hb} \Gamma_{\ib \cb \bb}
		\right) 
\nonumber \\ && \qquad
	+ 8 u^{\bb \cb} \left( u \cdot \Delta x \right)^2\left(
			\Gamma^{\ib}{}_{\fb [\hb} \Gamma_{|\ib| \cb] \bb} + \Gamma_{\fb \cb [\hb, \bb]}
		\right)
	\bigg]
	+ u^{\ab \bb} \Delta x^{i j} \Delta x_k \left( 2 g^{\cb \kb} + 3 u^{\cb \kb} \right)\Gamma_{\bb \fb \hb} \Gamma_{\cb \ib \jb}
\bigg\}
\nonumber \\ &&
-\frac{5 \Delta x^{d e f h i j} }{16 \rho^7} \Gamma _{\ab \db \eb} \Gamma _{\bb \fb \hb} \Gamma _{\cb \ib \jb}\left[
	\Delta x^{a b c} 
	+ 3 u^\ab \Delta x^{b c} \left( u \cdot \Delta x \right) 
	+ 3 u^{\ab \bb} \Delta x^{c} \left( u \cdot \Delta x \right)^2 
	+ u^{\ab \bb \cb} \left( u \cdot \Delta x \right)^3 
\right].
\nonumber
\end{eqnarray}


\subsection{The scalar self-force} \label{sec:dw:sf}

Once we have the singular scalar field, a simple partial derivative gives the required singular self-force of Eq.~\eqref{eqn:eom} in the form,
\begin{equation} \label{eqn:fasum}
F^{\rm{\sing}}_a \left(x\right) = \sum_{n=-1}^{\infty} \sum_{p=-n-2}^{\lfloor (n-1)/2 \rfloor}b^{[n]}_{\ab \bar{c}_1 \dots \bar{c}_{(n-2p)}}(\bar{x}) \Delta x^{c_1} \dots \Delta x^{c_{(n-2p)}} \zrho^{2 p - 1} \epsilon^{n-1},
\end{equation}
where one might note the different upper limit on the $p$-sum differs from that of the accelerated scalar self-force \cite{Heffernan:2017cad}. Explicitly, this looks like
\begin{eqnarray} 
F_{a \lnpow{1}}^{\rm{\sing}}&=&
\frac{-\Delta x^b}{\rho^3} \left(g_{\ab \bb}+u_{\ab \bb} \right), \label{eqn:FaSm1}\\
F_{a[0]}^{\rm{\sing}}&=&
- \frac{\Delta x^{b c}}{2\rho^3} \left[\left( \Gamma_{\db \bb \cb} u_\ab + 2 \Gamma_{\db \ab \bb} u_\cb \right) u^\db + \Gamma_{\ab \bb \cb} + 2 \Gamma_{\bb \ab \cb} \right] 
+\frac{3 \Delta x^{b c d e}}{2\rho^5} \left(g_{\ab \bb} +u_{\ab \bb}\right) \left(\Gamma_{\cb \db \eb} + \Gamma^\fb{}_{\cb \db} u_{\eb \fb} \right),\label{eqn:FaS0} \\
F_{a[1]}^{\rm{\sing}}&=& 
\frac1{3 \rho} u^{\bb \cb} \Delta x^d \left(
	2 \Gamma^{\eb}{}_{\ab [\db} \Gamma_{|\eb| \bb] \cb} + \Gamma_{\db \bb [\ab,\cb]} + \Gamma_{\ab \bb [\db, \cb]}
\right)
\label{eqn:FaS1} \\ &&
+\frac{\Delta x^d}{6\rho^3} \bigg[
	2 u^{\bb  \cb} \Delta x^{e f} \left(
		g_{\ab \fb} + 3 u_{\ab \fb} 
	\right) \left(
		\Gamma^{\hb}{}_{\bb [\db} \Gamma_{|\hb| \cb] \eb} + \Gamma_{\db \bb [\cb, \eb]}
	\right) 
	+ u_{\ab} u^{\bb} \Delta x^{ c e} \left(
		\Gamma^{\fb}{}_{\bb \cb} \Gamma_{\fb \db \eb} - \Gamma_{\bb \cb \db, \eb}
	\right)
\nonumber \\ && \quad
	+ \Delta x^{b c } \left(
		\Gamma^{\eb}{}_{\bb \ab} \Gamma_{\eb \cb \db} - 2 \Gamma_{\bb \cb \ab, \db} 
		- \Gamma_{\ab \bb \cb, \db} - \Gamma_{\bb \cb \db, \ab}
	\right)
	+ 2 u^{\bb \cb} \left( u \cdot \Delta x \right)^2 \left(
		2 \Gamma^{\eb}{}_{\bb [\ab} \Gamma_{|\eb| \cb] \db} + \Gamma_{\db \bb [\cb, \ab]} 
		+ \Gamma_{\ab \bb [\cb, \db]}
	\right)
\nonumber \\&& \quad
	+ u^{\bb} \Delta x^{c} \left( u \cdot \Delta x \right) \left(
		2 \Gamma^{\fb}{}_{\bb \db} \Gamma_{\fb \cb \ab} 
		+ \Gamma^{\fb}{}_{\bb \ab} \Gamma_{\fb \cb \db}
		- 2 \Gamma_{\bb \cb \ab, \db} - \Gamma_{\bb \cb \db, \ab}
	\right) - 3 u^{\bb \cb} \Delta x^{e f} \Gamma_{\bb \db \ab} \Gamma_{\cb \eb \fb}
\bigg]
\nonumber \\ &&
+\frac{\Delta x^{d e f}}{8 \rho^5} \bigg\{\left(
		g_{\ab \db}  + u_{\ab \db} 
	\right) \Big[
		\Delta x^{b c}  \left(
			4 \Gamma _{\bb \cb \eb}{}_{,\fb} - \Gamma ^{\hb}{}_{\bb \cb} \Gamma _{\hb \eb \fb} 
		\right)
		+ 4 u^{\bb} \Delta x^{c} \left( u \cdot \Delta x \right) \left(
			 \Gamma _{\bb \cb \eb}{}_{,\fb} -\Gamma ^{\hb}{}_{\bb \cb} \Gamma _{\hb \eb \fb} 
		\right)
\nonumber \\ && \qquad
		+ 8 u^{\bb \cb}  \left( u \cdot \Delta x \right)^2 \left(
			\Gamma ^{\hb}{}_{\bb [\cb} \Gamma _{|\hb| \eb] \fb} + \Gamma _{\eb \bb [\fb}{}_{,\cb]} 
		\right) 
	\Big] 
	+ 3 u^{\bb}  \Delta x^{c} \left[
		u^{\hb} \Delta x^i  \left( g_{\ab \ib} + 3 u_{\ab \ib} \right)
		+ 2 u_{\ab} \Delta x^{h} 
	\right] \Gamma _{\bb \cb \db} \Gamma _{\hb \eb \fb} 
\nonumber \\ && \quad	
	+ 6 u^{\bb} \Delta x^{c} \left( u \cdot \Delta x \right) \left[
		2 \Gamma _{\bb \cb \ab} \Gamma_{\db \eb \fb} + \Gamma _{\bb \eb \fb} \left(
			2 \Gamma _{\cb \db \ab} + \Gamma _{\ab \cb \db} 
		\right)
	\right]
	+ 12 u^{\bb \cb} \left( u \cdot \Delta x \right)^2 \Gamma _{\bb \db \ab} \Gamma _{\cb \eb \fb} 
	+ 6\Delta x^{b c} \Gamma _{\db \eb \fb} \left(
		2\Gamma _{\bb \cb \ab} + \Gamma _{\ab \bb \cb}
	\right) 
\bigg\}
\nonumber \\ &&
-\frac{15 \Delta x^{b e f h i}}{8 \rho^7} \left(g_{\ab \bb}+u_{\ab \bb} \right) \bigg[
	\Delta x^d\Delta x_j\left( g^{\cb \jb} + 2 u^{\cb \jb} \right)\Gamma _{\cb \db \eb} \Gamma _{\fb \hb \ib}
	+ u^{\cb \db} \left( u \cdot \Delta x \right)^2 \Gamma _{\cb \eb \fb} \Gamma _{\db \hb \ib} 
\bigg]
, \nonumber
\end{eqnarray}
\end{widetext}
where we have omitted the larger term of $F_{a[2]}^{\rm{\sing}}$ due to its size, however it is provided in a usable format as regularization parameters 
online \cite{heffernan_anna_2022_6282572, BlackHolePerturbationToolkit}. From Eqs.~\eqref{eqn:FaSm1}, \eqref{eqn:FaS0} and \eqref{eqn:FaS1}, one can simply read off the $b$ coefficients of Eq.~\eqref{eqn:fasum}, e.g.,
\begin{eqnarray} \label{eqn:b}
b^{[0]}_{\ab \bb \cb}(\bar{x})&=& -\frac12 \left[\left( \Gamma_{\db \bb \cb} u_\ab + 2 \Gamma_{\db \ab \bb} u_\cb \right) u^\db + \Gamma_{\ab \bb \cb} + 2 \Gamma_{\bb \ab \cb} \right], \nonumber \\
b^{[0]}_{\ab \bb \cb \db \eb}(\bar{x})&=&\frac{3}{2} \left(g_{\ab \bb} +u_{\ab \bb}\right) \left(\Gamma_{\cb \db \eb} + \Gamma^\fb{}_{\cb \db} u_{\eb \fb} \right).
\end{eqnarray}


\section{Mode-sum regularization} \label{sec:ms}
Once we have the singular self-force, regularization is obtained by simply subtracting the singular from the (numerically calculated) retarded self-force. However, subtracting an infinity from an infinity can prove problematic; this lead Barack and Ori \cite{Barack:1999wf} to propose a spherical harmonic decomposition of both, allowing subtraction mode by mode,
\begin{equation} \label{eqn:Fa}
F_a(\xb)=\sum^\infty_{\ell}\left[F_a^{\ell \rm{\ret}}(\xb) - F_a^{\ell \rm{\sing}} (\xb)\right],
\end{equation}
where
\begin{eqnarray} \label{eqn:FRS}
F_a^{\ell\rm{\sing}}(\xb) &=& \sum_{m} F_a^{\ell m \rm{\sing}} (\tb, \rb) Y_{\ell m} (\bar{\theta}, \bar{\phi}), \nonumber \\
&=& \sqrt{\frac{2\ell+1}{4\pi}} F_a^{\ell 0\rm{\sing}} (\tb, \rb), \nonumber \\
&=&\lim_{\Delta r \rightarrow 0} \frac{2\ell+1}{4 \pi}  \int F_a^{\rm{\sing}} (0,\Delta r,\alpha,\beta) \nonumber \\
&& \quad \times P_\ell(\cos{\alpha}) d \Omega,
\end{eqnarray}
Here, as has become standard in mode-sum calculations \cite{Barack:2002mha}, we are operating in a rotated coordinate system $(\theta,\phi) \rightarrow (\alpha,  \beta)$ where the particle is on the pole $(\bar{\alpha}=0=\bar{\beta})$, immediately reducing the sum of $m$-modes to $m=0$ in the second equality. The format of Eq.~\eqref{eqn:fasum} and the assumption that both $x$ and $\xb$ are on the same hypersurface allows us to write
\[ 
F_a^{\rm{\sing}}(x)=F^{\rm{\sing}}_a(\Delta t=0, \Delta r, \Delta \alpha=\alpha, \Delta \beta=\beta).
\]
In the third equality, it is necessary to take the limit $x\rightarrow\xb$, to avoid the discontinuity at the particle. As is typical, this limit is taken in the radial direction; this must be mirrored in the calculation of the retarded self-force to insure the singularities cancel exactly, in particular from which direction along the radial axis.\\


\subsection{Coordinate system} \label{sec:ms:coord}

\begin{figure} 
\includegraphics[width=6.5cm]{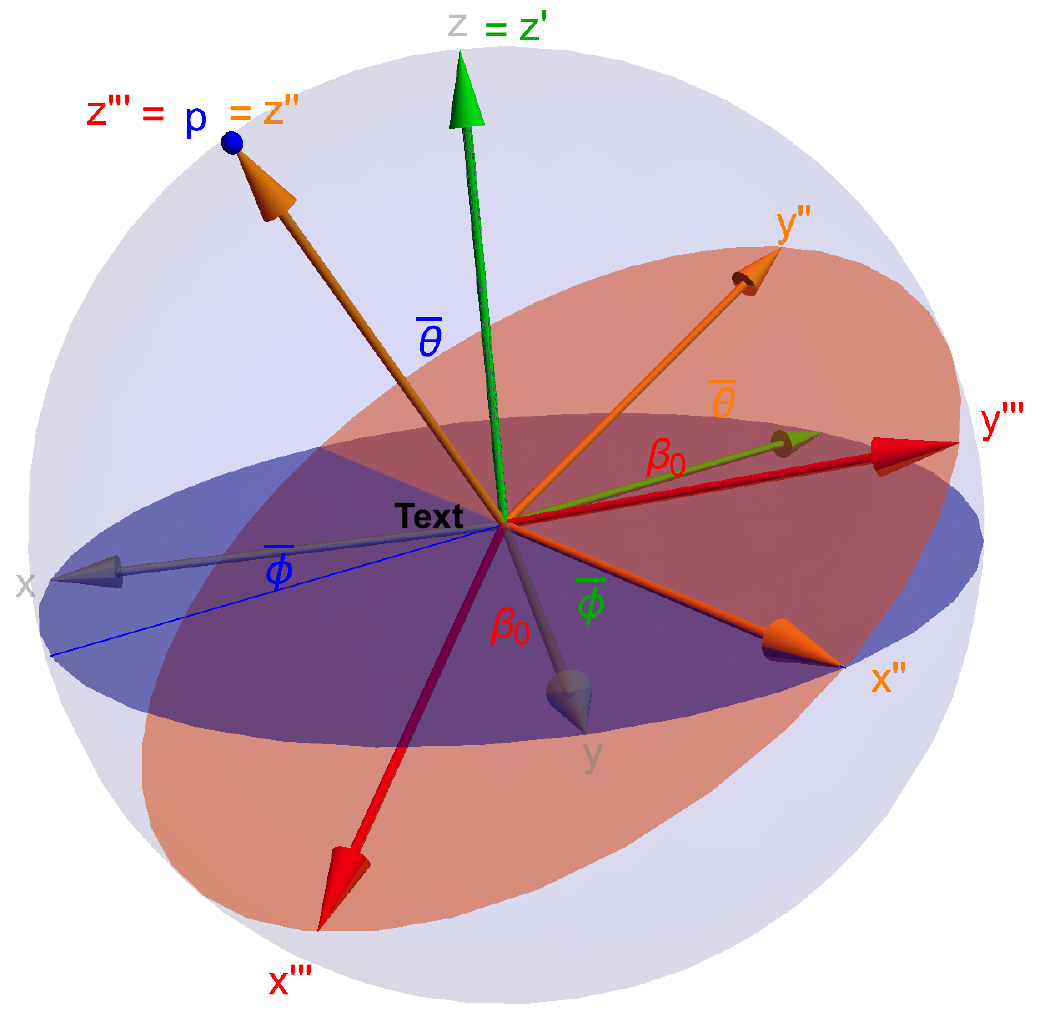} 
\caption{Taking the particle to have spherical coordinates $(\rb,\thb, \phb)$, we rotate through the Euler angles $(\phb+\frac{\pi}{2}, \thb, -\beta_0)$, that is $\phb+\frac{\pi}{2}$ and $\thb$ around the original $z$-axis and rotated $x'$-axis respectively (resulting in the green and orange axes). Our new coordinates have the particle on the pole. We rotate through a third Euler angle, $-\beta_0$, that is a clockwise rotation $\beta_0$ around the $z''$-axis resulting in the red axes.}
\label{fig:rot3}
\end{figure}

In order to carry out the integration in Eq.~\eqref{eqn:FRS}, it is important to choose suitable coordinates. As already mentioned, placing the particle on the pole is the first step; for Schwarzschild, this previously entailed a counter-clockwise rotation of $\pi/2$ around the $x$-axis \cite{Barack:2002mha}. When considering generic orbits in Kerr, we require two rotations usually obtained by the first two Euler angles as illustrated in Fig.~\ref{fig:rot3}. Indeed we choose the rotated coordinate system described by the Euler angles $(\phb+\frac{\pi}{2}, \thb, -\beta_0)$ or explicitly
\begin{eqnarray} \label{eqn:coordRotn}
\sin{\alpha} \cos{(\beta-\beta_0)}&=&\sin{\theta} \sin{(\phi-\phb)},\\
\sin{\alpha} \sin{(\beta-\beta_0)}&=&\cos{\theta} \sin{\thb}-\sin{\theta}  \cos{\thb} \cos{(\phi-\phb)},\nonumber \\
\cos{\alpha}&=&\cos{\theta} \cos{\thb}+\sin{\theta} \sin{\thb} \cos{(\phi-\phb)}, \nonumber
\end{eqnarray}
where we reserve $\beta_0$ for later simplification; for comparison Barack and Ori \cite{Barack:2002mh} used $(\phb+\frac{\pi}{2}, \thb, -\beta_0-\frac{\pi}{2})$ as described in Barack's later review \cite{Barack:2009}.

\begin{figure}
\includegraphics[width=6.5cm]{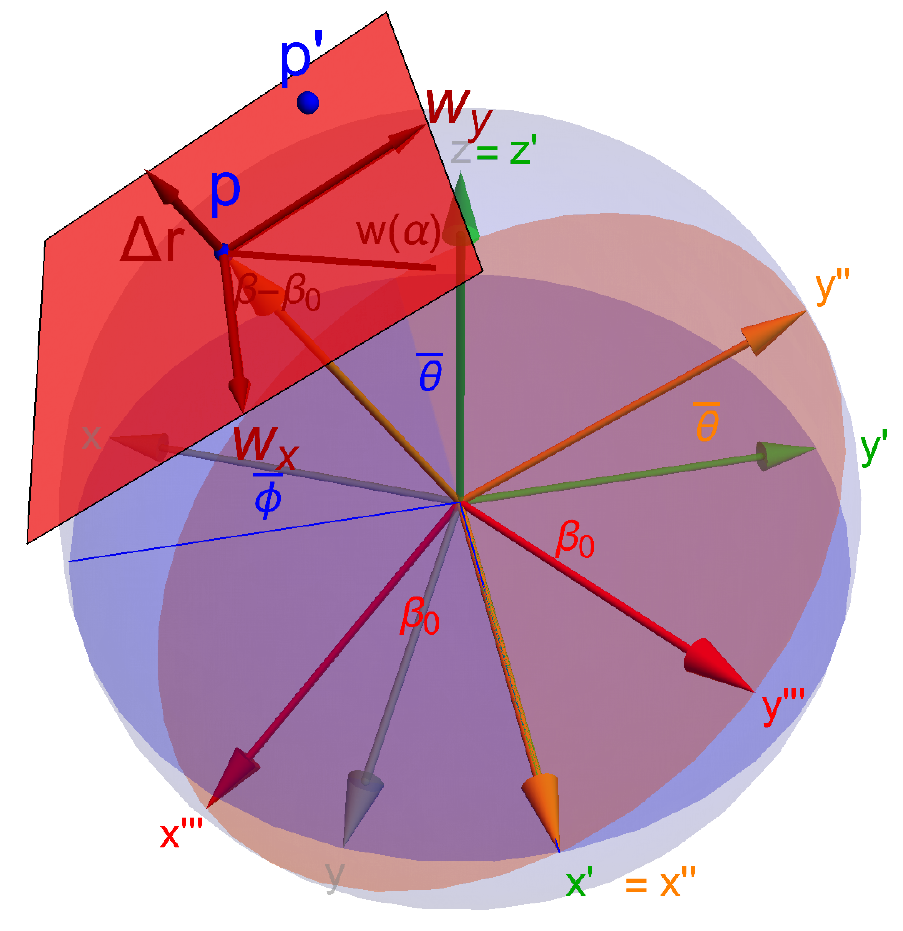}
\caption{Locally Cartesian coordinates $(w_x, w_y)$ align with our second rotation axes $(x'', y'')$ and allow coordinate expansions off the particle and its worldline. The $z$-axis of these coordinates is parallel to the rotated $z''$-axis. With the particle on $(0,0,0)$ in these coordinates, 
we note a neighbouring particle will have coordinates $(w_x,w_y,\Delta r)$.}
\label{fig:rotw}
\end{figure}

These rotated coordinates are not well behaved at the particle (as it is on the pole); to enable coordinate expansions off the particle's worldline, a further coordinate transformation to ``locally Cartesian coordinates'', or $(x, y)$ in \cite{Barack:2002mha}, is required. More recent literature \cite{Heffernan:2012su, Heffernan:2012vj, Heffernan:2017cad} have these labeled $(w_1, w_2)$; we therefore settle on $(w_x, w_y)$,
\begin{eqnarray} 
w_x&=w(\alpha) \cos(\beta-\beta_0)&=2 \sin \left(\frac{\alpha}{2}\right) \cos(\beta-\beta_0), \label{eqn:wx} \\
w_y&=w(\alpha) \sin(\beta-\beta_0)&=2 \sin \left(\frac{\alpha}{2}\right) \sin(\beta-\beta_0),\label{eqn:wy}
\end{eqnarray}
where $w(\alpha)$ is selected due to its regularity, that $w(\alpha)=\alpha+\mathcal{O}(\alpha)^3 \approx \sin\alpha$, and that it monotonically increases for $\alpha \in [0,\pi)$. These coordinates align with those of our second rotation $(x'', y'')$ as illustrated in Fig.~\ref{fig:rotw}. 

The singular scalar self-force from Eq.~\eqref{eqn:fasum}  of a field point `close' to that particle at $(\rb, \thb, \phb)$ on the same hypersurface can now be written 
\[
F_a^{\rm{\sing}}(x)=F_a^{\rm{\sing}}(\Delta t=0, \Delta r, \Delta w_x=w_x, \Delta w_y=w_y),
\]
where again we see the sign of $\Delta r$ will depend from which direction the limit is taken; above or below the plane. To obtain such an expression, the initial mode-sum papers \cite{Barack:2002mha, Detweiler:2002gi, Haas:2006ne} calculated $F_a^{\rm{\sing}}(x)$ in Cartesian coordinates, and reexpanded into the $w$ coordinates to then carry out the integration of Eq.~\eqref{eqn:FRS}. Here we follow Heffernan et al. \cite{Heffernan:2012su, Heffernan:2012vj} who illustrated this reexpansion was a limiting feature in pushing to higher orders. It is more efficient to calculate $F_a^{\rm{\sing}}(x)$ in the $w$ coordinates and carry out a coordinate transformation back to spherical coordinates for integration. The coordinate transformation does not affect $(t,r)$ components, however, for the angular components, one has
\begin{widetext}
\begin{eqnarray}
F_{\theta}&=&-F_{w_y [-1]} \epsilon^{-2} -\left(F_{w_y [0]} - w_x \cot \thz F_{w_x [-1]}  \right) \epsilon^{-1}\\
&& 
-\left\{F_{w_y [1]} - w_x \cot \thz F_{w_x [0]} +  \frac14 w_x w_y \left( 1-4 \csc^2 \thz \right) F_{w_x [-1]} 
+ \frac18 \left[ w_x^2 \left(5-4\csc^2 \thz \right) - w_y^2 \right] F_{w_y [-1]} \right\} \epsilon^0 
\nonumber \\
&&-\bigg\{
F_{w_y [2]} - w_x \cot \thz F_{w_x [1]}  +\frac14 w_x w_y\left( 1-4 \csc^2 \thz \right) F_{w_x [0]} + \frac18 \left[ w_x^2 \left(5-4\csc^2 \thz \right) - w_y^2 \right] F_{w_y [0]}  \nonumber \\
&& \quad 
+ \frac14 w_x^2 w_y \cot \thz \left( 3-4 \csc^2 \thz \right) F_{w_y [-1]} + \frac18 w_x \csc^2 \thz \cot \thz \left[ w_x^2 \left(3+2\cos 2 \thz \right) - 8 w_y^2 \right] F_{w_x [-1]} 
\bigg\} \epsilon + \mathcal{O}(\epsilon^2), \nonumber\\
F_{\phi}&=& \sin \thz F_{w_x [-1]} \epsilon^{-2} +\sin \thz  \left[F_{w_x [0]} + \cot \thz \left( w_x F_{w_y [-1]} -w_y F_{w_x [-1]}   \right) \right]\epsilon^{-1} \\
&& 
+ \sin \thz \left\{F_{w_x [1]} + \cot \thz \left( w_x F_{w_y [0]} -w_y F_{w_x [0]}   \right)+  \frac18  \left[2 w_x w_y F_{w_y [-1]}  - \left(w_x^2 +3 w_y^2\right) F_{w_x [-1]}  \right] \right\}\epsilon^0 \nonumber \\
&&+ \sin \thz \left\{
F_{w_x [2]} +\cot \thz \left( w_x F_{w_y [1]} -w_y F_{w_x [1]}   \right) +  \frac18  \left[2 w_x w_y F_{w_y [0]}  - \left(w_x^2 +3 w_y^2\right) F_{w_x [0]}  \right]  \right\}\epsilon + \mathcal{O}(\epsilon^2). \nonumber
\end{eqnarray}
\end{widetext}
In previous Kerr calculations, the constraint of an equatorial plane \cite{Heffernan:2012vj} or a lower order \cite{Barack:2002mh} meant one could discard $F_{a [1]}$ as it only appeared in the $\epsilon^0$ term (which always integrates to zero as outlined in the next section). It is therefore worth noting that $F_{a [1]}$ will be required due to its presence in the $\epsilon^1$ term.

In Eq.~\eqref{eqn:fasum}, all $b$ coefficients, which concern the metric, four-velocity, Christoffel symbols and their derivatives, are evaluated at the particle (see Eq.~\eqref{eqn:b} for example). It is still necessary to calculate the Christoffel symbols at $x$ to allow differentiation before evaluating at $\xb$. We use Boyer-Lindquist as our base coordinates, 
\begin{eqnarray}
g_{t b}^{(BL)}&=&\left\{-1+\frac{2 m r}{\Sigma}, 0, 0,-\frac{2 a m r \sin^2{\theta}}{\Sigma} \right\},\\
g_{r b}^{(BL)}&=&\left\{0, \frac{\Sigma}{\Delta},0,0\right\}, \nonumber \\
g_{\theta b}^{(BL)}&=&\left\{0,0,\Sigma,0\right\}, \nonumber \\
g_{\phi b}^{(BL)}&=&\left\{-\frac{2 a m r \sin^2{\theta}}{\Sigma}, 0, 0, \left[\Delta+2 m r \frac{r^2+a^2}{\Sigma}\right]\sin^2{\theta}\right\}, \nonumber
\end{eqnarray}
where $m$ is total mass, $a=J/m$, $J$ is the angular momentum and
\begin{eqnarray}
\Sigma=r^2+a^2 \cos{\theta}, \qquad \Delta=r^2-2 m r +a^2.
\end{eqnarray}
Those components of the metric that differ in the $w$ coordinates are
\begin{widetext}
\begin{eqnarray}
g_{t w_x}^{(w)}&=&
\frac{g_{t\phi}^{(BL)}}{4 \sin^2 \theta} \left[ 
	w_y \left( w_x^2+w_y^2-4\right) \cos{\thz} 
		+\frac{\left( 8-6 w_y^2+w_x^2 w_y^2 +w_y^4\right)\sin{\thz} }{\sqrt{4-w_x^2-w_y^2}} 
\right], \\
g_{t w_y}^{(w)}&=&
-\frac{w_x g_{t \phi}^{(BL)}}{4 \sin^2 \theta} \left[ 
	\left( w_x^2+w_y^2-4\right) \cos{\thz} 
		+\frac{w_y\left( w_x^2 +w_y^2-6 \right)\sin{\thz} }{\sqrt{4-w_x^2-w_y^2}} 
\right], \nonumber \\
g_{w_x w_x}^{(w)}&=&
\frac{w_x^2 g_{\theta \theta}^{(BL)}}{4 \sin^2 \theta} 
	\left( 2 \cos{\thz}+\frac{w_y \sin{\thz}}{\sqrt{4-w_x^2-w_y^2}} \right)^2
\nonumber\\&&\quad 
+ \frac{g_{\phi \phi}^{(BL)}}{16 \sin^4 \theta} \left[ 
	w_y \left( w_x^2+w_y^2-4\right) \cos{\thz} 
	+\frac{\left(8-6 w_y^2+w_x^2 w_y^2+ w_y^4 \right) \sin{\thz}}{\sqrt{4-w_x^2-w_y^2}}
\right]^2, \nonumber  \\
g_{w_x w_y}^{(w)}&=&
\frac{w_x g_{\theta \theta}^{(BL)}}{8 \sin^2 \theta} \left\{
	\frac{w_y\left[ 
		3 w_x^2-12+2 w_y^2+\left(5 w_x^2+6 w_y^2-20 \right) \cos{2\thz}
	\right]}{w_x^2+w_y^2-4}
	+ \frac{2 \left( w_x^2+3 w_y^2-4\right)\sin{2\thz}}{\sqrt{4-w_x^2-w_y^2}}
\right\} \nonumber \\ &&\quad
+\frac{w_x g_{\phi \phi}^{(BL)}}{16 \sin^4 \theta} \Bigg\{
	w_y \left[
		w_x^2\left( 8 -3 w_y^2 + \frac{4}{w_x^2+w_y^2-4}\right) - 2\left( 14-8w_y^2+w_y^4\right)-w_x^4
	\right] \cos{\thz}^2 \nonumber \\ && \qquad
	+ \frac{w_y \left(w_x^2+w_y^2-6\right) \left(8-6 w_y^2+w_x^2 w_y^2 +w_y^4\right)}{w_x^2+w_y^2-4}
	+\sqrt{4-w_x^2-w_y^2} \left[
		4-6w_y^2+w_x^2 w_y^2+w_y^4
	\right]\sin{2 \thz}
\Bigg\}, \nonumber \\
g_{w_y w_y}^{(w)}&=&
\frac{g_{\theta \theta}^{(BL)}}{\sin^2 \theta} \left[
	w_y \cos{\thz} + \frac{\left( w_x^2+2 w_y^2-4 \right) \sin{\thz}}{2 \sqrt{4-w_x^2-w_y^2}}
\right]^2
+\frac{w_x^2 g_{\phi \phi}^{(BL)}}{16 \sin^4 \theta} \left[
	\left(w_x^2+w_y^2-4\right) \cos{\thz} + \frac{ w_y \left(w_x^2+w_y^2-6\right) \sin{\thz}}{\sqrt{4-w_x^2-w_y^2}}
\right]^2,
\nonumber
\end{eqnarray}
where from Eqs.~\eqref{eqn:coordRotn} to \eqref{eqn:wy},
\begin{equation}
\sin^2 \theta=1-\frac14\left[\left(w_x^2+w_y^2-2\right) \cos{\thz}-w_y \sqrt{4-w_x^2-w_y^2} \sin{\thz} \right]^2.
\end{equation}
\end{widetext}
The four-velocity, $u^\ab$ which also arises in the $b$ coefficients are evaluated at the particle in our $w$ coordinates,
\begin{equation}
u^\ab = \left\{ 
	u^\tb, u^\rb, \sin{\thz} u^{\bar{\phi}}, -u^{\bar{\theta}} 
\right\},
\end{equation}
where the prerotated Boyer-Lindquist 4-velocities are given by
\begin{eqnarray}
u^\tb &=& E \left[ 
	\frac{\left(a^2 + \rb^2 \right)^2}{ \Delta \Sigma} - \frac{ a^2 \sin{^2\thb}}{\Sigma} \right]
	- \frac{2 a L m \rb}{\Delta \Sigma}, \nonumber \\
\left( u^\rb \right)^2&=&\frac1{\Sigma^2} \Big\{
	\left[ E \left(\rb^2+a^2 \right) - a L \right]^2 
	\nonumber \\ && \quad
	- \Delta \left[ \rb^2 + \left(L - a E \right)^2 + Q \right] 
\Big\}, \nonumber \\
\left( u^\thb \right)^2&=& \frac1{\Sigma^2} \left[
	Q - L^2 \cot{^2\thb}  - a^2 \left(1 - E^2 \right) \cos{^2\thb} 
\right], \nonumber \\
u^\phb &=& \frac{L \csc{^2\thb}}{\Sigma} + \frac{a \left( 2 E m r - a L \right)}{ \Delta \Sigma},
\end{eqnarray}
and $\{E, Q, L\}$ are the energy, Carter constant and azimuthal angular momentum, respectively.


\subsection{Integration} \label{sec:ms:int}
The first thing to note about the integration of Eq.~\eqref{eqn:FRS}, is that the leading order of the singular self-force, $F_{a[-1]}^{\sing}$ is treated differently than the higher terms. The regularity of the higher terms enables the interchanging of the limit $\Delta r \rightarrow 0$ and the integral (see Appendix B of \cite{Barack:2002mha}). 


\subsubsection{Format of $\rho^2$}
Regardless of the term, one can use the rotated coordinates to target problematic expressions. In particular, we adapt the method first introduced for Schwarzschild by Detweiler et al. \cite{Detweiler:2002gi}, later refined by Haas and Poisson \cite{Haas:2006ne} with eccentric orbits and extended to higher-order terms, equatorial Kerr and nongeodesic motion by Heffernan et al.\cite{Heffernan:2012su, Heffernan:2012vj, Heffernan:2017cad}. For higher terms, the key to the method is the rewriting of 
\begin{equation}
\rho(x)=\rho(\Delta t=0,\Delta r=0, \Delta w_x=w_x, \Delta w_y=w_y), \nonumber
\end{equation}
in the format,
\begin{eqnarray}
\rho(0,0,w_x,w_y)^2&=&(g_{\ab \bb}+u_{\ab \bb}) \Delta x^{a b}, \\
&=&4 \sin^2\left( \frac{\alpha}{2}\right) f(\xb) \chi (\xb, \beta) \nonumber,
\end{eqnarray}
where
\begin{equation} \label{eqn:chi}
\chi  (\xb, \beta)= 1-k(\xb) \sin^2 \beta.
\end{equation}
This translates to having no $\Delta w_x \Delta w_y$ cross terms. In the case of Schwarzschild and equatorial orbits in Kerr, this is automatic as $g_{\thb \phb}=0=u^\thb$; for generic orbits in Kerr, this is not the case.

Similar to \cite{Heffernan:2012vj, Heffernan:2012xlf, Heffernan:2017cad}, we rewrite $\rho$ as
\begin{equation}
\rho^2(0,0,w_x,w_y)= \zeta^2 w_x^2 + 2 \mu w_x w_y + \xi w_y^2,
\end{equation}
where
\begin{eqnarray}
\zeta^2&\equiv&g_{w_x w_x}+u_{w_x w_x}, \\
	&=& L^2 \csc^2\thb+\frac{2 a^2 m \rb + 2 m \rb^3 + \Delta \Sigma}{\Sigma},\nonumber\\
\xi^2&\equiv&g_{w_y w_y}+u_{w_y w_y}
= \Sigma+ \Sigma^2 (u^\thb)^2, \nonumber \\
\mu&\equiv&g_{w_x w_y}+u_{w_x w_y} 
=- L  \Sigma \csc \thb u^\thb. \nonumber
\end{eqnarray}
Recalling the definition of our $w$ coordinates from Eqs.~\eqref{eqn:wx} and \eqref{eqn:wy}, we have
\begin{widetext}
\begin{eqnarray} \label{eqn:rho00}
\rho(0,0,w_x,w_y)^2 &=& 4 \sin^2 \left(\frac{\alpha}{2}\right) \left[
	\zeta^2 \cos^2(\beta-\beta_0) 
	+\mu \sin2(\beta-\beta_0)+\xi^2\sin^2(\beta-\beta_0)
\right], \nonumber \\
&=& 4 \sin^2 \left(\frac{\alpha}{2}\right) \bigg\{
	\frac12 \left[
		\zeta^2+\xi^2 - \left(\xi^2 - \zeta^2\right) \cos 2\beta_0-2\mu \sin 2\beta_0
	\right] +\sin^2 \beta \left[
		\left(\xi^2 - \zeta^2\right) \cos 2\beta_0+2\mu \sin 2\beta_0
	\right] \nonumber \\ && \quad - 
	\cos \beta \sin \beta \left[
		\left(\xi^2 - \zeta^2\right) \sin 2\beta_0-2\mu \cos 2\beta_0
	\right]
\bigg\}.
\end{eqnarray}
\end{widetext}
The $\cos \beta \sin \beta$ term can now be eradicated using the freedom of $\beta_0$, that is we choose $\beta_0$ so that
\begin{equation} \label{eqn:tan}
\tan 2 \beta_0 =\frac{ 2\mu}{\xi^2 - \zeta^2}.
\end{equation}
Alternatively,
\begin{equation}\label{eqn:cossin}
\sin 2\beta_0 = -\frac{2\mu}{\eta}, \quad \cos 2 \beta_0 = \frac{ \zeta^2 - \xi^2}{\eta}, \\
\end{equation}
where
\begin{equation}
\eta^2=4\mu^2+(\xi^2 - \zeta^2)^2.
\end{equation}
Substituting our choice of $\beta_0$ back into $\rho^2$ via Eq.~\eqref{eqn:rho00}, we get
\begin{eqnarray}
\rho(w_x,w_y)^2 &=& 2 \sin^2  \frac{\alpha}{2}\left[
	\xi^2 + \zeta^2+\eta- 2 \eta \sin \beta
\right],  \\
 &=& 2 \sin^2  \frac{\alpha}{2} \left(\xi^2 + \zeta^2+\eta \right)\left[
	1-k(\xb) \sin^2 \beta
\right], \nonumber \\
&=& 4 \sin^2  \frac{\alpha}{2} \left(\frac{\eta}k \right)\chi,
\end{eqnarray}
where
\begin{equation} \label{eqn:k}
k(\xb) \equiv \frac{2 \eta}{\xi^2 + \zeta^2 +\eta}.
\end{equation}
This, combined with our $w$-coordinate definitions of Eqs.~\eqref{eqn:wx} and \eqref{eqn:wy}, allows us to rewrite $F^{\rm{\sing}}_a$ of Eq.~\eqref{eqn:fasum} as
\begin{widetext}
\begin{eqnarray} \label{eqn:fasum00}
F^{\rm{\sing}}_a \left(0,0,\alpha,\beta\right) 
&=&
 \sum_{n=-1}^{\infty} \sum_{p=-n-2}^{\lfloor (n-1)/2\rfloor} \sum_{q=0}^{n-2p} b^{[n]}_{\ab (n-2p,q)}(\bar{x}) \left(2\sin^2 \frac{\alpha}{2}\right)^{(n-1)/2} 2^{(n-1)/2} \binom{n-2p}{q}
\nonumber \\ && \qquad \qquad \qquad \times
\cos^{q}(\beta-\beta_0) \sin^{(n-2p-q)} (\beta-\beta_0) 
	\left(\frac{\eta \chi }k\right)^{(2 p - 1)/2} \epsilon^{n-1}, 
\end{eqnarray}
\end{widetext}
where we have relabeled the $b^{[n]}_{\ab (n-2p,q)}$ coefficient,
\begin{equation}
b^{[n]}_{\ab (n-2p, q)} \equiv b^{[n]}_{\ab \bar{c}_1 \dots \bar{c}_{(n-2p)}},
\end{equation}
where $n-2p$ is the total number of $w_x$'s and $w_y$'s in the sequence $\bar{c}_1 \dots \bar{c}_{(n-2p)}$ and $q$ the number of $w_x$'s.


\subsubsection{$\alpha$ integral}

For the $\alpha$ integral, we follow the techniques of Appendix D in \cite{Detweiler:2002gi}; we rewrite the integral in Eq.~\eqref{eqn:FRS} for the subleading $(n\geq0)$ orders as
\begin{align} \label{eqn:intAlpha}
\frac{2\ell+1}{4\pi}&\int_0^\pi F^{\rm{\sing}}_{a[n]} \left(0,0,\alpha,\beta\right) P_\ell \left( \cos \alpha \right) \sin \alpha d \alpha
\nonumber \\
&= f_{a[n]}(\beta)\epsilon^{n-1} \frac{2\ell+1}{4\pi}\int_{-1}^1 \left(1-y \right)^{(n-1)/2} P_\ell (y) d y, \nonumber \\
&= f_{a[n]}(\beta)\epsilon^{n-1} \frac1{2 \pi} \mathcal{A}_{[n]}^\ell,
\end{align}
where we have expanded $\left(1-y \right)^{(n-1)/2} $ in Legendre polynomials
\begin{equation} \label{eqn:yAl}
\left(1-y \right)^{(n-1)/2}=\sum_{\ell=0} \mathcal{A}_{[n]}^\ell P_\ell (y),
\end{equation}
and used their orthogonality,
\begin{equation}
\int_{-1}^{1} P_\ell (y) P_{\ell'} (y) dy = \frac{2 \delta_{\ell \ell'}}{2 \ell+1}.
\end{equation}

For odd $n$, the left-hand side of Eq.~\eqref{eqn:yAl} gives a finite polynomial and we observe the sum over $\ell$ will truncate at $\ell=(n-1)/2$. For $n=1$, we get $\mathcal{A}^\ell_{[1]} = \delta_{\ell 0}$ and by considering $y=1$ for odd $n\geq 3$, we have
\begin{equation}
\sum_{\ell=0}^{(n-1)/2} \mathcal{A}_{[n]}=0.
\end{equation}
Therefore, as long as we are summing $\ell\geq (n-1)/2$, odd $n\geq 3$ orders will not have a contribution. Indeed, for even $n>0$, we similarly have
\begin{equation}
\sum_{\ell=0}^{\infty} \mathcal{A}_{[n]}=0.
\end{equation}
However as summing over all $\ell$ is not practical, we require the parameters to the designated truncation $\ell$.

For $n=0$, taking the Legendre generating function for $|t|<1$,
\begin{equation}
(t^{-1}-2y+t)^{-1/2}=\sum_{\ell=0} t^{\ell+1/2} P_\ell (y),
\end{equation}
and making the substitution $t=e^{- d \delta} \rightarrow 1+\mathcal{O}(\delta)$ where $0<d\in\mathcal{R}$. In the limit $\delta \rightarrow 0$, one gets
\begin{equation}
(1-y)^{-1/2}=\sum_\ell \sqrt{2} P_\ell (y)  +\mathcal{O}( \ell \delta),
\end{equation}
giving $\mathcal{A}^\ell_{[0]}=\sqrt{2}+\mathcal{O}( \ell \delta)$ in Eq.~\eqref{eqn:yAl}. 

For even $n\geq2$, we make use of the formula,
\begin{align} \label{eqn:Ael}
	\mathcal{A}_{[n]}^{\ell} =& 
		 \frac{\mathcal{P}_{[n]} \left( 2 \ell + 1\right) }{\left(2 \ell - n +1\right)\left(2 \ell - n +3\right) \cdots \left(2 \ell + n -1\right) \left(2 \ell + n +1 \right)},
\end{align}
where
\begin{align} \label{eqn:Pn}
	\mathcal{P}_{[n]} = &\left(-1\right)^{n/2} 2^{(n + 1)/2} \left[(n-1)!! \right]^2. \nonumber \\
\end{align}
derived by induction in Appendix D of \cite{Detweiler:2002gi}. We now have as a refinement of Eq.~\eqref{eqn:FRS},
\begin{eqnarray} \label{eqn:FRS2}
F_a^{\ell\rm{\sing}}&&(\xb) = \lim_{\Delta r \rightarrow 0} \frac{2\ell+1}{4 \pi} \epsilon^{-2} \int F_{a[-1]}^{\rm{\sing}} (0,\Delta r,\alpha,\beta) \nonumber \\
&& \times P_\ell(\cos{\alpha}) d \Omega +\frac1{2 \pi}   \delta_{\ell 0}  \epsilon^{0} \int_0^{2\pi} f_{a[1]}(\beta) d\beta \nonumber \\
&& + \frac1{2 \pi}  \sum_{\substack{n=0\\ \text{even } n}} \mathcal{A}_{[n]}^\ell  \epsilon^{n-1} \int_0^{2\pi} f_{a[n]}(\beta) d\beta. 
\end{eqnarray}


\subsubsection{$\beta$ integral}

From Eqs.~\eqref{eqn:fasum00} and \eqref{eqn:intAlpha}, we read off
\begin{eqnarray} \label{eqn:fbeta}
f_{\ab[n]}(\beta)&&= \sum_{p=-n-2}^{\lfloor (n-1)/2 \rfloor} \sum_{q=0}^{n-2p} b^{[n]}_{\ab (n-2p,q)}(\bar{x}) 2^{(n-1)/2}  \nonumber \\ &&
\times \binom{n-2p}{q} \cos^{q}(\beta-\beta_0) \sin^{(n-2p-q)}   (\beta-\beta_0) 
\nonumber \\ &&  \times
\left(
	\frac{\eta \chi}k 
\right)^{(2 p - 1)/2}, 
\end{eqnarray}
where we have used Eq.~\eqref{eqn:k} to introduce $k$. For the remaining odd $n=1$ case, the powers of $\sin \beta$ and $\cos \beta$ will always sum to an odd number. For odd powers of $\sin \beta$ this will integrate to zero as $\chi$ is an even function of $\beta$. Odd powers of $\cos \beta$ will also integrate to zero; this can be shown with the standard trick of splitting the integral, 
\begin{eqnarray}
\int_0^{2\pi} = \int_0^{3\pi/2} +\int_{3\pi/2}^{2\pi}  = \int_0^{3\pi/2} +\int_{-\pi/2}^0=\int_{-\pi/2}^{3\pi/2} \nonumber 
\end{eqnarray}
where the second equality shifts the second integral by $2\pi$ without affecting the trigonometric functions. A shift of $\beta \rightarrow \beta - \pi/2$ now gives
\begin{equation}
\int_{-\pi/2}^{3\pi/2} \frac{\cos^a \beta \sin^b \beta}{ \chi[\sin^2(\beta)]^c} d\beta = \int_0^{2\pi} \frac{\sin^a \beta (-\cos \beta)^b } {\chi[\cos^2(\beta)]^c} d\beta
\end{equation}
which is now odd in $\beta$ for odd $a$ and will integrate to zero. We are now left with only our leading and even $n$ terms in Eq.~\eqref{eqn:fbeta}.

When expanding out the $(\beta-\beta_0)$ functions in the remaining even $n$ terms, it should be noted that any odd partitions of $\cos \beta$ and $\sin \beta$ will also integrate to zero as again $\chi$ is an even function of $\beta$.  We are therefore always dealing with even powers of $\cos \beta$ and $\sin \beta$. Recalling our definition of $\chi$, Eq.~\eqref{eqn:chi}, we have
\begin{eqnarray}
\cos^2 (\beta)&=&\frac1k\left(
	k-1+\chi \right) , \\
\sin^2 (\beta)&=&\frac1k\left(
	1-\chi
\right). \nonumber
\end{eqnarray}
Substituting this back into Eq.~\eqref{eqn:fbeta} gives
\begin{eqnarray} \label{eqn:fabeta}
f_{\ab[n]}(\beta)&=& \sum_{p=-n-2}^{\lfloor (n-1)/2 \rfloor} \sum_{q=0}^{n-2p} 
	b^{[n]}_{\ab (n-2p,q)}(\bar{x}) \left(\frac2k\right)^{(n-1)/2} \nonumber \\ &&
\times \sum_s^q \sum_t^{n-2p-q} \frac{(n-2p)!}{s! t! (q-s)! (n-2p-q-t)!}\nonumber \\&&
\times \left(k-1+\chi\right)^{(s+t)/2}\cos^{(n-2p-q-t+s)} \beta_0 \nonumber \\&&
\times (-1)^t\left(1-\chi \right)^{(n-2p-s-t)/2} \sin^{(q+t-s)} \beta_0
\nonumber \\  &&\times
\left(
	\eta \chi
\right)^{(2 p - 1)/2}, 
\end{eqnarray}
where the zero contribution of odd powers in $\cos \beta$ and $\sin \beta$ requires $(s+t)$ be even. With $n$ also even, we can multiply out the two brackets containing $\chi$ to give a polynomial in $\chi$.

To integrate Eq.~\eqref{eqn:fabeta}, we follow the methods of Appendix C in \cite{Detweiler:2002gi}, and use
\begin{eqnarray} \label{eqn:zeta}
\frac{1}{2 \pi} \int  \frac{d \beta}{\chi(\beta)^{n/2}} &=& 
\left<\chi^{-{n/2}}(\beta) \right> 
={}_2 F_1 \left(\frac{n}{2}, \frac{1}{2}; 1; k \right), \nonumber
\end{eqnarray}
where $(n+1)/2\in\mathbb{N}\cup \{0\}$ and ${}_2 F_1$ are hypergeometric functions. We use the recurrence
relation in Eq.~(15.2.10) of \cite{Abramowitz:Stegun},
\begin{equation}
\mathcal{F}_{p+1} (k) = \frac{p-1}{p \left(k - 1\right)} \mathcal{F}_{p-1}(k) + \frac{1 - 2p + \left(p - \frac{1}{2}\right) k}{p \left(k - 1\right)} \mathcal{F}_p(k),
\end{equation}
to reduce the number of hypergeometric functions to two, ${}_2F_1 \left(\pm\frac{1}{2}, \frac{1}{2};1;k \right)$. These in turn translate to elliptic integrals
\begin{IEEEeqnarray}{lClClCl}
\left<\chi^{-\frac{1}{2}}\right> 
&=& {}_2F_1 \left(\frac{1}{2}, \frac{1}{2};1;k \right) &=& \frac{2}{\pi} \mathcal{K}, \\
\left<\chi^{\frac{1}{2}}\right> 
&=& {}_2F_1 \left(-\frac{1}{2}, \frac{1}{2};1;k \right) &=& \frac{2}{\pi} \mathcal{E} ,
\end{IEEEeqnarray}
where
\begin{eqnarray}
\mathcal{K} &\equiv& \int_0^{\pi/2} (1 - k \sin^2 \beta)^{-1/2} d\beta,  \\
\mathcal{E} &\equiv& \int_0^{\pi/2} (1 - k \sin^2 \beta)^{1/2} d\beta,
\end{eqnarray}
are complete elliptic integrals of the first and second kinds, respectively. 


\section{Results} \label{sec:result}
\subsection{The leading term}
We have successfully carried out the integration of Eq.~\eqref{eqn:FRS} up to order $\epsilon$ or $F_{a [2]}$ ($D_a$ in the older notation of \cite{Detweiler:2002gi, Haas:2006ne}); all except the leading term as shown in  Eq.~\eqref{eqn:FRS2},
\begin{eqnarray}
F_{a [-1]}^{\ell \sing}(\xb) &=& -\frac{2\ell+1}{4 \pi} \epsilon^{-2} \left(g_{\ab \bb}+u_\ab u_\bb \right) \lim_{\Delta r \rightarrow 0} \int \frac{\Delta x^\bb}{\rho^3} \nonumber \\
&&  \qquad \times P_\ell(\cos{\alpha}) d \Omega,
\end{eqnarray}
where we have used Eq.~\eqref{eqn:FaSm1}. In our rotated coordinates without $\Delta r=0$, we have
\begin{eqnarray} \label{eqn:rhoS}
\rho^2&=&(g_{\ab \bb}+u_{\ab \bb}) \Delta x^{ab},  \\
&=& \Delta r\left( \nu_{r} \Delta r +2 \nu_{x} w_x+ 2\nu_{y} w_y \right) + 4 \sin^2  \frac{\alpha}{2} \left(\frac{\eta}k \right)\chi. \nonumber
\end{eqnarray}

All the techniques we exploited in integrating the higher orders were developed to tackle the cross term in $\rho^2$ \cite{Detweiler:2002gi, Haas:2006ne, Heffernan:2012su, Heffernan:2012vj, Heffernan:2017cad} for Schwarzschild and equatorial Kerr, $w_x \Delta r$. As $u^\theta=0$ in these scenarios, $w_x \Delta r$ is the only cross term that arises, and it promptly disappears in the higher orders. As the use of hypergeometric functions scales well to higher-order expansions (little hand holding required), we adapted these techniques and targeted the $w_x w_y$ terms in the higher orders. Unfortunately, for generic Kerr orbits, two rotations were used to bring the particle to the pole, leaving only one rotation $\beta_0$ to tackle the cross terms. We therefore do not have any more coordinate freedom to exploit in removing the cross terms $w_x \Delta r$ and $w_y \Delta r$ of Eq.~\eqref{eqn:rhoS}.

\begin{figure*}
        \centering
	\subfigure[]{\includegraphics[width=0.475\textwidth]{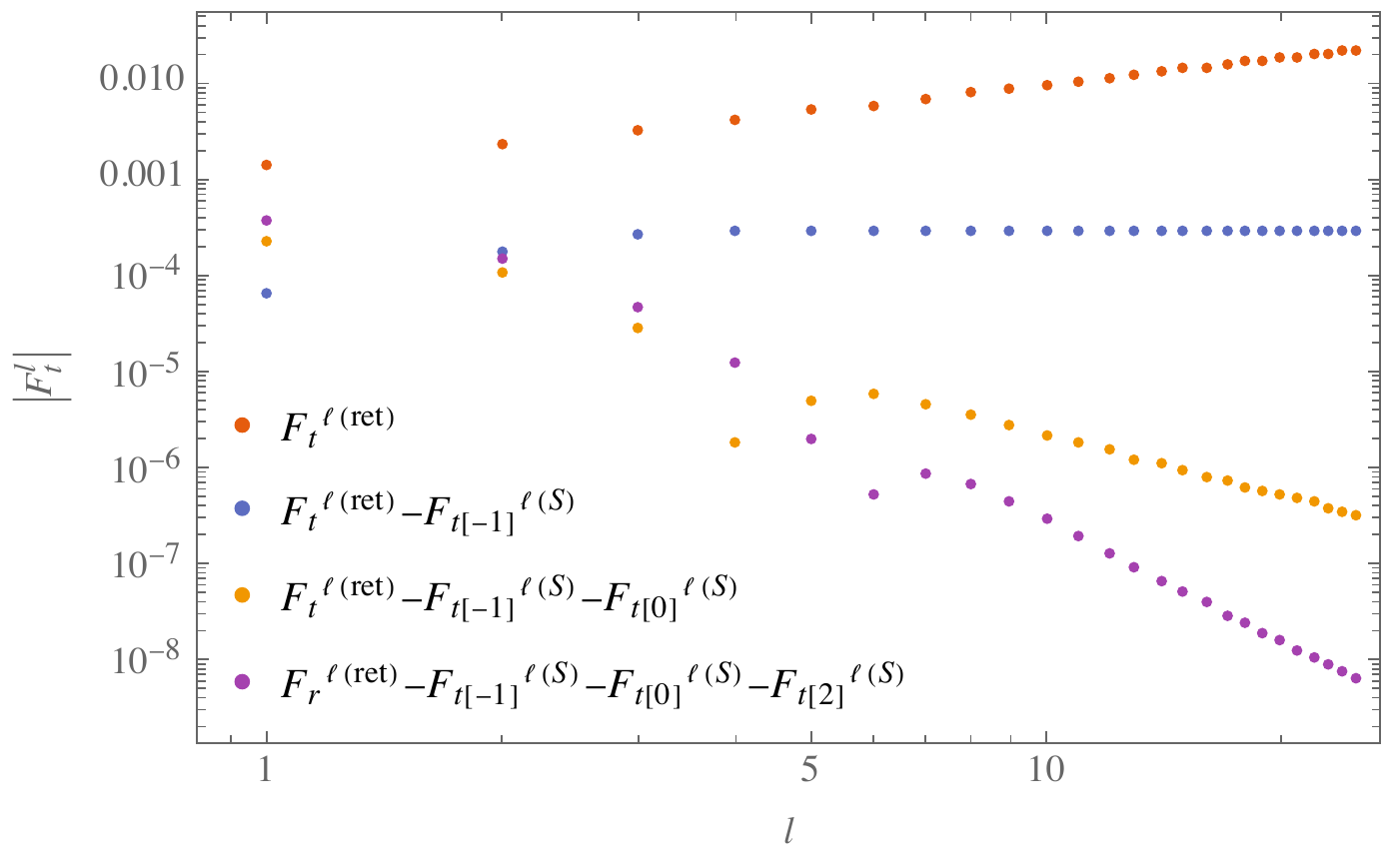}}
    \subfigure[]{\includegraphics[width=0.475\textwidth]{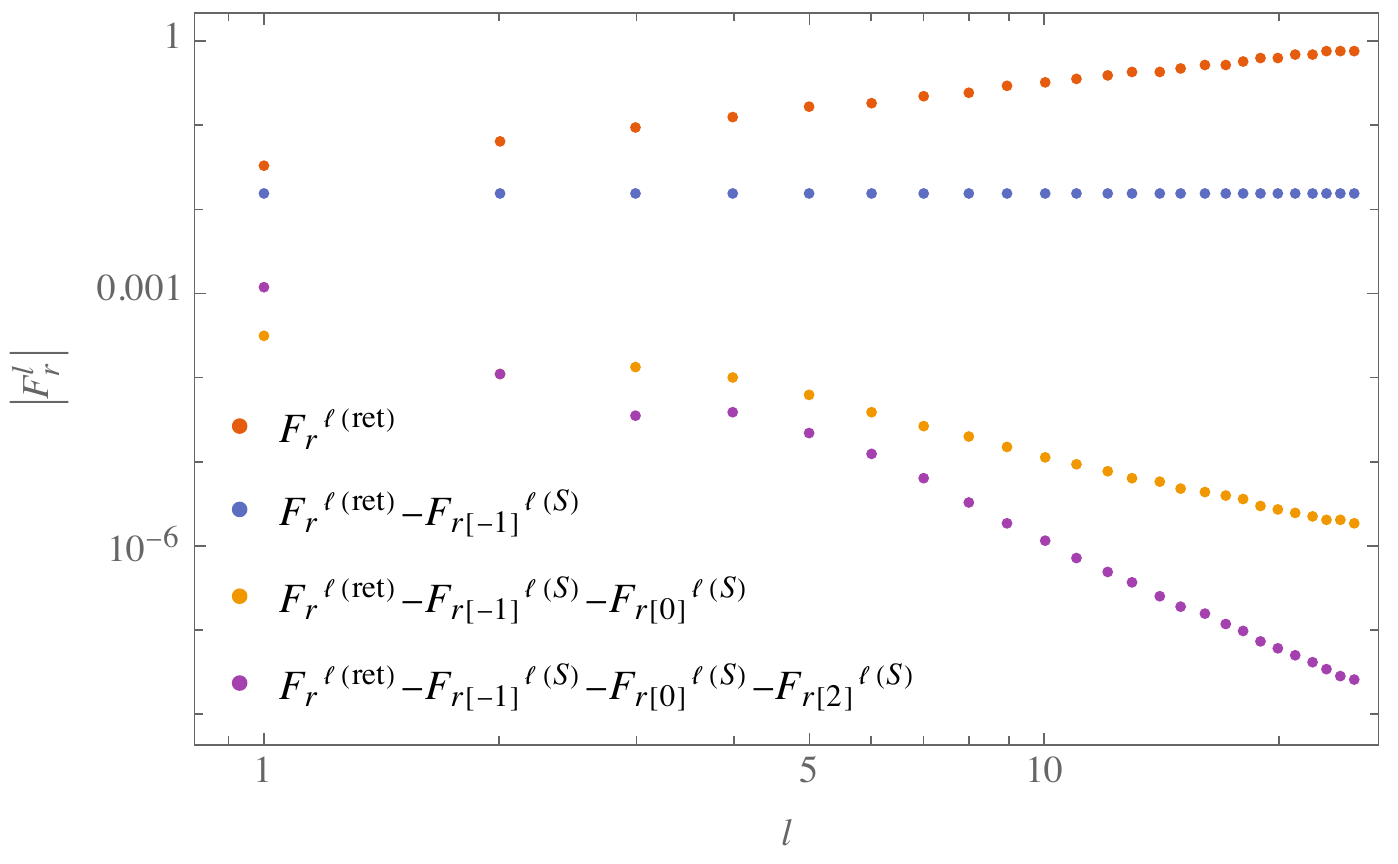}} 
    \subfigure[]{\includegraphics[width=0.475\textwidth]{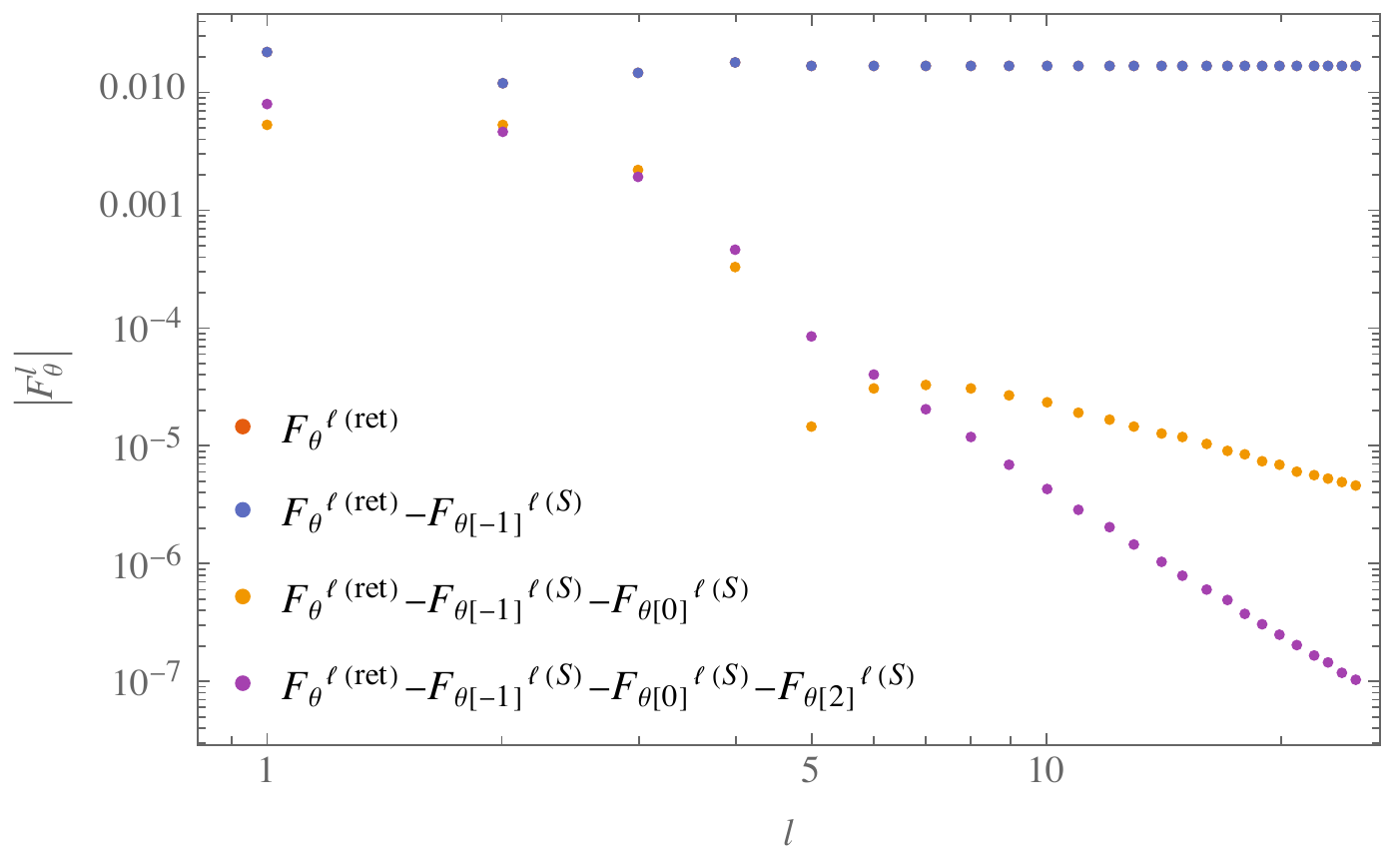}}
    \subfigure[]{\includegraphics[width=0.475\textwidth]{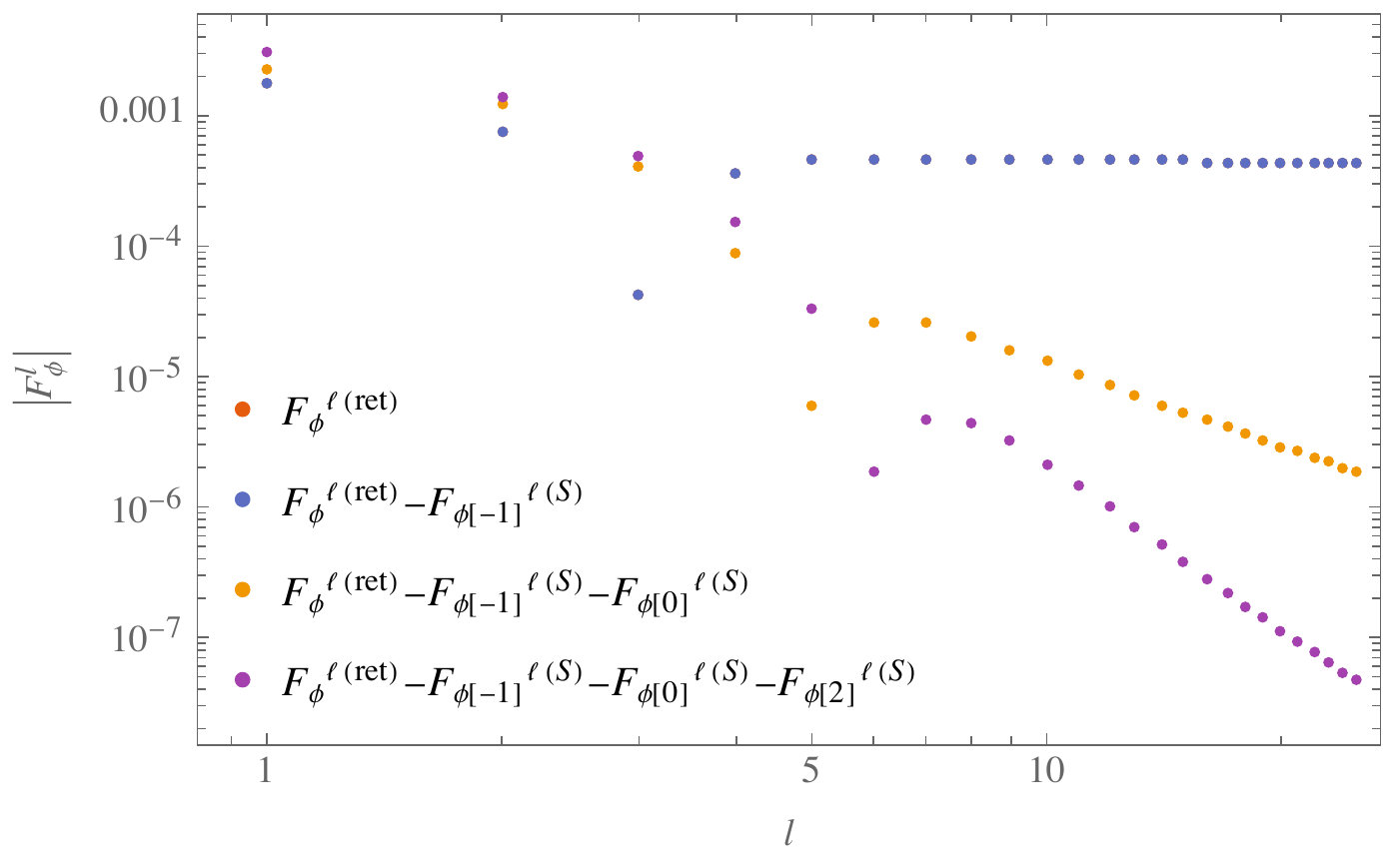}}
        \caption{Regularization of the $(t,r,\theta, \phi)$ component of the self-force for a scalar particle with initial orbit eccentricity of 0.2 and spin $a= 0.9$ with data from \cite{Nasipak:2019hxh}. In this log-log plot, one can observe by subtracting the new parameter $F_{a[2]}^\ell$ the progressing sum becomes more accurate with less $\ell$'s. We also see the expected reduction with $\ell^{-2}$ for $F_{a[0]}^\ell$ and $\ell^{-4}$ with the new $F_{a[2]}^\ell$. Fig.~(a) gives $t$ component $F_t$, Fig.~(b) gives $r$ component $F_r$, Fig.~(c) gives $\theta$ component $F_\theta$ and Fig.~(d) gives $\phi$ component $F_\phi$.}
        \label{fig:reg}
    \end{figure*}

All is not lost however, as we do not need to reinvent the wheel. Indeed, Barack and Ori tackled the leading term in their original paper \cite{Barack:2002mh} (with notation $A_a$) with more in-depth details appearing in Barack's review \cite{Barack:2009}. As previously mentioned, they used a slightly different coordinate system $(\phb+\frac{\pi}{2}, \thb, -\beta_0-\frac{\pi}{2})$ and used $\beta_0$ to set $u_{w_y} \rightarrow 0$. They proceeded by expanding $P_\ell (\sin \alpha)$ and scaling the coordinates by $\Delta r$. Due to a typo in \cite{Barack:2009}, we give the correct leading-order parameters here (in agreement with  \cite{Barack:2002mh}). Here and in all our parameters, the $\mathcal{P}_{[n]}$ term of Eq.~\eqref{eqn:Pn} is included; this means we have a factor of $1/2$ differing from \cite{Barack:2002mh} and Eq.~\eqref{eqn:Ael} gives the $\ell$ dependency of $\mathcal{A}^\ell_{[-1]}=1+2 \ell$,
\begin{eqnarray}
F_{t[-1]} &=& -\frac{u^\rb}{u^\tb}F_{r[-1]}, \quad F_{\theta [-1]} =0, \quad F_{\phi [-1]} = 0, \nonumber \\
F_{r[-1]} &=&-\frac{\sgn \Delta r}{2 V}
\sqrt{\frac{\sin ^2\thb }{g_{\phb \phb} \Delta}}
   \sqrt{V +\frac{\Delta  u_\rb^2}{\Sigma}}, \nonumber 
\end{eqnarray}
where
\begin{equation}
V = 1 + \frac{u_\thb^2}\Sigma + \frac{L^2}{g_{\phb \phb}}. \nonumber
\end{equation}


\subsection{Higher terms}

Carrying out the integration of Eq.~\eqref{eqn:fabeta} as outlined in Sec.~\ref{sec:ms:int} for the next-to-leading order, where we have included $\mathcal{P}_{[0]}=\sqrt{2}$ and Eq.~\eqref{eqn:Ael} gives $\mathcal{A}^\ell_{[0]}=1$,
\begin{equation}
 F_{a [0]}=\frac{\sqrt{k} \left[2 F_{a [0]}^{\mathcal{E}} \mathcal{E} +(k-1) F_{a [0]}^{\mathcal{K}} \mathcal{K}\right]} {12
   \pi  \eta ^{5/2} (k-1)^2},
\end{equation}
where
\begin{widetext}
\begin{eqnarray} \label{eqn:fB}
F_{a [0]}^{\mathcal{E}} &=&
(2-k) (k^2+4k-4) \sin
   \left(4 \beta _0\right) \left(b^{[0]}_{\bar{a}( 4, 1)}-b^{[0]}_{\bar{a}( 4, 3)}\right)
+(2-k) (k^2+4k-4) \cos \left(4 \beta _0\right) \left(b^{[0]}_{\bar{a}( 4, 0)}-b^{[0]}_{\bar{a}( 4,
   2)}+b^{[0]}_{\bar{a}( 4, 4)}\right) \nonumber \\
&& 
+\sin \left(2 \beta _0\right) \left[6 \eta  (k-2) (k-1) b^{[0]}_{\bar{a} (2, 1)}+2 k (k^2- k+1)
   \left(b^{[0]}_{\bar{a}( 4, 1)}+b^{[0]}_{\bar{a}( 4, 3)}\right)\right] \nonumber \\
&&
+\cos \left(2 \beta _0\right)
   \left[6 \eta  (k-2) (k-1) \left(b^{[0]}_{\bar{a}( 2, 0)}-b^{[0]}_{\bar{a}(2, 2)}\right)+4 k (k^2- k+1) \left(b^{[0]}_{\bar{a}( 4, 0)}-b^{[0]}_{\bar{a}( 4, 4)}\right)\right] \nonumber \\
&&
+k \left[k (2-k)\left(3 b^{[0]}_{\bar{a}( 4, 0)}+b^{[0]}_{\bar{a}(
   4, 2)}+3 b^{[0]}_{\bar{a} ( 4, 4)}\right)-6 \eta  (k-1) \left(b^{[0]}_{\bar{a} ( 2, 0)}+b^{[0]}_{\bar{a}
   (2, 2)}\right)\right], \nonumber \\
F_{a [0]}^{\mathcal{K}} &=& 
(k^2+16k-16) \sin \left(4
   \beta _0\right) \left(b^{[0]}_{\bar{a} ( 4, 1)}-b^{[0]}_{\bar{a} ( 4, 3)}\right)
+(k^2+16k-16) \cos
   \left(4 \beta _0\right) \left(b^{[0]}_{\bar{a} ( 4, 0)}-b^{[0]}_{\bar{a} ( 4, 2)}+b^{[0]}_{\bar{a} ( 4,
   4)}\right) \nonumber \\
&&
+k^2 \left(3 b^{[0]}_{\bar{a}( 4, 0)}+b^{[0]}_{\bar{a} ( 4, 2)}+3 b^{[0]}_{\bar{a} ( 4, 4)}\right)
-\sin
   \left(2 \beta _0\right) \left[24 \eta  (k-1) b^{[0]}_{\bar{a} ( 2, 1)}+2 k (k-2)
   \left(b^{[0]}_{\bar{a} ( 4, 1)}+b^{[0]}_{\bar{a} ( 4, 3)}\right)\right] \nonumber \\
&&
+\cos \left(2 \beta _0\right)
   \left[24 \eta  (k-1) \left(b^{[0]}_{\bar{a} ( 2, 2)}-b^{[0]}_{\bar{a} ( 2, 0)}\right)+4 k (k-2)
   \left(b^{[0]}_{\bar{a} ( 4, 4)}-b^{[0]}_{\bar{a} ( 4, 0)}\right)\right],
\end{eqnarray}
\end{widetext}
where we provide the $b^{[0]}_{\ab}$ coefficients in Appendix~\ref{sec:app:sf} below. The third order is always zero due to the odd functions arising during the integrations outlined in Sec.~\ref{sec:ms:int}, that is
\begin{equation}
F_{a[1]}=0.
\end{equation}
The fourth order, and the main result of this paper, where we have included $\mathcal{P}_{[2]}=-2 \sqrt{2}$ and Eq.~\eqref{eqn:Ael} gives
\begin{equation}
\mathcal{A}^\ell_{[2]}=\frac1{(2 \ell - 1)(2 \ell +3)},
\end{equation}
so we have
\begin{equation} \label{eqn:faD}
 F_{a [2]}=-\frac{2  F_{a [2]}^{\mathcal{E}} \mathcal{E} +(k-1)  F_{a [2]}^{\mathcal{K}} \mathcal{K}}{6720 \pi  \eta
   ^{9/2} \sqrt{k} (k-1)^4 },
\end{equation}
where $F_{a [2]}^{\mathcal{E}} $ and $F_{a [2]}^{\mathcal{K}}$ can be found in  Appendix~\ref{sec:app:sf}, however the $b^{[2]}_{\ab}$ coefficients are not provided here. Even at next-to-leading order, $F_{a [0]}$, one can see the exceptional increase in the size of the parameters compared to the Schwarzschild case \cite{Heffernan:2012su} or even the equatorial Kerr scenario \cite{Heffernan:2012vj}. As the higher terms are quite unwieldy we make them available  online in the form of an open source Mathematica \cite{Mathematica} package on Zenodo \cite{heffernan_anna_2022_6282572} and shortly on black hole perturbation Toolkit \cite{BlackHolePerturbationToolkit}. To ensure confidence in the resulting expressions, we set the spin $a=0$ and constrict orbital motion to the equatorial plane, $\theta=\pi/2$, and safely recovered the Schwarzschild counterparts in \cite{Heffernan:2012su} - this is explicitly illustrated in the readme notebook provided with the Mathematica package on Zenodo \cite{heffernan_anna_2022_6282572}. In addition, Nasipak and Evans generously shared their numerical data \cite{Nasipak:2019hxh, Nasipak:2021qfu,Nasipak:2022xjh} for generic orbits of a scalar particle in Kerr spacetime, which we use in Fig.~\ref{fig:reg} to successfully illustrate the increased $\ell$-mode convergence with the additional regularization parameters. The generation of these figures is also provided in the readme notebook on Zenodo \cite{heffernan_anna_2022_6282572}. From these one can clearly see by subtracting the new 
$F_{a[2]}^\ell$ the accuracy of the $\ell$ sum increases drastically. We also see the expected convergence with $\ell^{-2}$ for 
$F_{a[0]}^\ell$and $\ell^{-4}$ with the new 
$F_{a[2]}^\ell$.


\section{Discussion} \label{sec:discuss}
The regularization parameters produced successfully increased convergence in $\ell$ of the scalar self-force calculations. However this high-order expansion can also be employed in the effective source $m$-mode scheme \cite{Dolan:Barack:2010} as previously illustrated in the case of equatorial orbits in Kerr \cite{Heffernan:2012vj}. The expansion in Riemann normal coordinates can also be exploited for high-order tail expressions $V(x,x')$ which are required in the matched expansions method \cite{Casals:2013mpa}.

The work here also serves as the groundwork for the more physically interesting cases of an electric charge or mass on a generic orbit in Kerr spacetime. The decomposition and integration techniques are all viable, one just requires expansions for slightly more complex expressions. Another interesting avenue forward would be to consider the decomposition into tensor harmonics \cite{Wardell:2015ada}. One can also consider nongeodesic motion in Kerr spacetime building on work previously done in Schwarzschild \cite{Heffernan:2017cad}.

Lastly, it should be noted that this work comes with a caveat, in extending this work to electromagnetism or gravity, these methods are mainly adaptable to the Lorenz gauge. The current state of art in gravitational self-force for generic orbits in Kerr spacetime is calculated in the radiation gauge \cite{vandeMeent:2017bcc} where only the first two parameters, previously calculated by Barack and Ori \cite{Barack:2002mh}, are viable. Saying that, Thompson et al. were able to transform the higher-order regularization parameters in Schwarzschild from the Lorenz gauge to the Regge-Wheeler and Detweiler easy gauges \cite{Thompson:2018lgb}. This remains to be seen for the radiation gauge.

\section*{Acknowledgements}
The large expressions produced in this paper were enabled by the openly developed xAct package in Mathematica, in particular xTensor and xCoba \cite{Martin-Garcia:2007bqa, Martin-Garcia:2008yei}; the author is grateful to all those who participate in the development of this useful package. The author would also like to thank Adrian Ottewill, Barry Wardell, Bernard Whiting, Leor Barack, Sascha Husa and Marta Colleoni for insightful discussions, as well as Leo Stein in sharing his mastery of xTensor and Zachary Nasipak for sharing his data and beta-testing the regularization package on Zenodo \cite{heffernan_anna_2022_6282572}. The author gratefully acknowledges funding from the European Union's Horizon 2020 research and innovation programme under Grant Agreement No. 661705-GravityWaveWindow and the Natural Sciences and Engineering Research Council of Canada. Research at Perimeter Institute is supported in part by the Government of Canada through the Department of Innovation, Science and Economic Development Canada and in part by the Province of Ontario through the Ministry of Colleges and Universities. This work was supported by European Union FEDER funds, the Spanish Ministerio de Ciencia e Innovación, and the Spanish Agencia Estatal de Investigación grant PID2019-106416GB-I00/AEI/MCIN/10.13039/501100011033, as well as Comunitat Autònoma de les Illes Balears through the Conselleria de Fons Europeus, Universitat i Cultura and the
Direcció General de Política Universitaria i Recerca with funds from the Tourist Stay Tax Law ITS 2017-006
(PRD2018/24, PDR2020/11).
%
\appendix

\section{Regularization Parameter Coefficients for the Self-Force} \label{sec:app:sf}
We provide the coefficients that appear in Eq.~\eqref{eqn:fB}
with $t$ components,
\begin{widetext}
\begin{eqnarray} \label{eqn:Bt}
b^{[0]}_{\bar{t} ( 2, 0)} &=& 
	-\frac{1}{2} E \rb \ur-\frac{a \uth }{4 \Sigma } \left [
		a E \left(4 m \rb+\Sigma \right) \sin 2 \thb -8 L m \rb \cot \thb  
	\right], \nonumber \\
b^{[0]}_{\bar{t} ( 2, 1)} &=& 
	\frac{a m}{\Sigma ^2} \left[
		2 \rb \left(E^2-1\right)\left(a^2+\rb^2-\Sigma \right) \cos \thb -2 \rb  L^2 \cot \thb  \csc \thb 
		+2 \rb \uth{ }^2 \Sigma \left(a^2+\rb^2\right)  \cos \thb +\ur \uth \Sigma  
		\left(\Sigma -2 \rb^2\right) \sin \thb
	\right], \nonumber \\
b^{[0]}_{\bar{t} ( 2, 2)} &=&
   	\frac1{2 \Sigma ^2} \big(
		\ur \left[
			2 m  \rb^2 \left(a^2 E+2 a L +\rb^2 E\right)+E (m-r) \Sigma ^2
			-m \left(a^2 E+2 a L +3 \rb^2 E\right) \Sigma 
			\right] \nonumber \\
&& \quad
		-a \uth \left\{
			4 L m \rb \left(a^2+\rb^2\right) \cot \thb +a E \left[\Sigma ^2+2 m \rb \Sigma 
			+2 m \rb \left(a^2+\rb^2\right)\right] \cos \thb  \sin \thb 
		\right\}
	\big), \nonumber \\
b^{[0]}_{\bar{t} ( 4, 0)} &=& 
	\frac{3}{4} E \uth \Sigma  \left[
		a^2 \left(\Sigma  \uth{}^2+1\right) \sin 2 \thb  +2 \rb \ur \uth \Sigma 
	\right], \nonumber \\
b^{[0]}_{\bar{t} ( 4, 1)} &=& 
	-\frac{3 a  \cos \thb}{2 \Sigma } \left\{
		2 m \rb \left(a^2 +\rb^2-\Sigma \right)\left[
			\left(2 \Sigma E^2+\Sigma \right) \uth{}^2+1
		\right] +a E L \Sigma \left[
			2 (\Sigma -2 m \rb) \uth{}^2+1
		\right] 
	\right\}
\nonumber \\
&& \quad 
	 -3 \rb \ur \uth \left(
		E L \Sigma  \csc ^2 \thb +a m \rb 
	\right) \sin \thb , \nonumber \\
b^{[0]}_{\bar{t} ( 4, 2)} &=& 
	\frac{3}{4} \Big(a^2 
		E \uth{}^3 \left[
			2 m \rb a^2+\Sigma ^2+2 m \rb \left(\rb^2+\Sigma \right)
		\right] \sin 2 \thb  + 2 E \ur \uth{}^2 \left[
			(\rb-m) \Sigma ^2+m \left(a^2+3 \rb^2\right) \Sigma -2 m \rb^2 \left(a^2+\rb^2\right) 
		\right] \nonumber \\
&& \quad
   		+\frac{2 a^2\uth}{\Sigma^2} \big\{  
			m \rb \left[ 
				E a^2 (4 m \rb-\Sigma ) +L a \left(4 \Sigma  E^2-4 m \rb+\Sigma \right)
				+ E (4 m \rb-\Sigma ) \left(\rb^2-\Sigma \right)
			\right] \sin 2 \thb \nonumber \\
&& \qquad 
			+E\Sigma  \left[
				L^2 (\Sigma -8 m \rb) \cot \thb +\Sigma ^2 \cos \thb \sin \thb 
			\right] 
		\big\} 
		+ \frac{2 L \rb \ur }{\Sigma}\left(E L \Sigma  \csc ^2 \thb +2 a m \rb \right)
	\Big), \nonumber \\
b^{[0]}_{\bar{t} ( 4, 3)} &=& 
	\frac{3 }{2 \Sigma ^3}\Big\{
		2 \ur  \uth \Sigma  \left[
			2 m \rb^2 \left(a^2+\rb^2\right) +(m-\rb) \Sigma ^2-m \left(a^2+3 \rb^2\right) \Sigma 
		\right] \left(
			E L \Sigma  \csc ^2 \thb +a m \rb
		\right) \sin \thb \nonumber \\
&& \quad
		-2 a \Sigma  \uth{}^2 \left[
			m \rb a^2+ a E L \Sigma+m \rb \left(\rb^2-\Sigma \right)
		\right] \left[
			2 m \rb a^2+\Sigma ^2+2 m \rb \left(\rb^2+\Sigma \right)
		\right] \cos \thb   \nonumber \\
&& \quad
		+ a \big(
			4 a^2 m \rb L^2 \left(2 m \rb  -  E^2 \Sigma \right) \cos \thb 
			+2 m \rb \left(a^2+\rb^2-\Sigma \right) \left[
				  2 m \rb \left(a^2+\rb^2 - \Sigma \right) -\Sigma ^2
			\right] \cos \thb \nonumber \\
&& \qquad
			+a E L \left\{
				4  m \rb L^2\Sigma  \csc^2 \thb 
				-\Sigma ^3-2 m \rb \Sigma ^2
				+2 m \rb \Sigma \left[a^2+\rb (4 m+\rb)\right] 
   				-8 m^2 \rb^2 \left(a^2+\rb^2\right)
			\right\} \cos \thb
		\big)
	\Big\}, \nonumber \\
b^{[0]}_{\bar{t}( 4, 4)} &=& 
	\frac{3 L }{2 \Sigma ^3}\Big\{
		a^2 \uth \left[
			2 m \rb a^2+\Sigma ^2+2 m \rb \left(\rb^2+\Sigma \right)
		\right] \left( E L \Sigma  \cot \thb +a m \rb \sin 2 \thb \right) \nonumber \\
&& \quad
		+\ur \left[
			\Sigma ^2(\rb-m) +m \Sigma\left(a^2+3 \rb^2\right)  -2 m \rb^2\left(a^2+\rb^2\right) 
		\right] \left(E L \Sigma  \csc ^2 \thb +2 a m \rb\right)
	\Big\},
\end{eqnarray}
$r$ components, 
\begin{eqnarray} \label{eqn:Br}
b^{[0]}_{\bar{r} (2, 0)} &=& 
	\frac{1}{2 \Delta } \left[
		3 a^2 \ur \uth \Sigma  \cos \thb \sin \thb +\rb \left(\ur{}^2 \Sigma  -\Delta -2 \uth{}^2 \Delta  
		\Sigma \right)
	\right], \nonumber \\
b^{[0]}_{\bar{r} (2, 1)} &=& 
	\frac{1}{\Delta  \Sigma } \Big(
		\uth \csc \thb \Big\{
			a m E \left[
				a^2\left(2 \rb^2-\Sigma \right) +\rb^2 \left(2 \rb^2+\Sigma \right)
			\right] \sin ^2 \thb 
			-2 m \rb^2 L\left(a^2+\rb^2\right) 
\nonumber \\
&& \qquad
			+L \Sigma  \left[
				a^2 \left(m+\rb \right)- \left(m-\rb \right) \left(\rb^2+\Sigma \right)
			\right]
		\Big\}
		- a \ur \left[
			a L \left(\Sigma -2 m \rb \right)+2 m \rb E \left( a^2+\rb^2-\Sigma \right)
		\right] \cos \thb
	\Big), \nonumber \\
b^{[0]}_{\bar{r} (2, 2)} &=&
	\frac{1}{2 \Delta  \Sigma ^2} \Big\{
		a^2 \ur \uth \Sigma  \left[
			\Sigma ^2+2 m \rb \Sigma +2 m \rb \left(a^2+\rb^2\right)
		\right] \cos \thb \sin \thb 
		-2 m r^2 \left(a^2+\rb^2\right) \left(2 a E L-\Delta \right)
\nonumber \\
&& \quad
		+2 L^2 \left[
			m \left(2 \rb^2 - \Sigma \right)\left(a^2+\rb^2\right) +\left(m-\rb \right) \Sigma ^2
		\right] \csc^2 \thb
		-2 m \Sigma \left[
			\rb^2 \left(a^2+\rb^2\right) \ur{}^2+a E L \left(\rb^2-a^2\right)
		\right] 
\nonumber \\
&& \quad
		+\Sigma \left(m a^2+3 m \rb^2-m \Sigma +\rb \Sigma \right) \left(\ur{}^2 \Sigma -\Delta \right)
	\Big\}, 
\nonumber \\
b^{[0]}_{\bar{r} (4, 0)}&=& 
	-\frac{3 \ur \uth \Sigma ^2 }{4 \Delta } \left[
		a^2\left(\Sigma  \uth{}^2+1\right) \sin 2 \thb +2 \rb \ur \uth \Sigma 
	\right], 
\nonumber \\
b^{[0]}_{\bar{r} (4, 1)} &=& 
	\frac{3 \ur \Sigma }{2 \Delta }  \left(
		a \left\{
			4 m \rb E \uth{}^2 \left(
				a^2 + \rb^2-\Sigma 
			\right) 
			+a L \left[
				2  \uth{}^2 \left(\Sigma -2 m \rb \right)+1
			\right]
		\right\} \cos \thb
		+2 \rb L \ur \uth \Sigma  \csc \thb
	\right), \nonumber \\
b^{[0]}_{\bar{r} (4, 2)} &=& 
	\frac{3 \ur }{4 \Delta } \Big(
		a^2 \uth \left\{
			2 m \rb \left(a^2-4 E L a+\rb^2\right)
			-\uth{}^2 \Sigma^3-\left(2 m \rb \uth{}^2+1\right) \Sigma ^2
			-2 m \rb  \Sigma  \left[
				\left(a^2+\rb^2\right) \uth{}^2+1
			\right]
		\right\} \sin 2 \thb  
\nonumber \\
&& \quad
		+2 a^2 L^2 \uth \left(8 m \rb-\Sigma \right) \cot \thb 
		+2 \ur \Sigma  \left\{
			\uth{}^2 \left[
				2 m  \rb^2 \left(a^2+\rb^2\right)+\Sigma ^2 \left(m-\rb \right) 
				-m \Sigma  \left(a^2+3 \rb^2\right) 
			\right]
			-\rb L^2  \csc^2 \thb
		\right\}
	\Big), 
\nonumber \\
b^{[0]}_{\bar{r} ( 4, 3)} &=& 
	\frac{3 L \ur}{2\Delta  \Sigma } \Big(
		a^2\left\{
			2 \uth{}^2 \Sigma ^3+\Sigma^2 \left(4 m \rb \uth{}^2+1\right) 
			+2 m \rb \Sigma \left[
				2 \left(a^2+\rb^2\right) \uth{}^2+1
			\right] 
			-2 m \rb \left(a^2-2 E L a+\rb^2\right)
		\right\} \cos \thb 
\nonumber \\
&& \quad
		-2 \left\{
			\ur \uth \Sigma  \left[
				2 m \rb^2 \left(a^2+\rb^2\right) -\Sigma ^2\left(\rb-m\right) 
				-m \Sigma  \left(a^2+3\rb^2\right) 
			\right]
			+2 a^2 L^2 m \rb \cot \thb
		\right\} \csc \thb
	\Big), 
\nonumber \\
b^{[0]}_{\bar{r} (4, 4)} &=& 
	\frac{3 L^2 \ur \csc \thb}{2 \Delta  \Sigma }\Big\{
		\ur \left[
			2 m \rb^2 \left(a^2+\rb^2\right) +\Sigma ^2 \left(m-\rb\right) 
			-m  \Sigma \left(a^2+3 \rb^2\right) 
		\right] \csc \thb
\nonumber \\
&& \quad
		-a^2 \uth \left[
			\Sigma ^2+2 m \rb \Sigma +2 m \rb \left(a^2+\rb^2\right)
		\right] \cos \thb
	\Big\},
\end{eqnarray}
$\theta$ components, 
\begin{eqnarray} \label{eqn:Bth}
b^{[0]}_{\bar{\theta} (2,0)} &=&
	\frac{3}{4} \left[
		a^2 \sin 2 \thb \left(\Sigma  \uth{}^2+1\right)+2 \rb \Sigma  \ur \uth
	\right],
\nonumber \\
b^{[0]}_{\bar{\theta} (2,1)} &=& 
	\frac{L \csc \thb}{\Sigma } \left(
		\uth \cot \thb \left\{
			4 m \rb \left(a^2+\rb^2\right)-\Sigma  \left[a^2+\rb \left(4 m+\rb\right) \right]+2 \Sigma^2
		\right\}
		-r \ur \Sigma
	\right)
	-\frac{4 a m \rb E \uth \cos \thb}{\Sigma } \left(a^2+\rb^2-\Sigma \right),
\nonumber \\
b^{[0]}_{\bar{\theta} (2,2)} &=&  
	\frac{1}{4 \Sigma ^2} \Big(
		a^2 \sin 2 \thb \left\{
			2 a^2 m \rb \Sigma  \uth{}^2+4 a m \rb E L +\Sigma  \uth{}^2 \left[
				2 m \rb \left(r^2+\Sigma \right)+\Sigma ^2
			\right]
		\right\}
\nonumber \\
&&\quad 
		-2 \cot \thb \left[
			2 m r \left(a^2+\rb^2\right)-2 m \rb \Sigma +\Sigma ^2
		\right] \left(
			a^2+2 L^2 \csc ^2\thb+\rb^2+\Sigma 
		\right)
\nonumber \\
&&\quad 
		+2 \Sigma  \ur \uth \left[
			m \Sigma  \left(a^2+3\rb^2\right)-2 m \rb^2 \left(a^2+\rb^2\right)+\Sigma ^2 (\rb-m)
		\right]
	\Big),
\nonumber \\
b^{[0]}_{\bar{\theta} (4,0)} &=&  
	-\frac{3 \Sigma  }{4}\left(
		\Sigma \uth{}^2+1
	\right) \left[
		a^2 \sin 2 \thb \left( \Sigma  \uth{}^2+1\right)+2 \rb \Sigma  \ur \uth
	\right],
\nonumber\\
b^{[0]}_{\bar{\theta} (4,1)} &=&  
	3 a \uth \cos \thb \left(
		\Sigma \uth{}^2+1
	\right) \left[
		2 a^2 m \rb E +a L \left(\Sigma -2 m \rb \right)+2  m r E \left(\rb^2-\Sigma \right)
	\right]
	+\frac{3}{2}  \rb \Sigma L  \ur \csc \thb \left(2 \Sigma  \uth{}^2+1\right),
\nonumber \\
b^{[0]}_{\bar{\theta} (4,2)} &=&  
	-\frac{3}{4 \Sigma } \Big[
		2 a^2 L^2 \cot \thb \left[
			\Sigma ^2 \uth{}^2-4 m \left(2 \rb \Sigma \uth{}^2+\rb\right)
		\right]
		+ a^2 \sin 2 \thb \Big(
			2 a^2 m \rb \left(\Sigma ^2 \uth{}^4-1\right)
			+4 a m \rb  E L \left(2 \Sigma \uth{}^2+1\right)
\nonumber \\
&& \qquad 
			+\left(\Sigma  \uth{}^2+1\right) \left\{
				2 m \rb \left[
					\rb^2 \left(\Sigma \uth{}^2-1\right)
					+\Sigma +\Sigma ^2 \uth{}^2
				\right]
				+\Sigma ^2 \left(\Sigma  \uth{}^2+1\right)
			\right\}
		\Big)
\nonumber \\
&& \quad
		+2 \Sigma  \ur \uth \left\{
			\left(\Sigma  \uth{}^2+1\right) \left[
				m \Sigma  \left(a^2+3 \rb^2\right)
				-2 m \rb^2 \left(a^2+\rb^2\right)
				+\Sigma ^2 \left(\rb-m\right)
			\right]
			+L^2 \rb \Sigma  \csc ^2\thb
		\right\}
	\Big],
\nonumber \\
b^{[0]}_{\bar{\theta} (4,3)} &=&  
	\frac{3 L }{2 \Sigma } \Big\{
		4 a^3 m \rb E L \uth \cos \thb  +2 a^2 \uth \left(
			\Sigma  \cos \thb \left\{
				\uth{}^2 \left[
					2 m \rb \left(a^2+\rb^2+\Sigma \right)+\Sigma ^2
				\right]
				+2 m \rb+\Sigma 
			\right\}
			-2  m \rb L^2 \cot \thb \csc \thb
		\right)
\nonumber \\
&& \quad
		-\ur \csc \thb \left(
			2 \Sigma  \uth{}^2+1 
		\right) \left[
			\Sigma ^2 \left(m-\rb\right)-m \Sigma  \left(a^2+3 \rb^2\right)
			+2 m \rb^2 \left(a^2+\rb^2\right)
		\right]
	\Big\},
 \\
b^{[0]}_{\bar{\theta} (4,4)} &=&  
	\frac{3 L^2 \uth \csc \thb}{2 \Sigma } \Big\{
		\ur \csc \thb \left[
			\Sigma ^2 \left(m-\rb\right)-m \Sigma  \left(a^2+3 \rb^2\right)
			+2 m \rb^2 \left(a^2+\rb^2\right)
		\right]
		-a^2 \uth \cos \thb \left[
			2 m \rb \left(a^2+\rb^2 + \Sigma\right) +\Sigma ^2
		\right]
	\Big\},
\nonumber 
\end{eqnarray}
and $\phi$ components,
\begin{eqnarray} \label{eqn:Bph}
b^{[0]}_{\bar{\phi} (2,0)} &=& 
	\frac{1}{2 \Sigma } \left\{
		2 a  m \rb E \uth \sin 2\thb \left(a^2+\rb^2-\Sigma \right)
		+L \uth \cot \thb \left[
			-4 m \rb \left(a^2+\rb^2\right)+\Sigma  \left(a^2+4 m \rb +\rb^2\right)-3 \Sigma ^2
		\right]
		+L \rb \Sigma \ur
	\right\}
\nonumber \\
b^{[0]}_{\bar{\phi} (2,1)} &=& 
	\frac{1}{\Sigma ^2} \Big(
		L^2 \cot \thb \csc \thb \left[
			4 m \rb \left(a^2+\rb^2\right)-4 m \rb \Sigma +\Sigma ^2
		\right]
		+\Sigma  \ur \uth \sin \thb \left[
			m \left(2 \rb^2-\Sigma \right) \left(a^2+\rb^2-\Sigma \right)-\rb \Sigma^2
		\right]
\nonumber \\
&& \quad
		-4 a m \rb E L  \cos \thb \left(a^2+\rb^2-\Sigma \right)
		+\cos\thb \left\{
			2 m \rb \left(a^2+\rb^2\right) \left(a^2+\rb^2-\Sigma \right)
			-\Sigma  \uth{}^2 \left[
				2 m \rb \left(a^2+\rb^2\right)^2+\Delta  \Sigma^2
			\right]
		\right\}
	\Big)
\nonumber \\
b^{[0]}_{\bar{\phi} (2,2)} &=& 
	\frac{ L}{2 \Sigma ^2} \left\{
		3 \ur \left[
			m \Sigma  \left(a^2+3 \rb^2\right)
			-2 m \rb^2 \left(a^2+\rb^2\right)
			+\Sigma ^2 \left(\rb-m\right)
		\right]
		+\uth \cot \thb \left[
			6 m \rb \left(a^2+\rb^2\right)^2
			+3 \Delta \Sigma^2-\Sigma^3 
		\right]
	\right\},
\nonumber \\
b^{[0]}_{\bar{\phi} (4,0)} &=& 
	-\frac{3}{4} \Sigma  L  \uth \left[
		a^2 \sin 2 \thb \left(\Sigma  \uth{}^2+1\right)+2 \rb \Sigma  \ur \uth
	\right],
\nonumber \\
b^{[0]}_{\bar{\phi} (4,1)} &=& 
	\frac{3}{2} \Big(
		\cos \thb \left\{
			a^2 L^2 \left[
				2 \uth{}^2 \left(\Sigma -2 m \rb \right)+1
			\right]
			+\left(\frac{1}{\Sigma }+\uth{}^2 \right) \left(a^2+\rb^2-\Sigma \right)
   			\left[
				a^2 \left(2 m \rb+\Sigma \right)+2 m \rb^3+\rb \Sigma  \left(\rb-2 m\right)
			\right]
		\right\}
\nonumber \\
&& \quad
		+ 4 a m \rb E L \uth{}^2 \cos \thb \left(a^2+\rb^2-\Sigma \right)
		+r \ur \uth \sin \thb \left[
			2 m \rb \left(a^2+\rb^2\right)+\Delta  \Sigma +2  \Sigma L^2  \csc^2\thb
		\right]
	\Big),
\nonumber \\
b^{[0]}_{\bar{\phi} (4,2)} &=& 
		-\frac{3}{2 \Sigma ^2} \Big[
			4 a m \rb E \uth \left\{
				a^2 L^2 \Sigma  \sin 2 \thb+\sin \thb\cos \thb \left(a^2+\rb^2-\Sigma \right) 
				\left[
					a^2 \left(2 m \rb+\Sigma \right)+2 m \rb^3+\rb \Sigma  \left(\rb-2 m\right)
				\right]
			\right\}
\nonumber \\
&& \quad
			+L \Big(
				a^2 \uth \Big\{
					\sin \thb \cos \thb \left[
						\left(a^2+\rb^2\right) \left(
							2 m \rb \Sigma ^2 \uth{}^2+\Sigma ^2 -8 m^2 \rb^2 
						\right)
						-4 \Delta  m \rb \Sigma +\Sigma ^3 \left(2 m \rb \uth{}^2+1\right)
						+\Sigma ^4 \uth{}^2
					\right]
\nonumber \\
&& \qquad \quad
					+L^2 \Sigma  \cot \thb\left(\Sigma -8 m \rb\right)
				\Big\}
				+\Sigma  \ur \Big\{
					m \left(a^2+\rb^2-\Sigma \right) \left[
						\rb^2 \left(2-2 \Sigma  \uth{}^2\right)
						+\Sigma ^2 \uth{}^2
					\right]
\nonumber \\
&& \qquad \quad
					+r \Sigma  \left(a^2+\rb^2+\Sigma ^2 \uth{}^2 + L^2 \csc ^2\thb \right)
				\Big\}
			\Big)
		\Big],
\nonumber \\
b^{[0]}_{\bar{\phi} (4,3)} &=& 
	\frac{3}{2 \Sigma ^3} \Big(
		4 a m \rb  E L \cos \thb \left\{
			a^2 L^2 \Sigma 
			+\left(a^2+\rb^2-\Sigma \right) \left[
				a^2 \left(2 m \rb+\Sigma\right)+2 m \rb^3+\rb \Sigma  \left(\rb-2 m\right)
			\right]
		\right\}
		-4 a^2 m \rb \Sigma  L^4 \cot  \thb  \csc  \thb 
\nonumber \\
&& \quad
		-\Sigma  \ur \uth \sin \thb \left[
			\Sigma ^2 \left(m-\rb\right) -m \Sigma  \left(a^2+3 \rb^2\right)
			+2 m \rb^2 \left(a^2+\rb^2\right)
		\right] \left[
			2 m \rb \left(a^2+\rb^2\right)+\Delta  \Sigma +2 L^2 \Sigma  \csc ^2 \thb 
		\right]
\nonumber \\
&& \quad
		+a^2 L^2 \cos  \thb  \Big\{
			8 m^2 \rb^2 \left(\Sigma -\rb^2\right)
			-2 a^2 m \rb \left[
				4 m \rb+\Sigma  \left(3-2 \Sigma  \uth{}^2\right)
			\right]
			+2 m r \Sigma  \left[
				\rb^2 \left(2 \Sigma \uth{}^2-3\right)+\Sigma +2 \Sigma ^2 \uth{}^2
			\right]
\nonumber \\
&& \qquad
			+\Sigma ^3 \left(2 \Sigma \uth{}^2+1\right)
		\Big\} +\cos  \thb \left(a^2+\rb^2-\Sigma \right) \left[
			a^2 \left(2 m \rb+\Sigma \right)+2 m \rb^3+\rb \Sigma  \left(\rb-2m\right)
		\right]
\nonumber \\
&& \quad
		\times \left\{
			2 m \rb \Sigma  \left[
				\uth{}^2 \left(a^2+\rb^2\right)+1
			\right]
			-2 m \rb \left(a^2+\rb^2\right)+\Sigma ^2 \left(2 m \rb \uth{}^2+1\right)
			+\Sigma ^3 \uth{}^2
		\right\}
	\Big),
\nonumber \\
b^{[0]}_{\bar{\phi} (4,4)} &=& 
	\frac{3 L}{2 \Sigma ^3} \Big\{
		\ur\left[
			\Sigma ^2 \left(m-\rb \right)-m \Sigma  \left(a^2+3 \rb^2\right)
			+2 m \rb^2 \left(a^2+\rb^2\right)
		\right] \left[
			2 m \rb \left(a^2+\rb^2\right)+\Delta  \Sigma +L^2 \Sigma  \csc ^2 \thb
		\right]
\nonumber \\
&& \quad
	 	-a^2 \uth \left[
			2 a^2 m \rb+2 m \rb \left(r^2+\Sigma \right)+\Sigma ^2
		\right] \left[
			m \rb \left(a^2+\rb^2\right) \sin 2\thb+\Delta  \Sigma  \sin \thb \cos \thb 
			+L^2 \Sigma  \cot \thb
		\right]
	\Big\}.
\end{eqnarray}
The fourth-order coefficients of Eq.~\eqref{eqn:faD} are
\begin{eqnarray}
F_{a [2]}^{\mathcal{E}} &=&
	3 \cos 10 \beta _0 (8k^8+36 k^7+169 k^6+1254 k^5-37555 k^4+136064 k^3-198272 k^2+131072 k-32768)
\nonumber \\
&& 
	\times \left( 
		b^{[2]}_{\bar{a} (10, 0)}- b^{[2]}_{\bar{a} (10, 2)}+ b^{[2]}_{\bar{a} (10, 4)}- b^{[2]}_{\bar{a} (10, 6)}
		+ b^{[2]}_{\bar{a} (10, 8)}- b^{[2]}_{\bar{a} (10, 10)}
	\right)
\nonumber \\
&& 
	-\cos 8 \beta _0 \Big[
		14 \eta (k-1) (8 k^6+41 k^5+279 k^4-6784 k^3+18752 k^2-18432 k+6144)   
\nonumber \\
&& \quad
		\times \left( 
			b^{[2]}_{\bar{a} (8, 0)}- b^{[2]}_{\bar{a} (8, 2)}+ b^{[2]}_{\bar{a} (8, 4)}- b^{[2]}_{\bar{a} (8, 6)}
			+ b^{[2]}_{\bar{a} (8, 8)}
		\right)
\nonumber \\
&& \quad
		+24 k (k-2) (k^2+4 k-4) (2 k^4-k^3+33 k^2-64 k+32) 
\nonumber \\
&& \quad
		\times \left(
			5  b^{[2]}_{\bar{a} (10, 0)}-3  b^{[2]}_{\bar{a} (10, 2)}+ b^{[2]}_{\bar{a} (10, 4)}
			+ b^{[2]}_{\bar{a} (10, 6)}-3  b^{[2]}_{\bar{a} (10, 8)}+5  b^{[2]}_{\bar{a} (10, 10)}
		\right)
	\Big]
\nonumber \\
&& 
	+\cos 6 \beta_0 \Big[
		(24 k^6-20 k^5-21 k^4-46 k^3+343 k^2-384 k+128) 
\nonumber \\
&& \quad
		\times \left(
			45  b^{[2]}_{\bar{a} (10, 0)}-13  b^{[2]}_{\bar{a} (10, 2)}-3  b^{[2]}_{\bar{a} (10, 4)}
			+3  b^{[2]}_{\bar{a} (10, 6)}+13  b^{[2]}_{\bar{a} (10, 8)}-45  b^{[2]}_{\bar{a} (10, 10)}
		\right) k^2
\nonumber \\
&& \quad
		+56 \eta (k-2) (k-1) (8 k^4+29 k^3+99 k^2-256 k+128)   \left(
			2  b^{[2]}_{\bar{a} (8, 0)}- b^{[2]}_{\bar{a} (8, 2)}+ b^{[2]}_{\bar{a} (8, 6)}-2  b^{[2]}_{\bar{a} (8, 8)}	
		\right) k
\nonumber \\
&& \quad
		+560\eta ^2 (k-1)^2 (k^4+7 k^3-135 k^2+256 k-128)  \left( 
			b^{[2]}_{\bar{a} (6, 0)}- b^{[2]}_{\bar{a} (6, 2)}+ b^{[2]}_{\bar{a} (6, 4)}- b^{[2]}_{\bar{a} (6, 6)}
		\right)
	\Big]
\nonumber \\
&& 
	-8 \cos 4 \beta _0 \Big[
		12 (k-2) (k+1) (2 k-1) (k^2-k+1)
\nonumber \\
&& \quad
		\times \left(
			15  b^{[2]}_{\bar{a} (10, 0)}- b^{[2]}_{\bar{a} (10, 2)}- b^{[2]}_{\bar{a} (10, 4)}
			- b^{[2]}_{\bar{a} (10, 6)}- b^{[2]}_{\bar{a} (10, 8)}+15  b^{[2]}_{\bar{a} (10, 10)}
		\right) k^3
\nonumber \\
&& \quad
		+7 \eta (k-1) (8 k^4-7 k^3-9 k^2+32 k-16)   \left(
			7 b^{[2]}_{\bar{a} (8, 0)}- b^{[2]}_{\bar{a} (8, 2)}- b^{[2]}_{\bar{a} (8, 4)}
			- b^{[2]}_{\bar{a} (8, 6)}+7 b^{[2]}_{\bar{a} (8, 8)}
		\right) k^2
\nonumber \\
&& \quad
		+140\eta ^2 (k-2) (k-1)^2 (k^2+4 k-4)  \left(
			3  b^{[2]}_{\bar{a} (6, 0)}- b^{[2]}_{\bar{a} (6, 2)}- b^{[2]}_{\bar{a} (6, 4)}
			+3  b^{[2]}_{\bar{a} (6, 6)}
		\right) k
\nonumber \\
&& \quad
		+420\eta ^3 (k-1)^3 (k^2-16 k+16)  \left( 
			b^{[2]}_{\bar{a} (4, 0)}- b^{[2]}_{\bar{a} (4, 2)}+ b^{[2]}_{\bar{a} (4, 4)}
		\right)
	\Big]
\nonumber \\
&& 
	+\cos 2 \beta _0 \Big[
		6 (8 k^4-28 k^3+33 k^2-10 k+5) 
\nonumber \\
&& \quad
		\times \left(
			105  b^{[2]}_{\bar{a} (10, 0)}+  7b^{[2]}_{\bar{a} (10, 2)} +b^{[2]}_{\bar{a} (10, 4)}
			- b^{[2]}_{\bar{a} (10, 6)}- 7b^{[2]}_{\bar{a} (10, 8)}- 105  b^{[2]}_{\bar{a} (10, 10)} 
		\right) k^4
\nonumber \\
&& \quad
		+56 \eta (k-2) (k-1) (8 k^2-3k+3)  \left(
			14  b^{[2]}_{\bar{a} (8, 0)}+ b^{[2]}_{\bar{a} (8, 2)}- b^{[2]}_{\bar{a} (8, 6)}
			-14  b^{[2]}_{\bar{a} (8, 8)}
		\right) k^3
\nonumber \\
&& \quad
		+560 \eta ^2 (k-1)^2 (k^2- k+1) \left(
			15  b^{[2]}_{\bar{a} (6, 0)}+ b^{[2]}_{\bar{a} (6, 2)}- b^{[2]}_{\bar{a} (6, 4)}
			-15  b^{[2]}_{\bar{a} (6, 6)}
		\right) k^2
\nonumber \\
&& \quad
		+13440 \eta ^3  (k-2) (k-1)^3\left( 
			b^{[2]}_{\bar{a} (4, 0)}- b^{[2]}_{\bar{a} (4, 4)}
		\right) k
		+26880  \eta ^4 (k-1)^4\left( 
			b^{[2]}_{\bar{a} (2, 2)}- b^{[2]}_{\bar{a} (2, 0)}
		\right)
	\Big]
\nonumber \\
&& 
	-2 k^2 \Big[
		4 (k-2) (6 k^2-11 k+11) \left(
			63 b^{[2]}_{\bar{a} (10, 0)}+7 b^{[2]}_{\bar{a}(10, 2)}+3 b^{[2]}_{\bar{a}(10, 4)}
			+3  b^{[2]}_{\bar{a} (10, 6)}+7  b^{[2]}_{\bar{a} (10, 8)}+63  b^{[2]}_{\bar{a} (10, 10)}
		\right) k^3
\nonumber \\
&& \quad
		+7 \eta (k-1) (8 k^2-23 k+23) \left(
			35  b^{[2]}_{\bar{a} (8, 0)}+5  b^{[2]}_{\bar{a} (8, 2)}+3  b^{[2]}_{\bar{a} (8, 4)}
			+5  b^{[2]}_{\bar{a} (8, 6)}+35  b^{[2]}_{\bar{a} (8, 8)}
		\right) k^2
\nonumber \\
&& \quad
		+560 \eta ^2 (k-2) (k-1)^2 \left(
			5  b^{[2]}_{\bar{a} (6, 0)}+ b^{[2]}_{\bar{a} (6, 2)}+ b^{[2]}_{\bar{a} (6, 4)}
			+5  b^{[2]}_{\bar{a} (6, 6)}
		\right) k
\nonumber \\
&& \quad
		+1680 \eta ^3 (k-1)^3  \left(
			3  b^{[2]}_{\bar{a} (4,0)}+ b^{[2]}_{\bar{a} (4,2)}+3  b^{[2]}_{\bar{a} (4,4)}
		\right)
	\Big] 
\nonumber \\
&& 
	+\sin 2 \beta _0 \Big[
		6 (8 k^4-28 k^3+33 k^2-10 k+5) \left(
			21 b^{[2]}_{\bar{a} (10, 1)}+7  b^{[2]}_{\bar{a} (10, 3)}+5  b^{[2]}_{\bar{a} (10, 5)}
			+7  b^{[2]}_{\bar{a} (10, 7)}+21  b^{[2]}_{\bar{a} (10, 9)}
		\right) k^4
\nonumber \\
&& \quad
		+28 \eta (k-2) (k-1) (8 k^2-3 k+3)   \left(
			7 b^{[2]}_{\bar{a} (8, 1)}+3  b^{[2]}_{\bar{a} (8, 3)}+ 3 b^{[2]}_{\bar{a} (8, 5)}
			+7  b^{[2]}_{\bar{a} (8, 7)}
		\right) k^3
\nonumber \\
&& \quad
		+560 \eta ^2 (k-1)^2 (k^2-k+1)  \left(
			5  b^{[2]}_{\bar{a} (6, 1)}+3 b^{[2]}_{\bar{a} (6, 3)}+5  b^{[2]}_{\bar{a} (6, 5)}
		\right) k^2
\nonumber \\
&& \quad
		+6720\eta ^3 (k-2) (k-1)^3  \left( 
			b^{[2]}_{\bar{a} (4, 1)}+ b^{[2]}_{\bar{a} (4, 3)}
		\right) k
		-26880 \eta ^4  (k-1)^4  b^{[2]}_{\bar{a} (2, 1)}
	\Big] 
\nonumber \\
&& 
	-4 \sin 4 \beta _0 \Big[
		24 (k-2) (k+1) (2 k-1) (k^2-k+1) \left(
			6  b^{[2]}_{\bar{a} (10, 1)}+ b^{[2]}_{\bar{a} (10, 3)}- b^{[2]}_{\bar{a} (10, 7)}
			-6  b^{[2]}_{\bar{a} (10, 9)}
		\right) k^3
\nonumber \\
&& \quad
		+7 \eta (k-1) (8 k^4-7 k^3-9 k^2+32 k-16)   \left(
			7  b^{[2]}_{\bar{a} (8, 1)}+ b^{[2]}_{\bar{a} (8, 3)}- b^{[2]}_{\bar{a} (8, 5)}
			-7  b^{[2]}_{\bar{a} (8, 7)}
		\right) k^2
\nonumber \\
&& \quad
		+560 \eta ^2  (k-2) (k-1)^2 (k^2+4 k-4) \left( 
			b^{[2]}_{\bar{a} (6, 1)}-b^{[2]}_{\bar{a} (6, 5)}
		\right) k
\nonumber \\
&& \quad
		+840 \eta ^3 (k-1)^3 (k^2-16 k+16) \left( 
			b^{[2]}_{\bar{a} (4, 1)}-b^{[2]}_{\bar{a} (4, 3)}
		\right)
	\Big] 
\nonumber \\
&& 
	+ \sin 6 \beta_0 \Big[
		(24 k^6-20 k^5-21 k^4-46 k^3+343 k^2-384 k+128) 
\nonumber \\
&& \quad
		\times \left(
			27  b^{[2]}_{\bar{a} (10, 1)}-3  b^{[2]}_{\bar{a} (10, 3)}-5  b^{[2]}_{\bar{a} (10, 5)}
			-3  b^{[2]}_{\bar{a} (10, 7)}+27  b^{[2]}_{\bar{a} (10, 9)}
		\right) k^2
\nonumber \\
&& \quad
		+28 \eta(k-2) (k-1) (8 k^4+29 k^3+99 k^2-256 k+128)   \left(
			3  b^{[2]}_{\bar{a} (8, 1)}- b^{[2]}_{\bar{a} (8, 3)}- b^{[2]}_{\bar{a} (8, 5)}
			+3  b^{[2]}_{\bar{a} (8, 7)}
		\right) k
\nonumber \\
&& \quad
		+560 \eta ^2  (k-1)^2 (k^4+7 k^3-135 k^2+256 k-128) \left( 
			b^{[2]}_{\bar{a} (6, 1)}- b^{[2]}_{\bar{a} (6, 3)}+ b^{[2]}_{\bar{a} (6, 5)}
		\right)
	\Big]
\nonumber \\
&& 
	-\sin 8 \beta _0 \Big[
		48 k (k-2) (k^2+4k-4) (2 k^4-k^3+33 k^2-64 k+32) \left(
			2  b^{[2]}_{\bar{a} (10, 1)}- b^{[2]}_{\bar{a} (10, 3)}+ b^{[2]}_{\bar{a} (10, 7)}
			-2  b^{[2]}_{\bar{a} (10, 9)}
		\right)
\nonumber \\
&& \quad
		+14 \eta (k-1) (8 k^6+41 k^5+279 k^4-6784 k^3+18752 k^2-18432 k+6144)  \left( 
			b^{[2]}_{\bar{a} (8, 1)}- b^{[2]}_{\bar{a} (8, 3)}+ b^{[2]}_{\bar{a} (8, 5)}- b^{[2]}_{\bar{a} (8, 7)}
		\right)
	\Big] 
\nonumber \\
&& 
	+3 \sin \left(10 \beta _0\right) (8k^8+36 k^7+169 k^6+1254 k^5-37555 k^4+136064 k^3-198272 k^2+131072 k-32768)
\nonumber \\
&& 
	\times \left( 
		b^{[2]}_{\bar{a} (10, 1)}- b^{[2]}_{\bar{a} (10, 3)}+ b^{[2]}_{\bar{a} (10, 5)}- b^{[2]}_{\bar{a} (10, 7)}
		+ b^{[2]}_{\bar{a} (10, 9)}
	\right) ,
\nonumber \\
F_{a [2]}^{\mathcal{K}} &=&
	3 \cos 10 \beta _0  (k-2) (8 k^6+59 k^5+325k^4+32000 k^3-97920 k^2+98304 k-32768) 
\nonumber \\
&&
	\times \left(
		-b^{[2]}_{\bar{a} (10, 0)}+b^{[2]}_{\bar{a} (10, 2)}-b^{[2]}_{\bar{a} (10, 4)}+b^{[2]}_{\bar{a} (10, 6)}
		-b^{[2]}_{\bar{a} (10, 8)}+b^{[2]}_{\bar{a} (10, 10)}
	\right)
\nonumber \\
&&
	+\cos 8 \beta _0 \Big[
		112 \eta (k-2) (k-1) (k^4+8 k^3+760 k^2-1536 k+768) \left(
			b^{[2]}_{\bar{a} (8, 0)}-b^{[2]}_{\bar{a} (8, 2)}+b^{[2]}_{\bar{a} (8, 4)}-b^{[2]}_{\bar{a} (8, 6)}
			+b^{[2]}_{\bar{a} (8, 8)}
		\right) 
\nonumber \\
&& \quad
		+6 k (8 k^6+19 k^5+45 k^4+1920 k^3-6080 k^2+6144 k-2048)
\nonumber \\
&& \quad
		\times \left(
			5 b^{[2]}_{\bar{a} (10, 0)}-3 b^{[2]}_{\bar{a} (10, 2)}+b^{[2]}_{\bar{a} (10, 4)}
			+b^{[2]}_{\bar{a} (10, 6)}-3 b^{[2]}_{\bar{a} (10, 8)}+5 b^{[2]}_{\bar{a} (10, 10)}
		\right)
	\Big]
\nonumber \\
&& 
	-\cos 6 \beta _0 \Big[
		(k-2) (24 k^4+49 k^3+79 k^2-256 k+128) 
\nonumber \\
&& \quad
		\times \left(
			45 b^{[2]}_{\bar{a} (10, 0)}-13 b^{[2]}_{\bar{a} (10, 2)}-3 b^{[2]}_{\bar{a} (10, 4)}
			+3 b^{[2]}_{\bar{a} (10, 6)}+13 b^{[2]}_{\bar{a} (10, 8)}-45 b^{[2]}_{\bar{a} (10, 10)}
		\right) k^2
\nonumber \\
&& \quad
		+224 \eta  (k-1) (2k^4+5 k^3+123 k^2-256 k+128)  \left(
			2 b^{[2]}_{\bar{a} (8, 0)}-b^{[2]}_{\bar{a} (8, 2)}+b^{[2]}_{\bar{a} (8, 6)}-2 b^{[2]}_{\bar{a} (8, 8)}	
		\right) k
\nonumber \\
&& \quad
		+560 \eta ^2 (k-2) (k-1)^2 (k^2+128 k-128) \left(
			b^{[2]}_{\bar{a} (6, 0)}-b^{[2]}_{\bar{a} (6, 2)}+b^{[2]}_{\bar{a} (6, 4)}
			-b^{[2]}_{\bar{a} (6, 6)}
		\right)
	\Big]
\nonumber \\
&& 
	+8 \cos 4 \beta_0 \Big[
		3 (8 k^4-13 k^3-3 k^2+32 k-16) \left(
			15 b^{[2]}_{\bar{a} (10, 0)}-b^{[2]}_{\bar{a} (10, 2)}-b^{[2]}_{\bar{a} (10, 4)}
			-b^{[2]}_{\bar{a} (10, 6)}-b^{[2]}_{\bar{a} (10, 8)}+15 b^{[2]}_{\bar{a} (10, 10)}
		\right) k^3 
\nonumber \\
&& \quad
		+56 \eta (k-1) \left(k^3-6 k+4\right)   \left(
			7 b^{[2]}_{\bar{a} (8, 0)}-b^{[2]}_{\bar{a} (8, 2)}-b^{[2]}_{\bar{a} (8, 4)}-b^{[2]}_{\bar{a} (8, 6)}
			+7 b^{[2]}_{\bar{a} (8, 8)}
		\right) k^2 
\nonumber \\
&& \quad
		+140 \eta ^2  (k-1)^2 (k^2+16 k-16) \left(
			3 b^{[2]}_{\bar{a} (6, 0)}-b^{[2]}_{\bar{a} (6, 2)}-b^{[2]}_{\bar{a} (6, 4)}+3 b^{[2]}_{\bar{a} (6, 6)}	
		\right) k
\nonumber \\
&& \quad
		+6720 \eta ^3 (k-2) (k-1)^3  \left(
			b^{[2]}_{\bar{a} (4, 0)}-b^{[2]}_{\bar{a} (4, 2)}+b^{[2]}_{\bar{a} (4, 4)}
		\right)
	\Big]
\nonumber \\
&& 
	-\cos 2 \beta _0 \Big[
		6 (k-2) (8 k^2-5 k+5) \left(
			105 b^{[2]}_{\bar{a} (10, 0)}+7 b^{[2]}_{\bar{a} (10, 2)}+b^{[2]}_{\bar{a} (10, 4)}
			-b^{[2]}_{\bar{a} (10, 6)}-7 b^{[2]}_{\bar{a} (10, 8)}-105 b^{[2]}_{\bar{a} (10, 10)}
		\right) k^4
\nonumber \\
&& \quad
		+224 \eta (k-1) (2 k^2-3 k+3)   \left(
			14 b^{[2]}_{\bar{a} (8, 0)}+b^{[2]}_{\bar{a} (8, 2)}-b^{[2]}_{\bar{a} (8, 6)}
			-14 b^{[2]}_{\bar{a} (8, 8)}
		\right) k^3
\nonumber \\
&& \quad
		+560 \eta ^2 (k-2) (k-1)^2  \left(
			15 b^{[2]}_{\bar{a} (6, 0)}+b^{[2]}_{\bar{a} (6, 2)}-b^{[2]}_{\bar{a} (6, 4)}
			-15 b^{[2]}_{\bar{a} (6, 6)}
		\right) k^2 
\nonumber \\
&& \quad
		+53760 \eta ^3 (k-1)^3  \left(
			b^{[2]}_{\bar{a} (4, 0)}-b^{[2]}_{\bar{a} (4, 4)}
		\right) k
		+26880 \eta ^4 (k-2) (k-1)^3  \left(
			b^{[2]}_{\bar{a} (2, 0)}-b^{[2]}_{\bar{a} (2, 2)}
		\right)
	\Big]
\nonumber \\
&& 
	+2 k \Big[
		(24 k^2-71 k+71) \left(
			63 b^{[2]}_{\bar{a} (10, 0)}+7 b^{[2]}_{\bar{a} (10, 2)}+3 b^{[2]}_{\bar{a} (10, 4)}
			+3 b^{[2]}_{\bar{a} (10, 6)}+7 b^{[2]}_{\bar{a} (10, 8)}+63b^{[2]}_{\bar{a} (10, 10)}
		\right) k^4
\nonumber \\
&& \quad
		+56 \eta (k-2) (k-1) \left(
			35 b^{[2]}_{\bar{a} (8, 0)}+5 b^{[2]}_{\bar{a} (8, 2)}+3 b^{[2]}_{\bar{a} (8, 4)}
			+5 b^{[2]}_{\bar{a} (8, 6)}+35 b^{[2]}_{\bar{a} (8, 8)}
		\right) k^3
\nonumber \\
&& \quad
		+560 \eta ^2 (k-1)^2  \left(
			5 b^{[2]}_{\bar{a} (6, 0)}+b^{[2]}_{\bar{a} (6, 2)}+b^{[2]}_{\bar{a} (6, 4)}+5 b^{[2]}_{\bar{a} (6, 6)}
		\right) k^2
		+13440 \eta ^4 (k-1)^3 \left(
			b^{[2]}_{\bar{a} (2, 0)}+b^{[2]}_{\bar{a} (2, 2)}
		\right)
	\Big]
\nonumber \\
&& 
	- \sin 2 \beta _0 \Big[
		6 (k-2) (8 k^2-5 k+5) \left(
			21 b^{[2]}_{\bar{a} (10, 1)}+7 b^{[2]}_{\bar{a} (10, 3)}+5 b^{[2]}_{\bar{a} (10, 5)}
			+7 b^{[2]}_{\bar{a} (10, 7)}+21 b^{[2]}_{\bar{a} (10, 9)}
		\right) k^4
\nonumber \\
&& \quad
		+112 \eta (k-1) (2 k^2-3 k+3) \left(
			7 b^{[2]}_{\bar{a} (8, 1)}+3 b^{[2]}_{\bar{a} (8, 3)}+3 b^{[2]}_{\bar{a} (8, 5)}
			+7 b^{[2]}_{\bar{a} (8, 7)}
		\right) k^3
\nonumber \\
&& \quad
		+560 \eta ^2 (k-2) (k-1)^2  \left(
			5 b^{[2]}_{\bar{a} (6, 1)}+3 b^{[2]}_{\bar{a} (6, 3)}+5 b^{[2]}_{\bar{a} (6, 5)}
		\right) k^2
		+26880 \eta ^3  (k-1)^3 \left(
			b^{[2]}_{\bar{a} (4, 1)}+b^{[2]}_{\bar{a} (4, 3)}
		\right) k
\nonumber \\
&& \quad
		+26880 \eta ^4  (k-2) (k-1)^3 b^{[2]}_{\bar{a} (2, 1)}
	\Big]
\nonumber \\
&& 
	+8  \sin 4 \beta _0 \Big[
		3 (8 k^4-13 k^3-3 k^2+32 k-16) \left(
			6 b^{[2]}_{\bar{a} (10, 1)}+b^{[2]}_{\bar{a} (10, 3)}-b^{[2]}_{\bar{a} (10, 7)}
			-6 b^{[2]}_{\bar{a} (10, 9)}
		\right) k^3
\nonumber \\
&& \quad
		+28  \eta (k-1) \left(k^3-6 k+4\right)  \left(
			7 b^{[2]}_{\bar{a} (8, 1)}+b^{[2]}_{\bar{a} (8, 3)}-b^{[2]}_{\bar{a} (8, 5)}-7 b^{[2]}_{\bar{a} (8, 7)}
		\right) k^2 
\nonumber \\
&& \quad
		+280 \eta ^2 (k-1)^2 (k^2+16 k-16)  \left(
			b^{[2]}_{\bar{a} (6, 1)}-b^{[2]}_{\bar{a} (6, 5)}
		\right) k
		+6720 \eta ^3 (k-2) (k-1)^3  \left(
			b^{[2]}_{\bar{a} (4, 1)}-b^{[2]}_{\bar{a} (4, 3)}
		\right)
	\Big]
\nonumber \\
&& 
	-\sin 6 \beta_0 \Big[
		560 \eta ^2 (k-2) (k-1)^2 (k^2+128 k-128)  \left(
			b^{[2]}_{\bar{a} (6, 1)}-b^{[2]}_{\bar{a} (6, 3)}+b^{[2]}_{\bar{a} (6, 5)}
		\right)
\nonumber \\
&& \quad
		+112\eta (k-1) (2 k^4+5 k^3+123 k^2-256 k+128)  \left(
			3 b^{[2]}_{\bar{a} (8, 1)}-b^{[2]}_{\bar{a} (8, 3)}-b^{[2]}_{\bar{a} (8, 5)}+3 b^{[2]}_{\bar{a} (8, 7)}
		\right) k
\nonumber \\
&& \quad
		+ (k-2) (24 k^4+49 k^3+79 k^2-256 k+128) \left(
			27 b^{[2]}_{\bar{a} (10, 1)}-3 b^{[2]}_{\bar{a} (10, 3)}-5 b^{[2]}_{\bar{a} (10, 5)}
			-3 b^{[2]}_{\bar{a} (10, 7)}+27 b^{[2]}_{\bar{a} (10, 9)}
		\right) k^2
	\Big]
\nonumber \\
&& 
	+\sin 8 \beta_0 \Big[
		112 \eta (k-2) (k-1) (k^4+8 k^3+760 k^2-1536 k+768)   \left(
			b^{[2]}_{\bar{a} (8, 1)}-b^{[2]}_{\bar{a} (8, 3)}+b^{[2]}_{\bar{a} (8, 5)}-b^{[2]}_{\bar{a} (8, 7)}
		\right)
\nonumber \\
&& \quad
		+12 k (8 k^6+19 k^5+45 k^4+1920 k^3-6080 k^2+6144 k-2048) \left(
			2 b^{[2]}_{\bar{a} (10, 1)}-b^{[2]}_{\bar{a} (10, 3)}+b^{[2]}_{\bar{a} (10, 7)}
			-2 b^{[2]}_{\bar{a} (10, 9)}
		\right)
	\Big] 
\nonumber \\
&& 
	-3  \sin 10 \beta _0 (k-2) (8 k^6+59 k^5+325 k^4+32000 k^3-97920 k^2+98304 k-32768) 
\nonumber \\
&& 
	\times \left(
		b^{[2]}_{\bar{a} (10, 1)}-b^{[2]}_{\bar{a} (10, 3)}+b^{[2]}_{\bar{a} (10, 5)}-b^{[2]}_{\bar{a} (10, 7)}
		+b^{[2]}_{\bar{a} (10, 9)}
	\right),
\end{eqnarray}
where, due to their large format, we have made the higher-order $b^{[2]}$ coefficients available online via a mathematica package on Zenodo \cite{heffernan_anna_2022_6282572} and shortly on the black hole perturbation toolkit \cite{BlackHolePerturbationToolkit}.
\end{widetext}
\bibliographystyle{apsrev4-1}
\bibliography{references}

\end{document}